\documentclass{aa}
\usepackage{longtable}
\usepackage{graphicx}
\usepackage{booktabs}
\usepackage{array}
\usepackage{supertabular}
\usepackage{siunitx}
\usepackage{subfig}
\usepackage{txfonts}
\usepackage{hyperref}
\newcommand{\placefigsimsDMtwoobjects}{
\begin{figure*}
    \includegraphics[width=\hsize, height=7cm]{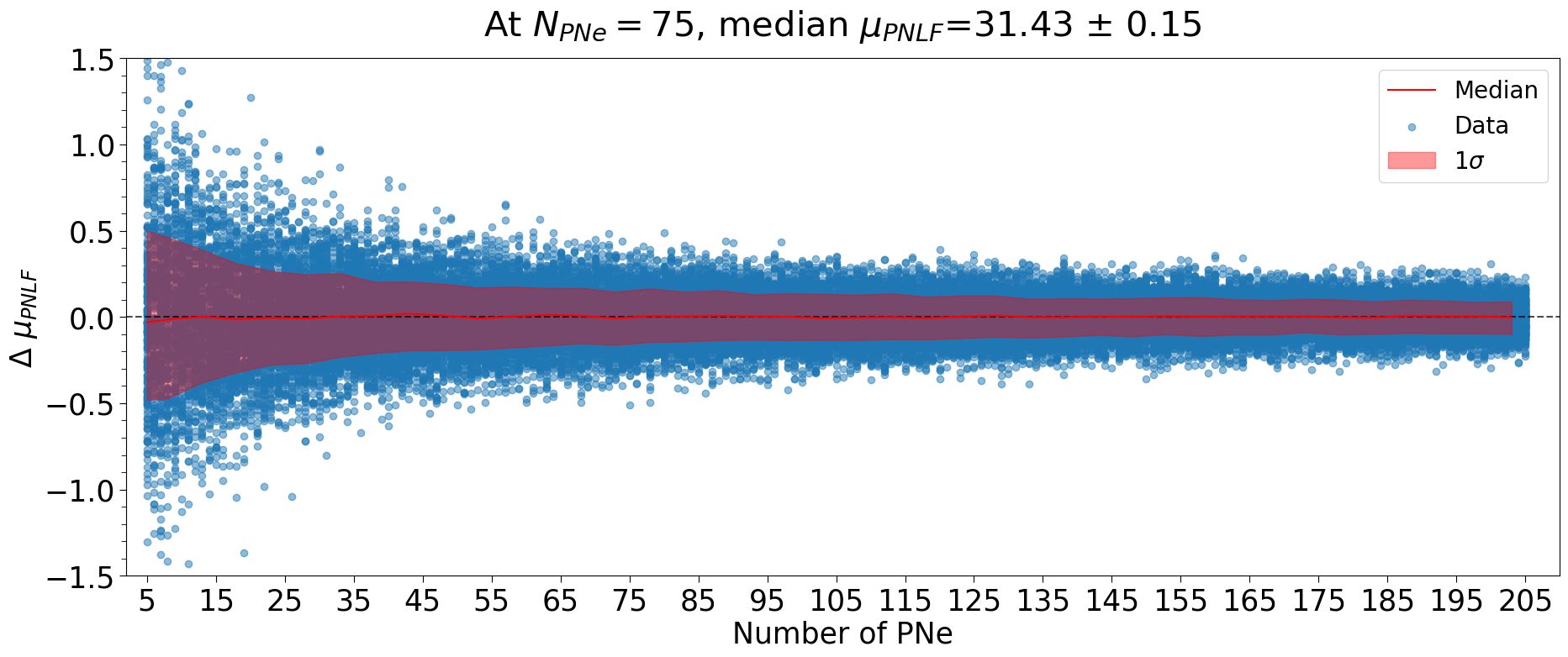}
    \includegraphics[width=\hsize, height=7cm]{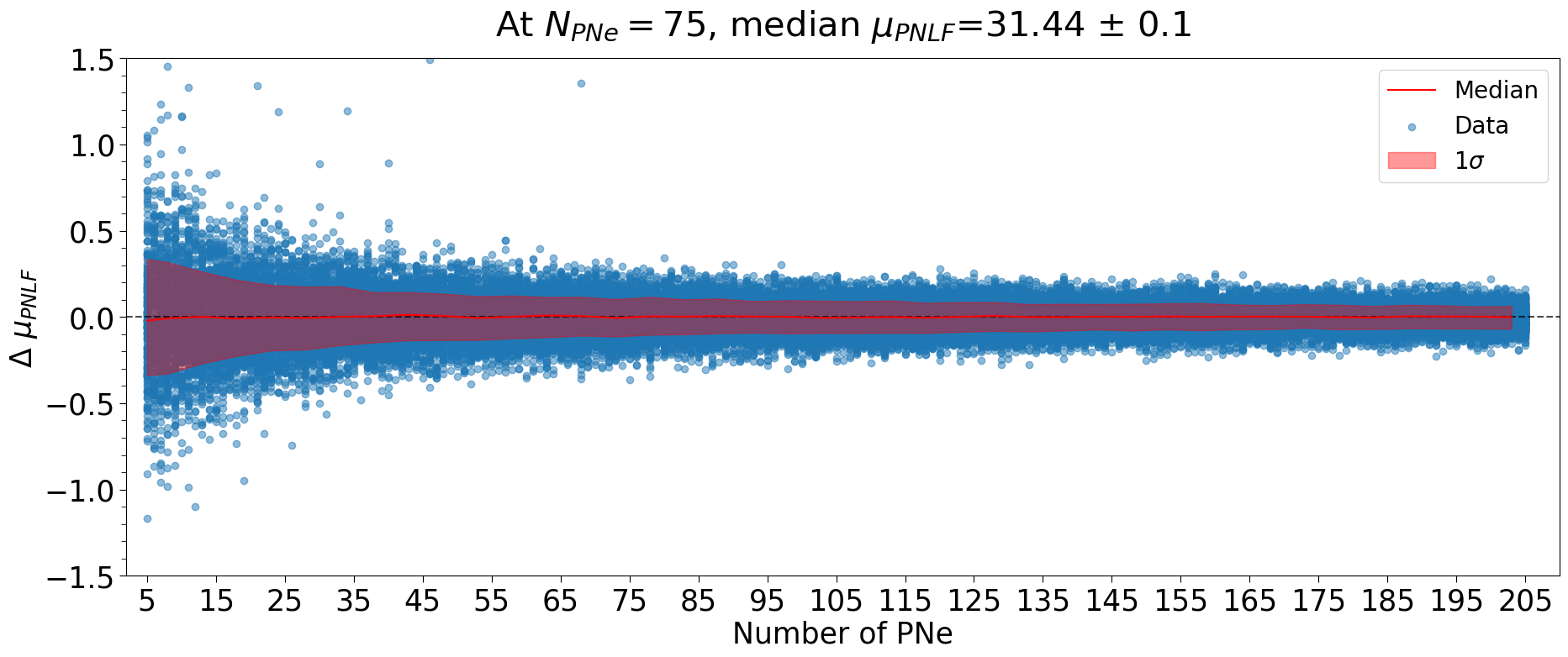}
    \caption{Results of our simulations assessing the accuracy of our PNLF modelling, while holding to the standard form of the PNLF with a fixed $c_{2}=0.307$ value and optimising the distance modulus $\mu_\mathrm{{PNLF}}$. Each blue dot shows the difference $\Delta \mu_\mathrm{PNLF} = \mu_\mathrm{PNLF \, best-fit} - \mu_\mathrm{PNLF \, in}$ between the returned best-fit $\mu_\mathrm{{PNLF}}$ value and the input value when fitting synthetic PNLFs drawn from a parent incompleteness-corrected PNLF model with a varying $N_\mathrm{{PNe \, sim}}$ number of PNe. In this case, the parent distribution assumes a distance modulus $\mu_\mathrm{PNLF \, in} = 31.45$ and the same completeness function as observed in FCC~193 (top panel) and FCC~147 (bottom panel), which reaches about 2.5 and 1.5 magnitude below the apparent magnitude of the PNLF cutoff, respectively (see FCC147 FCC193 PNLF in Fig~\ref{fig:all_PNLF}). As $N_\mathrm{{PNe \, sim}}$ increases, the scatter in the $\Delta \mu_\mathrm{PNLF}$ values decreases while overall returning an unbiased answer, even at low $N_\mathrm{{PNe \, sim}}$, as traced also by the binned median red line. The red shaded region traces the 1$\sigma$ confidence limit.}
    \label{fig:simulations_dM_FCC193_FCC147}
\end{figure*}
}

\newcommand{\placefigsimsDM}{
\begin{figure*}
    \includegraphics[width=\hsize]{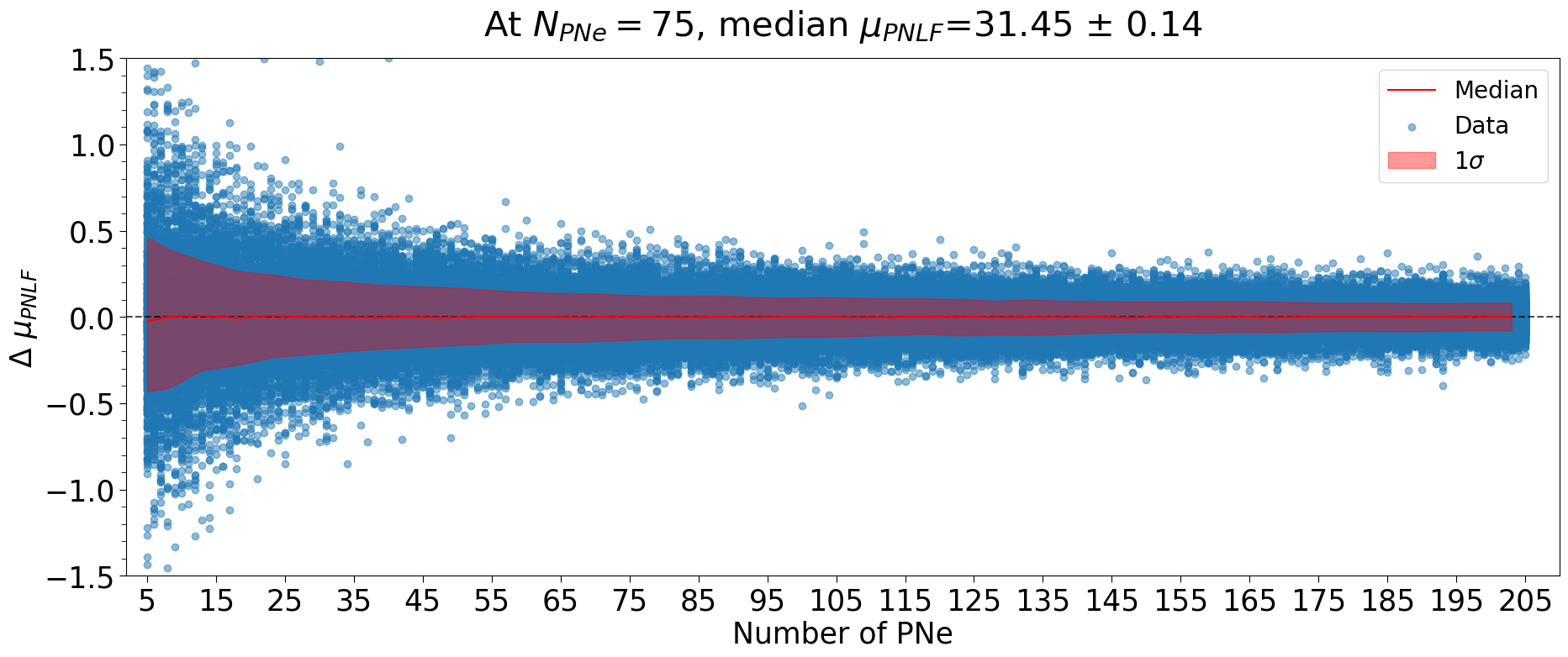}
    \caption{Same as in Fig.~\ref{fig:simulations_dM_FCC193_FCC147}, but now accounting in the simulations for the whole range of different completeness profiles observed across our sample.}
    \label{fig:simulations_dM}
\end{figure*}
}

\newcommand{\placefigerrorcomparison}{
\begin{figure}
    \centering
    \includegraphics[width=\hsize]{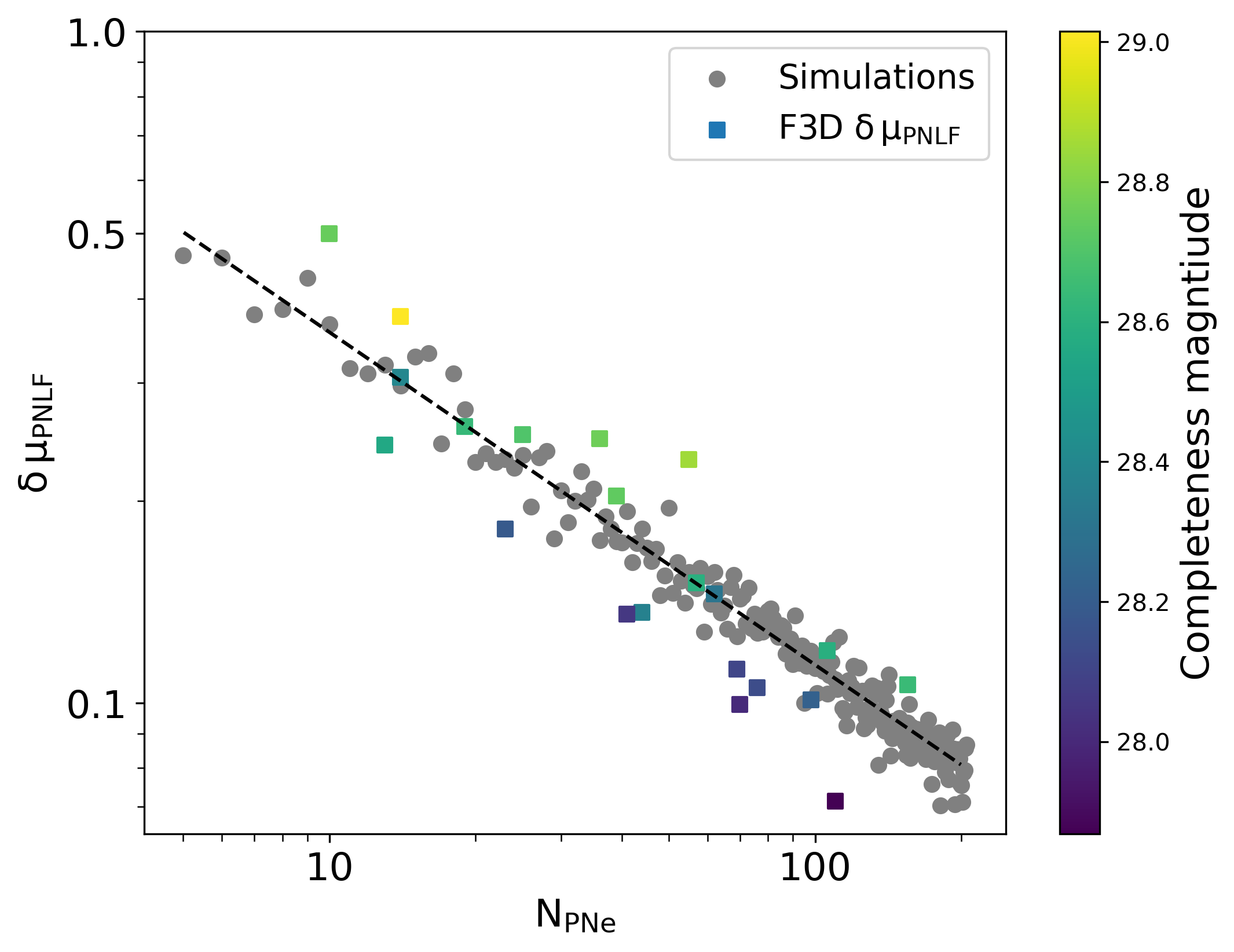}
    \caption{Comparison between the average scatter in the recovered distance magnitude values from our simulations when accounting for the entire range of completeness profiles observed across our sample (grey circles) and the errors on the individual distance magnitude measurements in our sample galaxies (squares, colour-coded by completeness magnitude).}
    \label{fig:sim_data_error_comparison}
\end{figure}
}

\newcommand{\placefigvelocitycomparison}{
\begin{figure}
    \centering
    \includegraphics[width=\hsize]{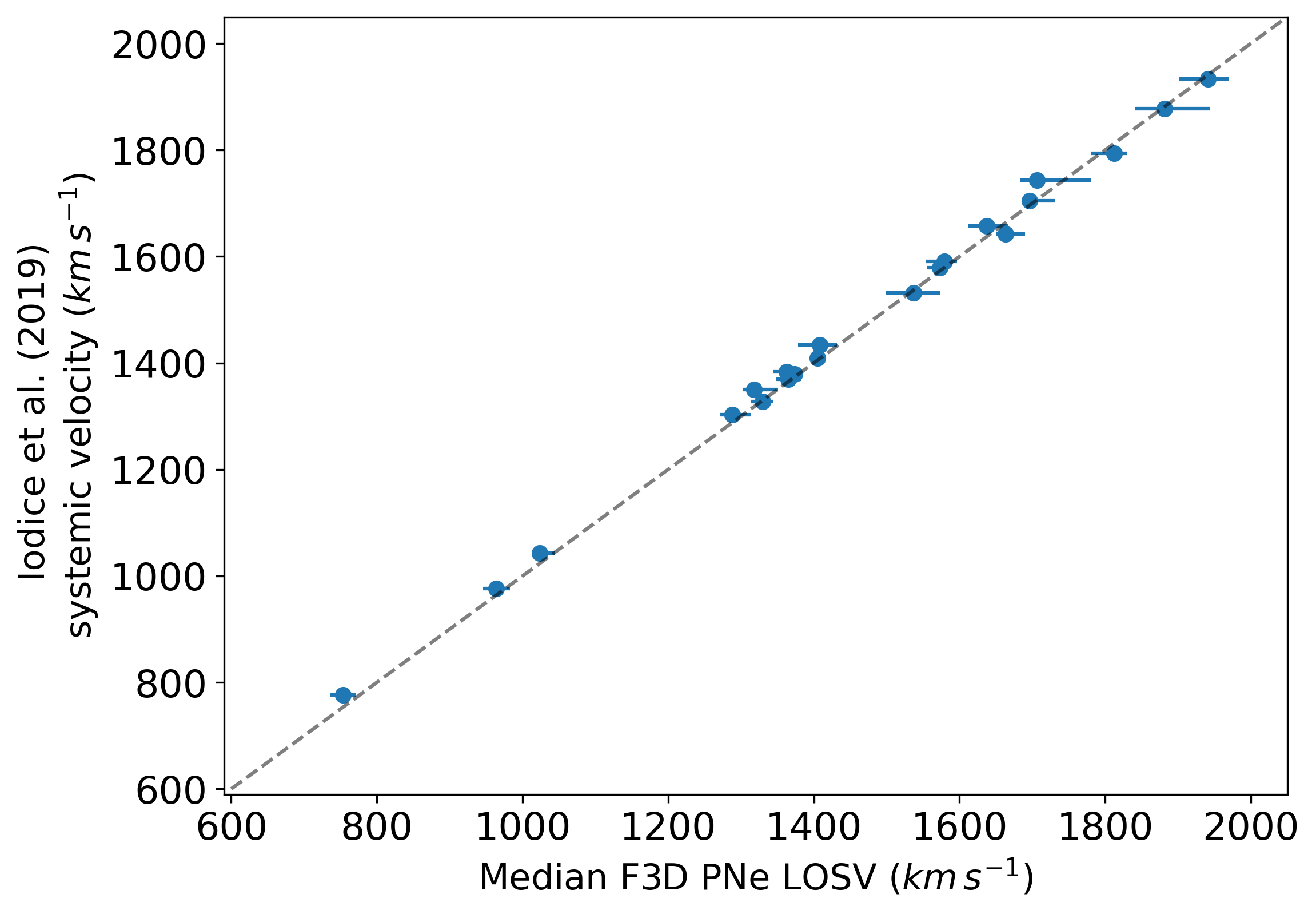}
    \caption{Comparison between the systemic velocities of our sample galaxies as measured by \citet{iodice_fornax3d_2019} from the stellar kinematic maps and as derived from the median PNe LOSV observed in the MUSE central pointing. The diagonal blue line indicates the one to one relation. Small residual offset may reflect an asymmetric sampling of the PNe around the centre of the galaxies.}
    \label{fig:velocity_comparison}
\end{figure}
}

\newcommand{\placefigSBFminPNLF}{
\begin{figure}
    \includegraphics[width=\hsize]{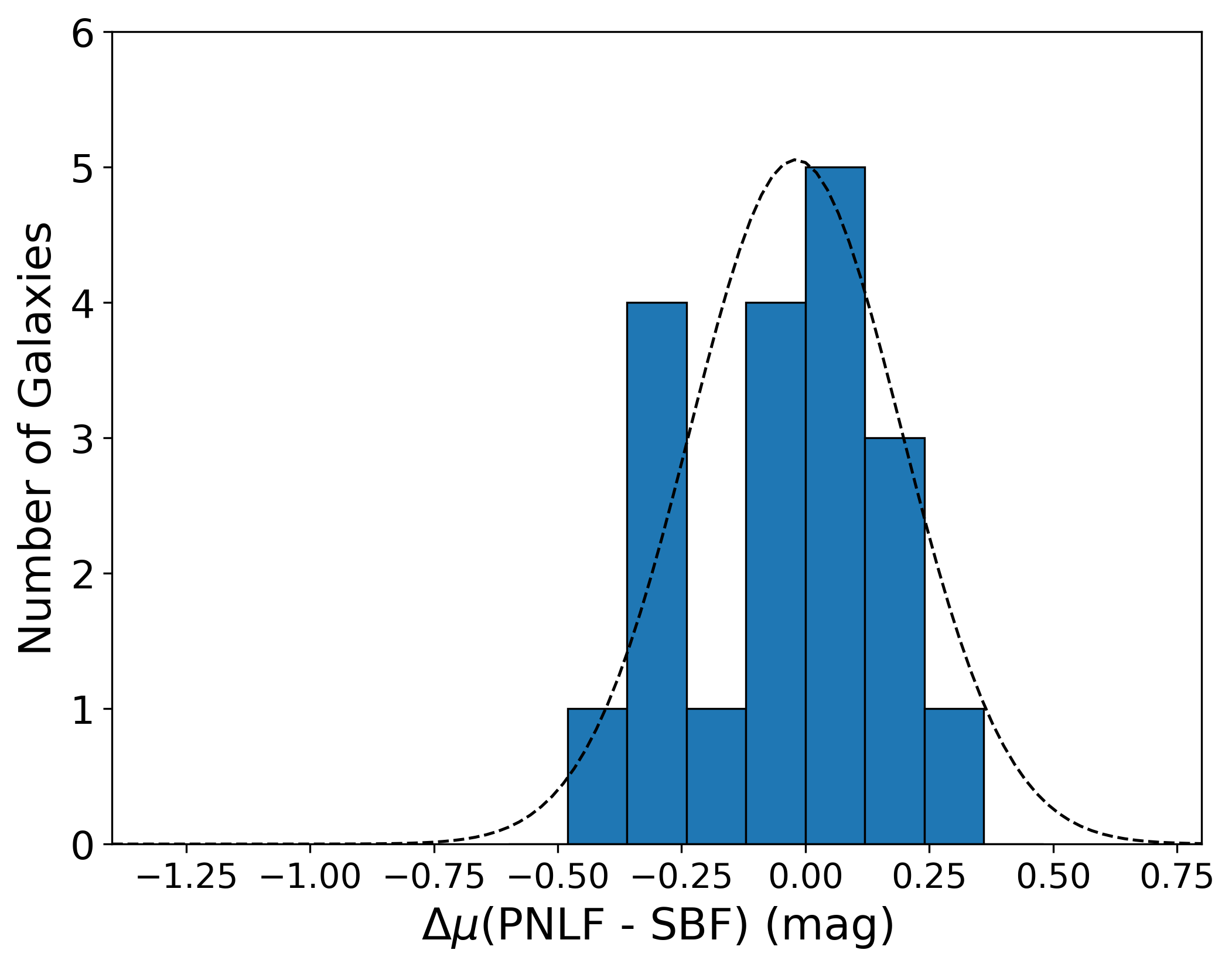}
    \caption{Distribution of the difference ($\Delta \mu$) between our PNLF values of the distance modulus and those derived using SBF by \citet[][]{blakeslee_acs_2009}. The $\Delta \mu$ distribution has a measured scatter of 0.21 mag around a median of 0.003 mag, as also shown by the reference Gaussian distribution with such median and standard deviation values. We adopt 0.12 mag bins to match Fig. 6 from \citet{ciardullo_planetary_2012}.}
    \label{fig:PNLF_minus_SBF}
\end{figure}
}

\newcommand{\placefigTensionSBFminPNLF}{
\begin{figure}
    \includegraphics[width=\hsize]{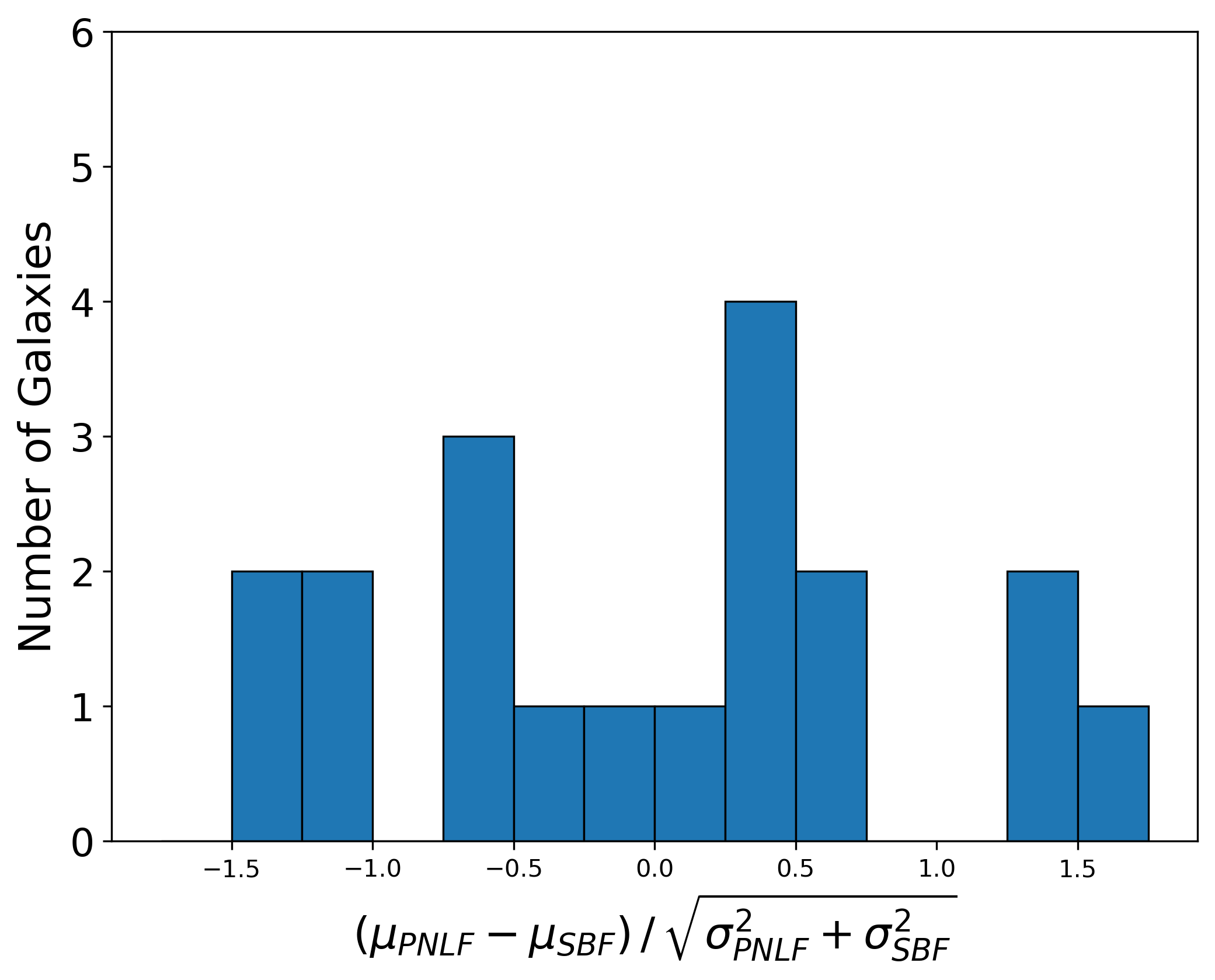}
    \caption{Distribution for tension values between PNLF and SBF distance modulus. It results that 13 out of the 20 objects with both PNLF and SBF distance modulus show less than a $1\sigma$ tension between these measurements.}
    \label{fig:tension_PNLF_minus_SBF}
\end{figure}
}

\newcommand{\placefigDistBl}{
\begin{figure*}
    \includegraphics[width=\hsize]{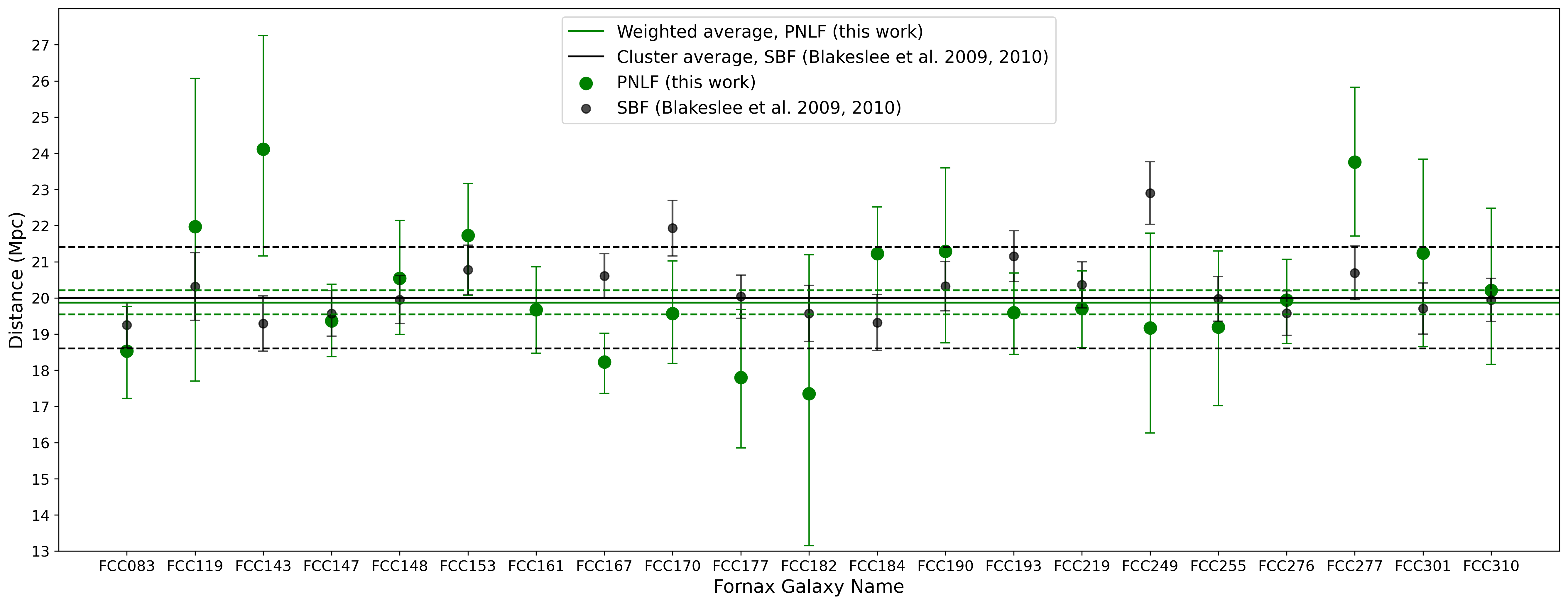}
    \caption{Comparison between individual distance measurements (in Mpc) as obtained here using the PNLF (green circles and error bars) and as derived through SBF by \citet{blakeslee_acs_2009, blakeslee_surface_2010}.
    The solid and dashed green lines show the weighted average of our PNLF measurements and its associated error, as an estimate for the distance to the Fornax cluster. The black solid and dashed line indicate instead the SBF estimate distance to Fornax provided by \citet{blakeslee_acs_2009} with its uncertainty, based on their entire sample of SBF distance measurements.}
    \label{fig:Dist_comp_Bl}
\end{figure*}
}

\newcommand{\placefigDistCF}{
\begin{figure*}
    \includegraphics[width=\hsize]{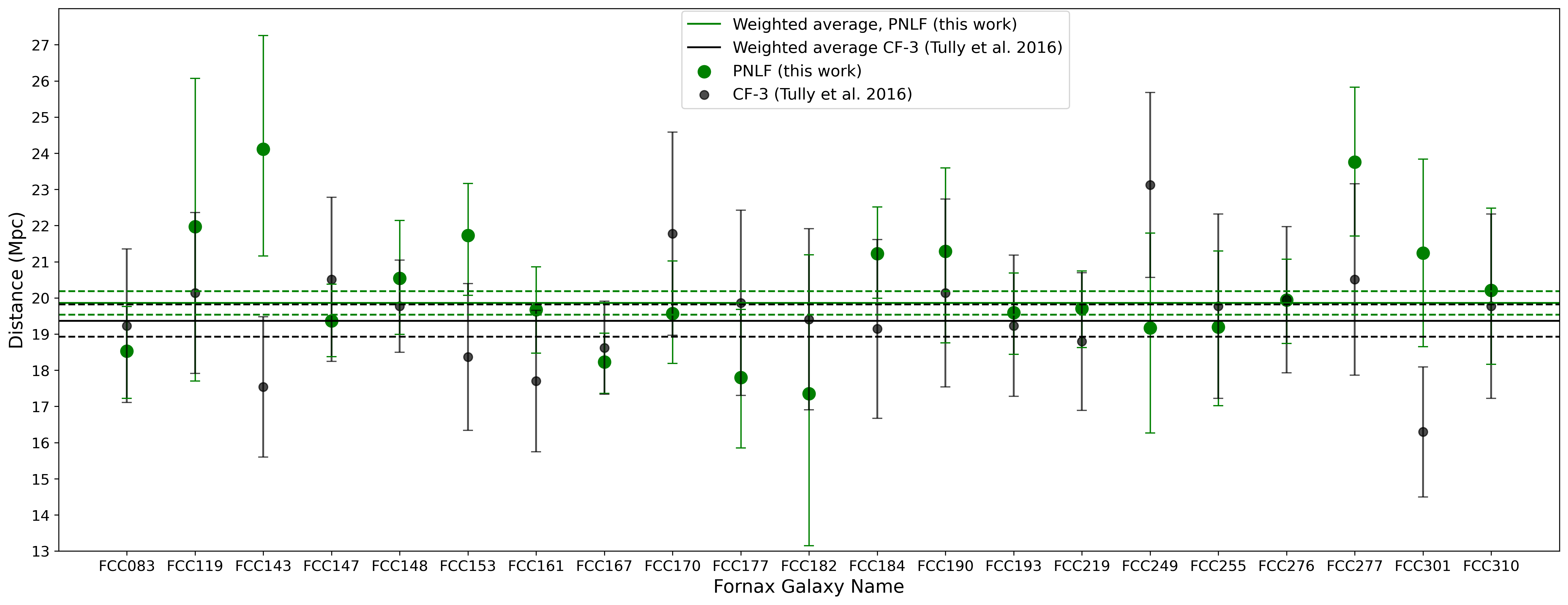}
    \caption{Same as in to Fig.~\ref{fig:Dist_comp_Bl}, but now comparing our PNLF distance estimates (green circles) with those reported in CosmicFlows-3 \citep[][black circles]{tully_cosmicflows-3_2016}. The black solid and dotted lines show the weighted average and error from the CosmicFlows-3 data points.}
    \label{fig:Dist_comp_CF3}
\end{figure*}
}

\newcommand{\placefigtangRmag}{
    \begin{figure}
        \includegraphics[width=\hsize]{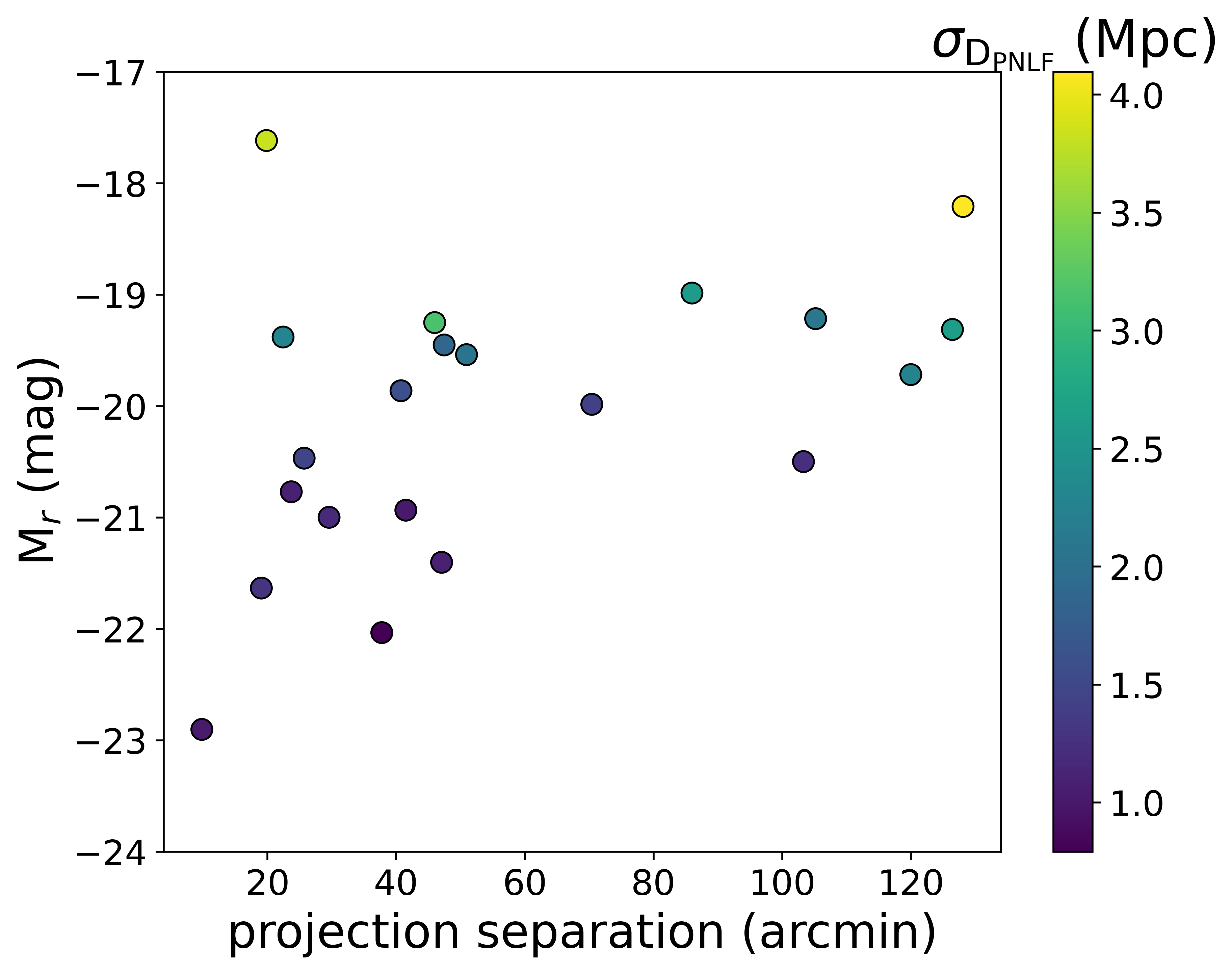}
        \caption{Projected distance (in arcmin), from the BCG of the Fornax cluster NGC~1399, to the ETGs of our sample, as a function of their absolute $r$-band magnitude. Colour scale follows the calculated uncertainties in $\mu_\mathrm{PNLF}$ for each galaxy. The r-band magnitudes are from \citet{iodice_fornax_2019} except for FCC~119, FCC~249, and FCC~255, which we estimated to be 13.5, 12.1 and 12.2 mag respectively. This was obtained starting from the $B$-band magnitude reported in \citet{sarzi_fornax3d_2018} and by applying a colour-correction based on the spectral modelling of the integrated MUSE spectra for this target.}
        \label{fig:tangential_dist_RMag}
    \end{figure}
}

\newcommand{\placefigstructure}{
    \begin{figure*}
        \includegraphics[width=\hsize]{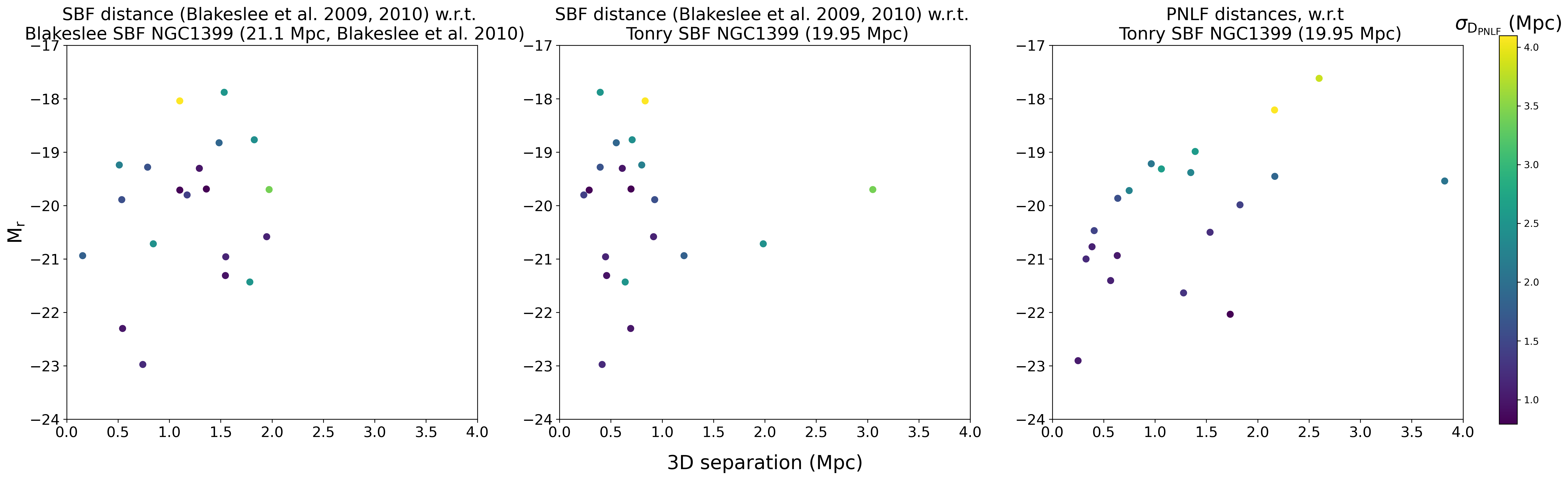}
        \caption{Comparison between the distance (in Mpc) from NGC\,1399 of our sample galaxies versus their absolute $r$-band magnitude, using different sources for the distance estimates to our sample galaxies and NGC~1399. Left panel: adopting the \citet{blakeslee_acs_2009} distances to the individual F3D galaxies and the \citet{blakeslee_surface_2010} distance to NGC\,1399 (21.1Mpc). Middle panel: same as in the left panel but using the SBF distance to NGC\,1399 derived by \citet[][19.95 Mpc]{tonry_sbf_2001}. Right panel: Still using the \citeauthor{tonry_sbf_2001} distance to NGC~1399 but using our PNLF-derived distances to our sample galaxies. The $r$-band magnitudes are from \citet{iodice_fornax_2019} except for FCC~119, FCC~249, and FCC~255, which are estimated to be 13.5, 12.1 and 12.2 mag respectively.}
        \label{fig:fornax_structure_comp}
    \end{figure*}
}

\newcommand{\placefigRmagNPNe}{
    \begin{figure}
        \includegraphics[width=\hsize]{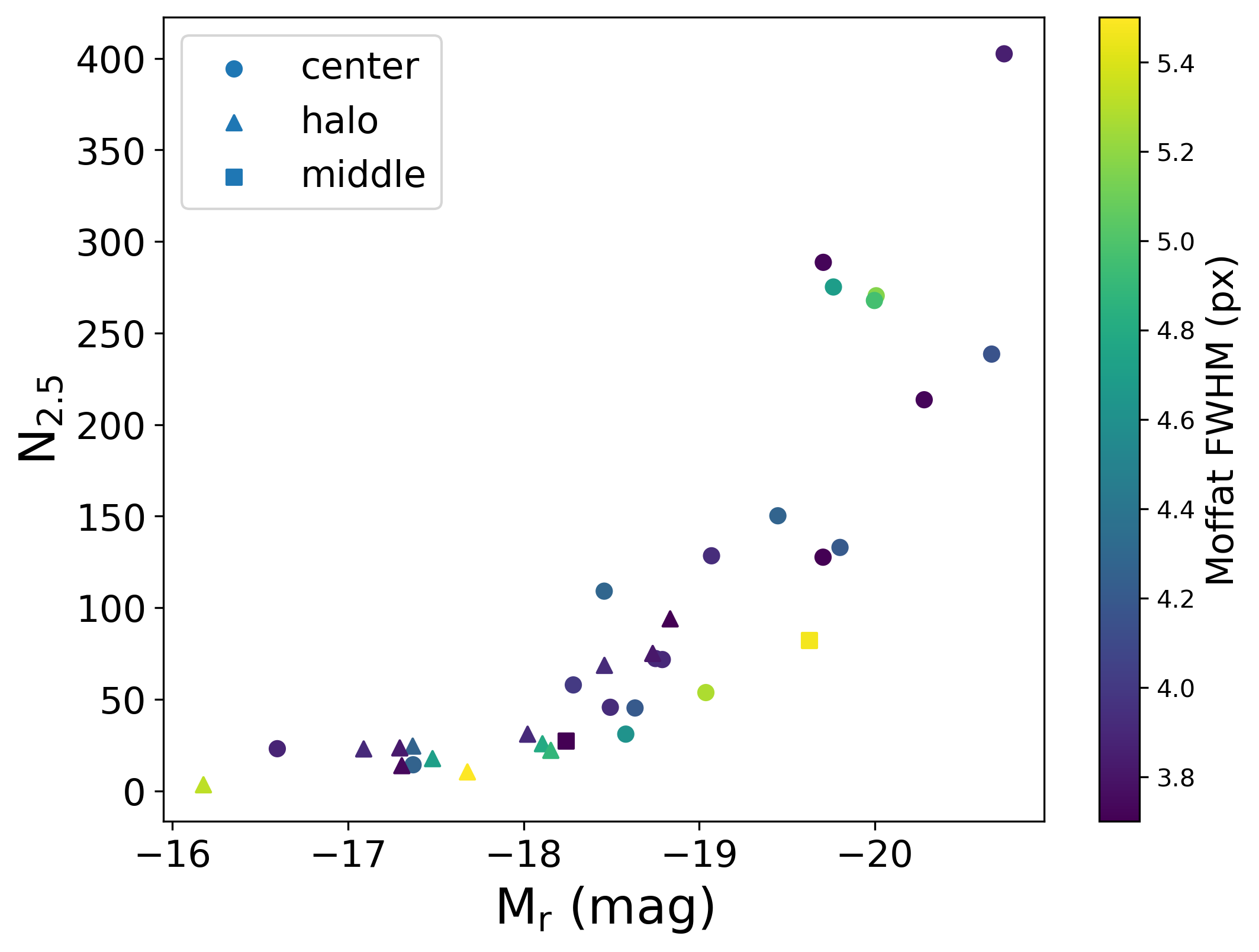}
        \caption{Number of expected PNe within each galaxy pointing (Table~\ref{tab:galaxy_info}) plotted against their $r$-band absolute magnitude, as derived from the MUSE data covering exactly the same regions within each pointing where the presence of PNe is investigated, thus excluding masked out regions. Different symbols denote central, middle, and halo pointing, and the colour-coding indicates the FWHM value of our Moffat PSF model, as derived within each pointing through the simultaneous fit of the brightest PNe.}
    \label{fig:N_PNe_vs_r_mag}
    \end{figure}
}

\newcommand{\placefigsimsDMctwo}{
\begin{figure*}
    \includegraphics[width=\hsize]{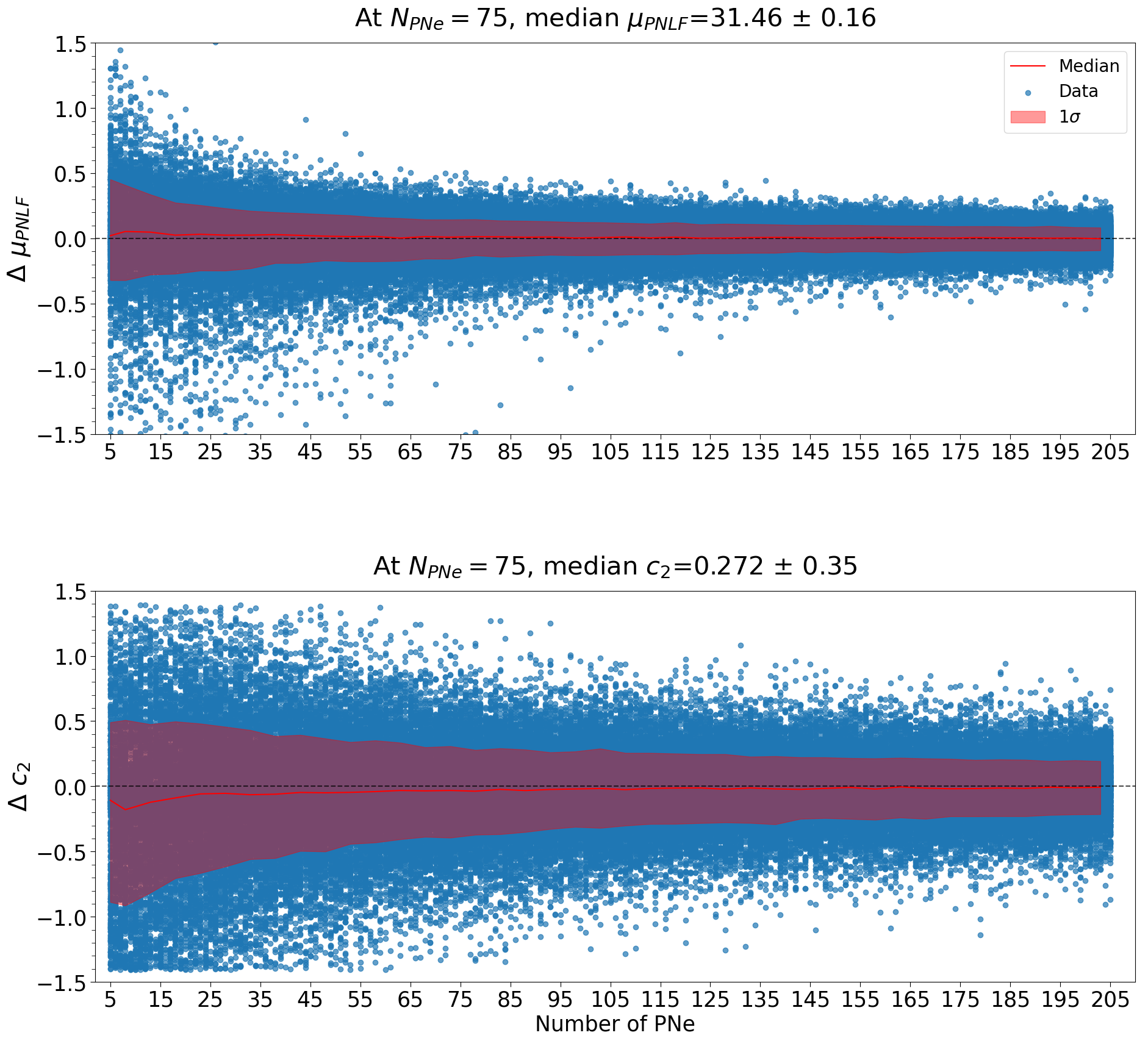}
    \caption{Same as Fig~\ref{fig:simulations_dM}, but now while optimising both the distance modulus $\mu_\mathrm{{PNLF}}$ and the PNLF shape parameter $c_{2}$. The parent distribution assumes a distance modulus $\mu_\mathrm{PNLF \, in}$ = 31.45 and a standard $c_{2}$=0.307 value. As in Fig~\ref{fig:simulations_dM}, the simulations also account for the whole range of completeness profiles observed across our sample.}
    \label{fig:simulations_dM_c2}
\end{figure*}
}

\newcommand{\placefigsimsDMctwoA}{
\begin{figure*}
    \includegraphics[width=\hsize]{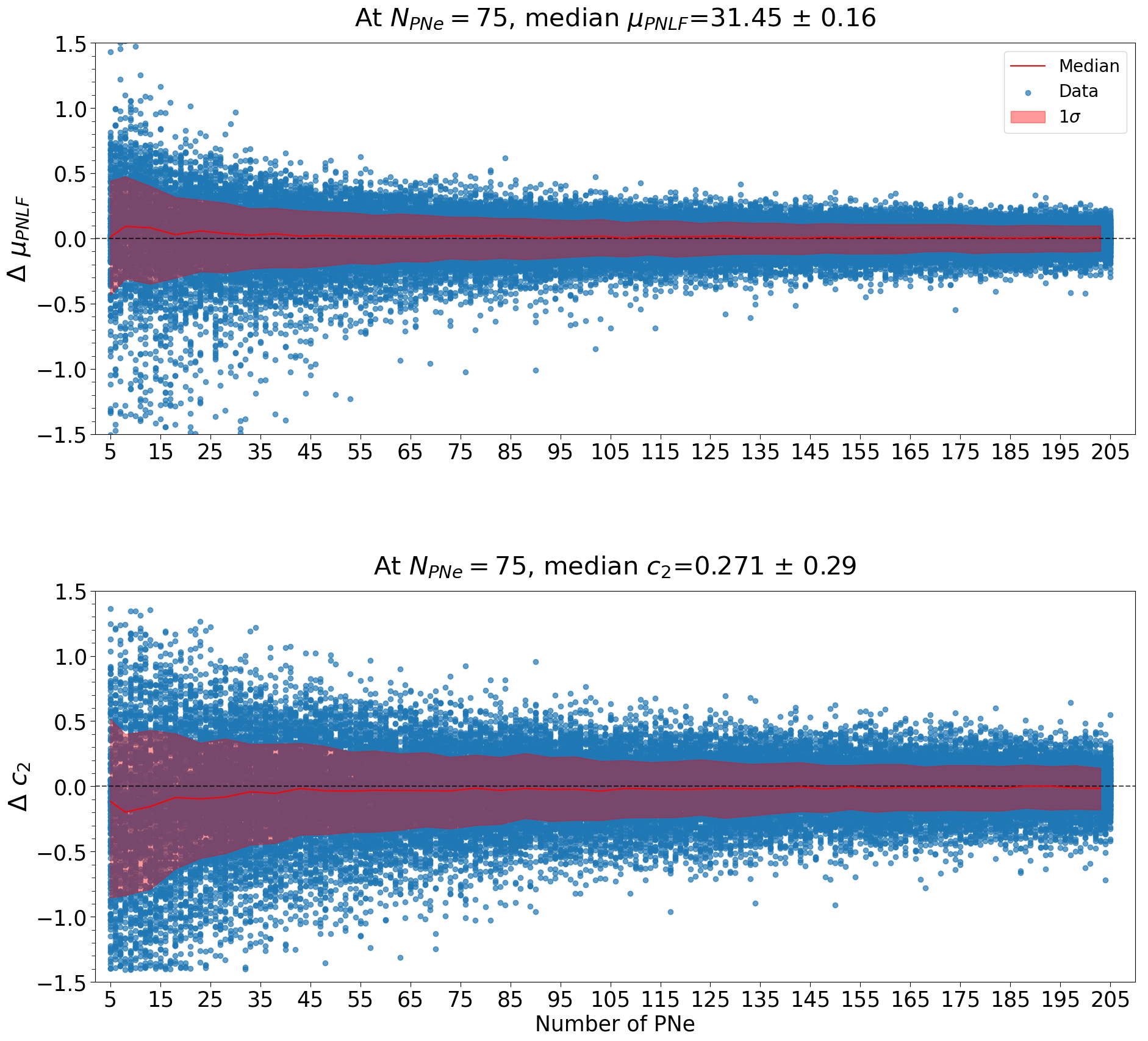}
    \caption{Same as Fig.~\ref{fig:simulations_dM_c2}, but now adopting the completeness function observed in FCC~193, which extends 2.5 magnitudes from the apparent magnitude of the PNLF cutoff.}
    \label{fig:simulations_dM_c2_FCC193}
\end{figure*}
}

\newcommand{\placefigsimsDMctwoB}{
\begin{figure*}
    \includegraphics[width=\hsize]{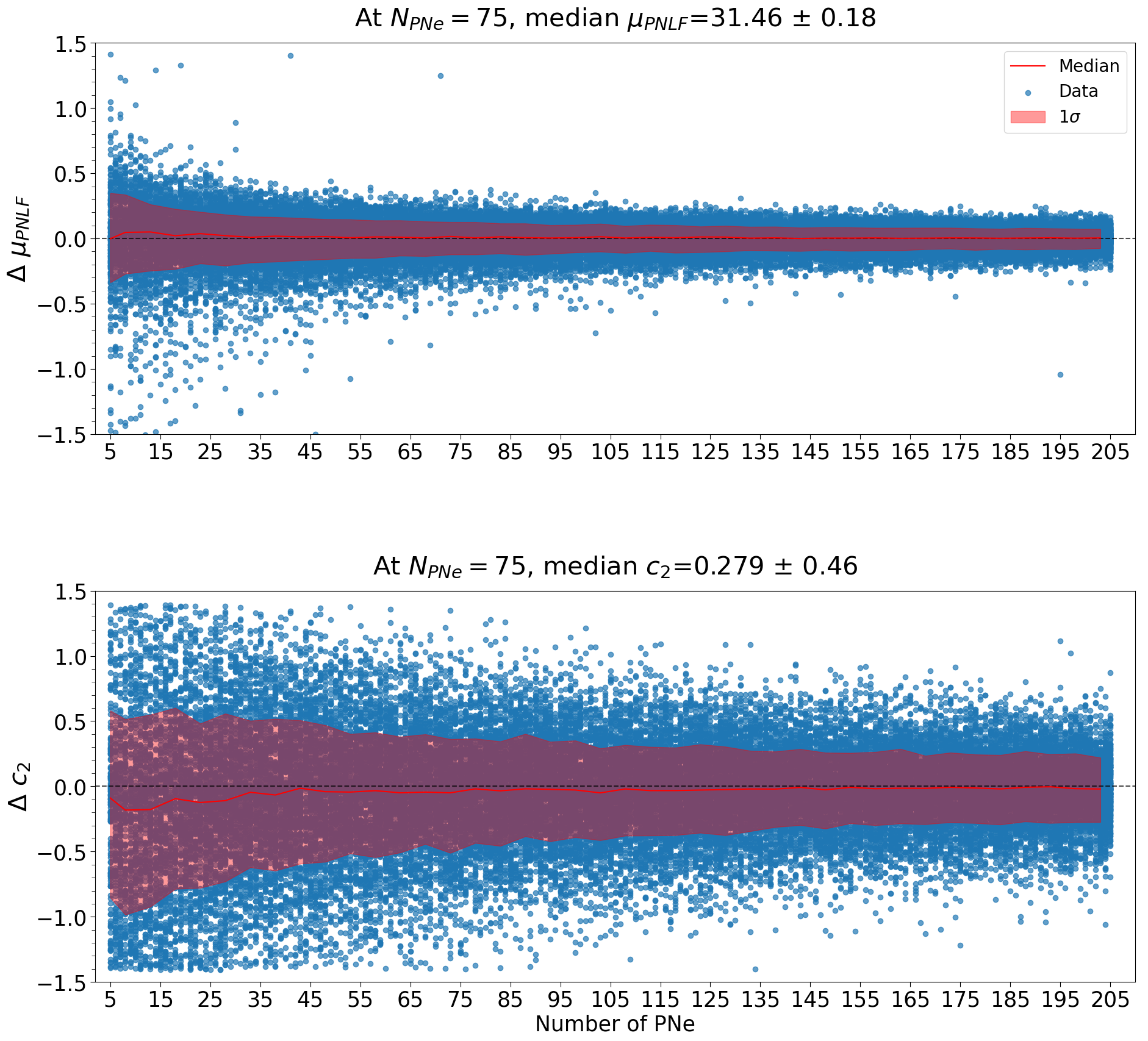}
    \caption{Same as Fig.~\ref{fig:simulations_dM_c2}, but now adopting the completeness function observed in FCC~147, which extends only 1.5 magnitudes from the apparent magnitude of the PNLF cutoff.}
    \label{fig:simulations_dM_c2_FCC147}
\end{figure*}
}

\newcommand{\placefigPNLFgrid}{
\begin{figure}
    \centering
    \begin{tabular}{ccc}
    \includegraphics[width=5.9cm]{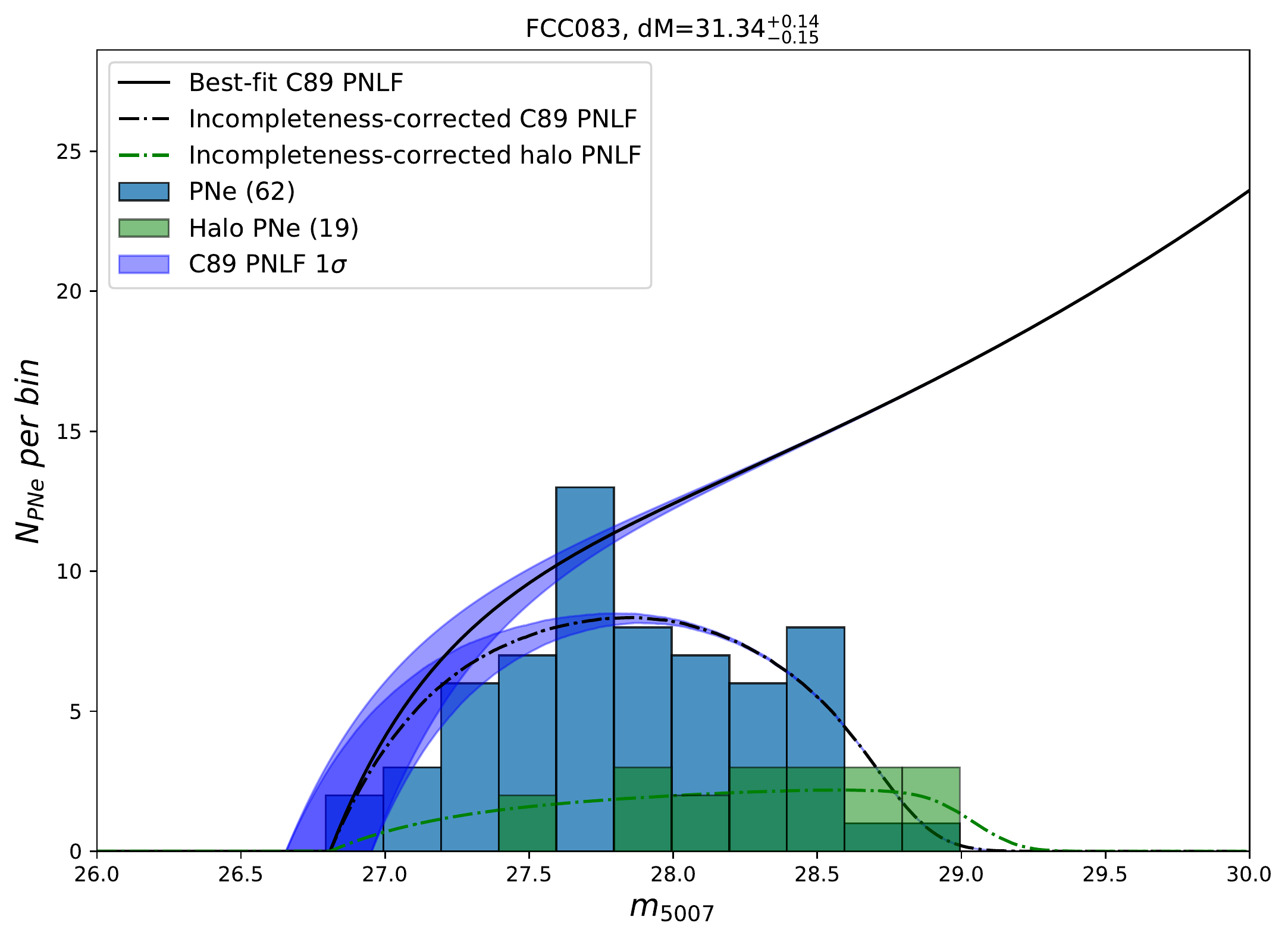} & \includegraphics[width=5.9cm]{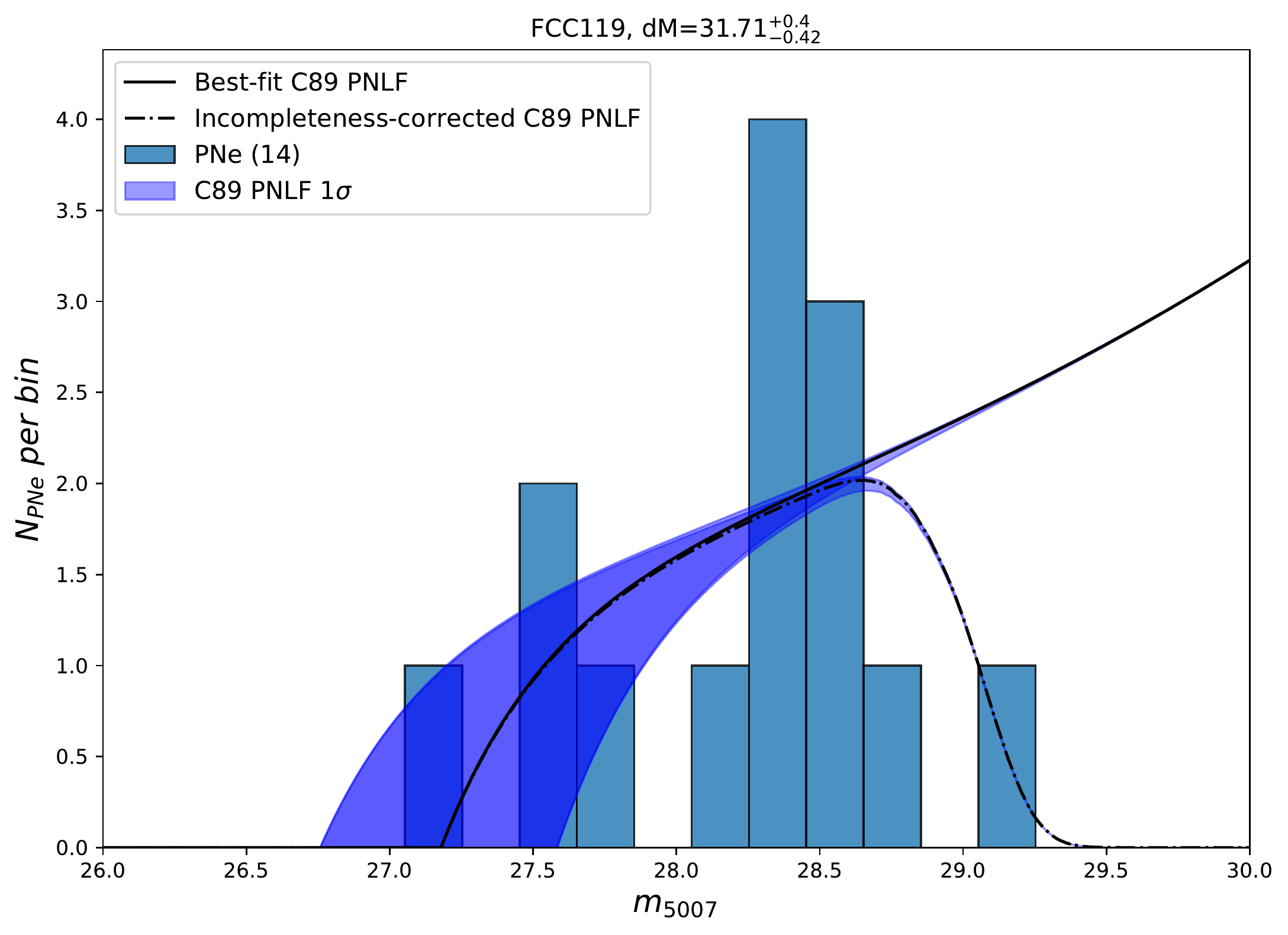} & \includegraphics[width=5.9cm]{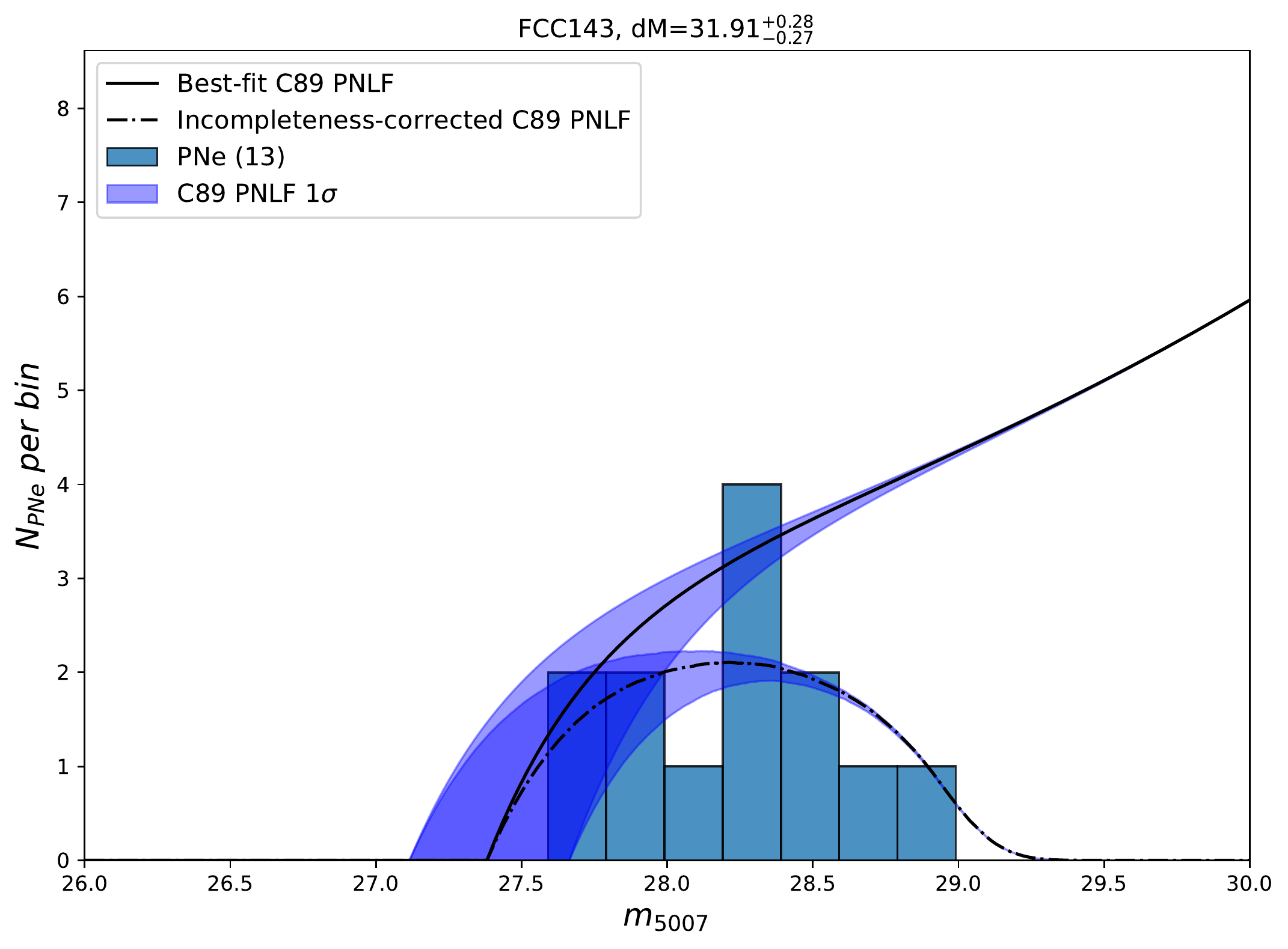} \\
    \tiny FCC083 & \tiny FCC119 & \tiny FCC143 \\
    \includegraphics[width=5.9cm]{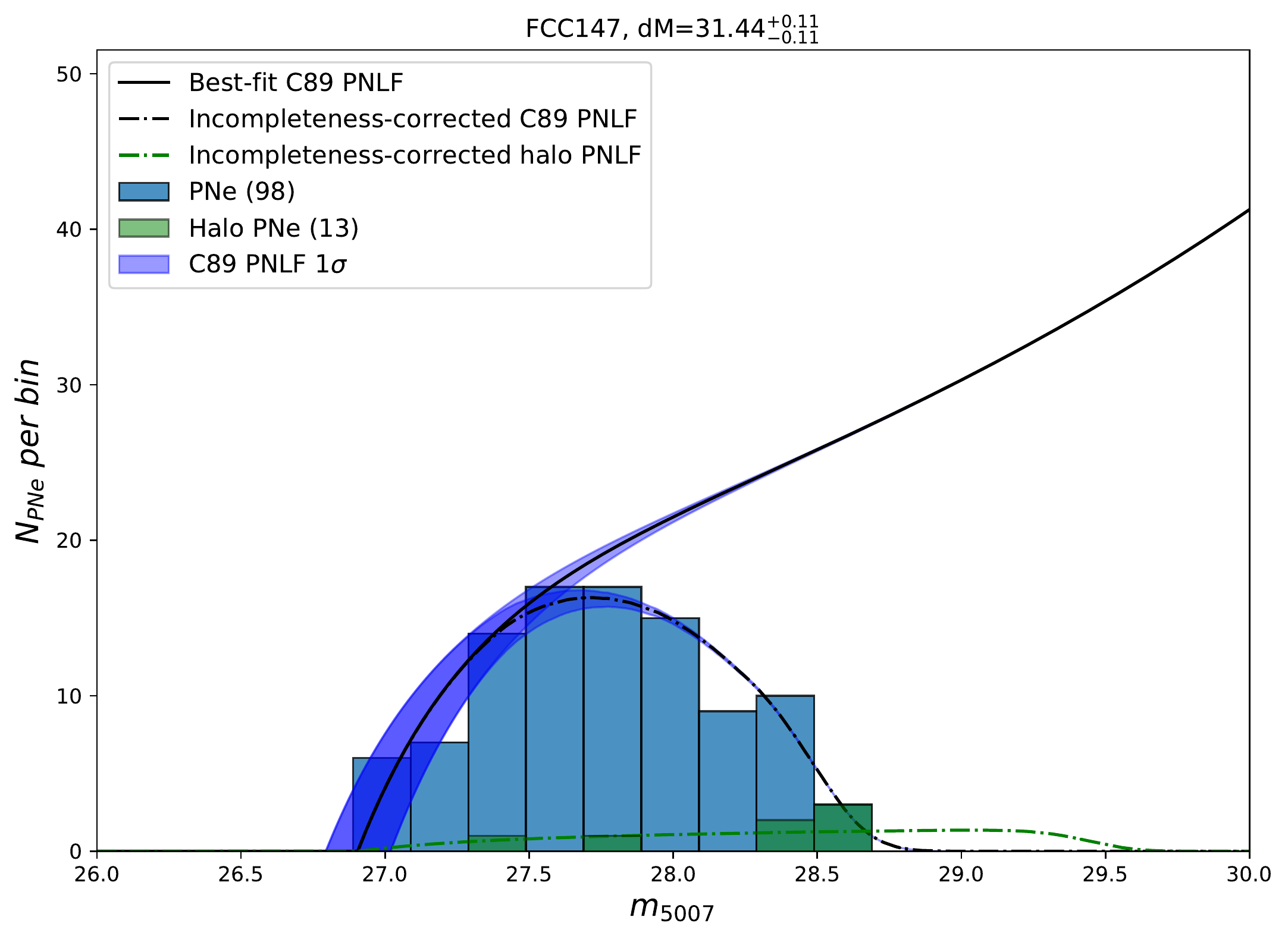} &
    \includegraphics[width=5.9cm]{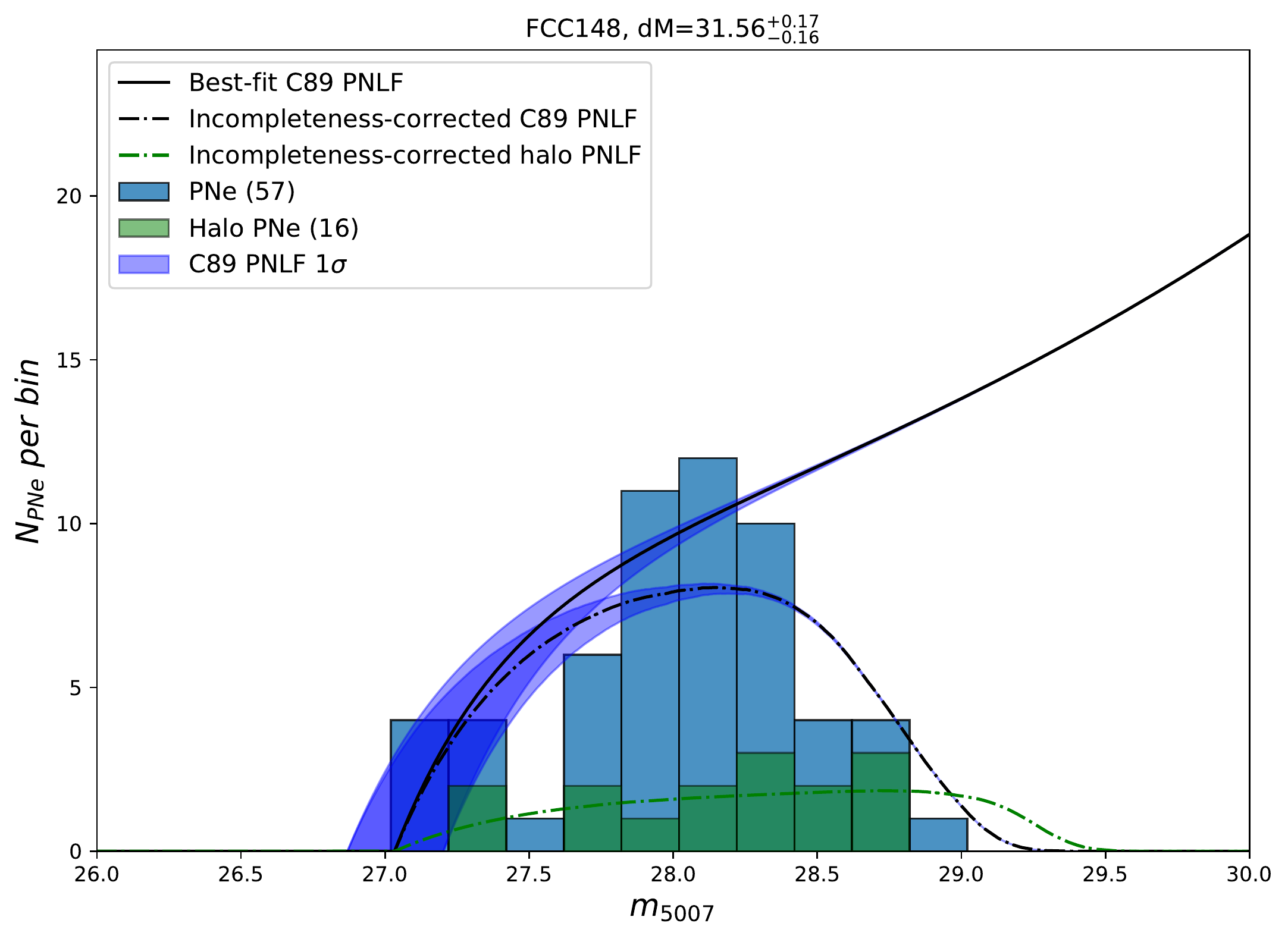} & \includegraphics[width=5.9cm]{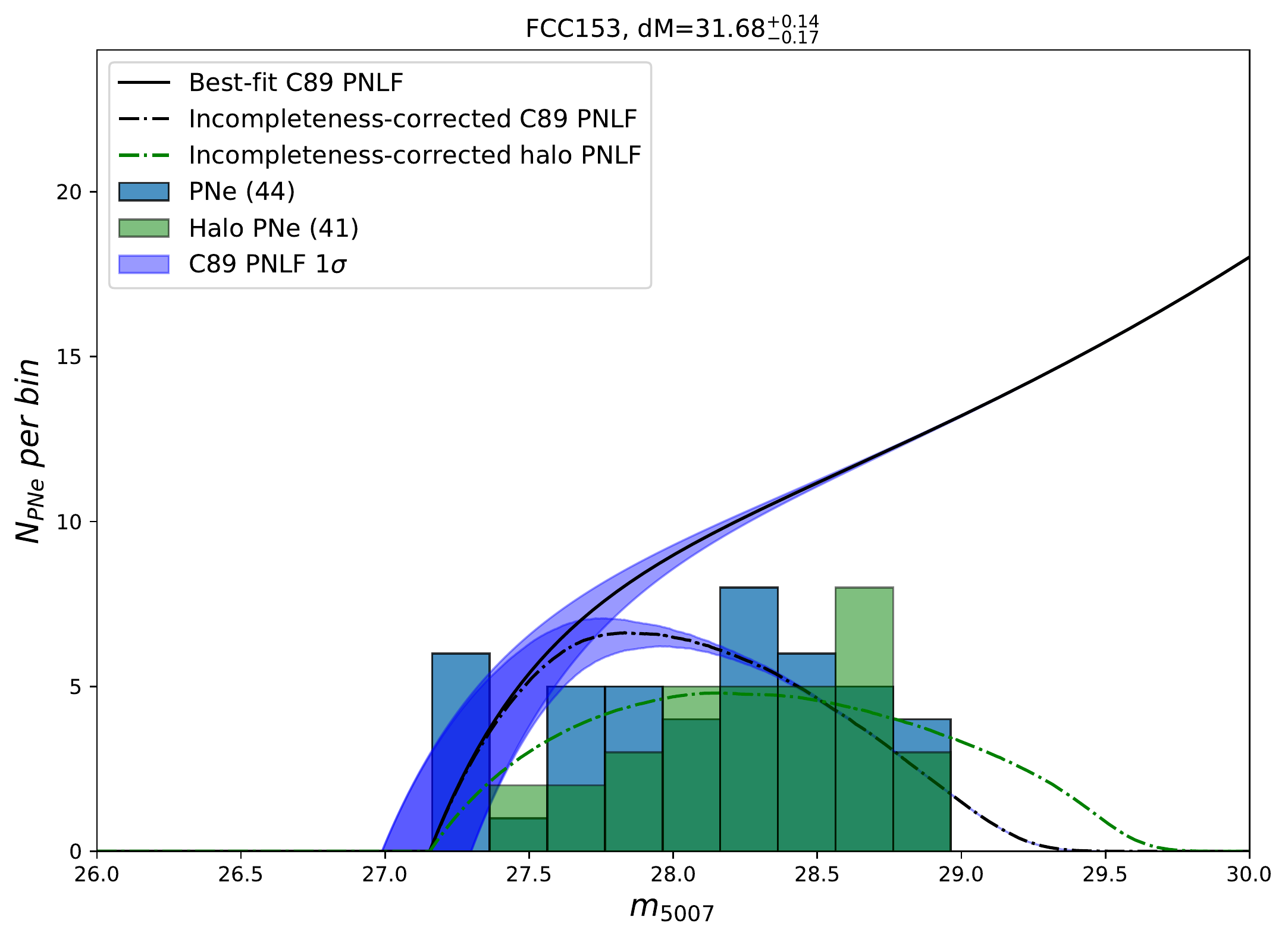} \\ 
    \tiny FCC147 & \tiny FCC148 & \tiny FCC153 \\
    \includegraphics[width=5.9cm]{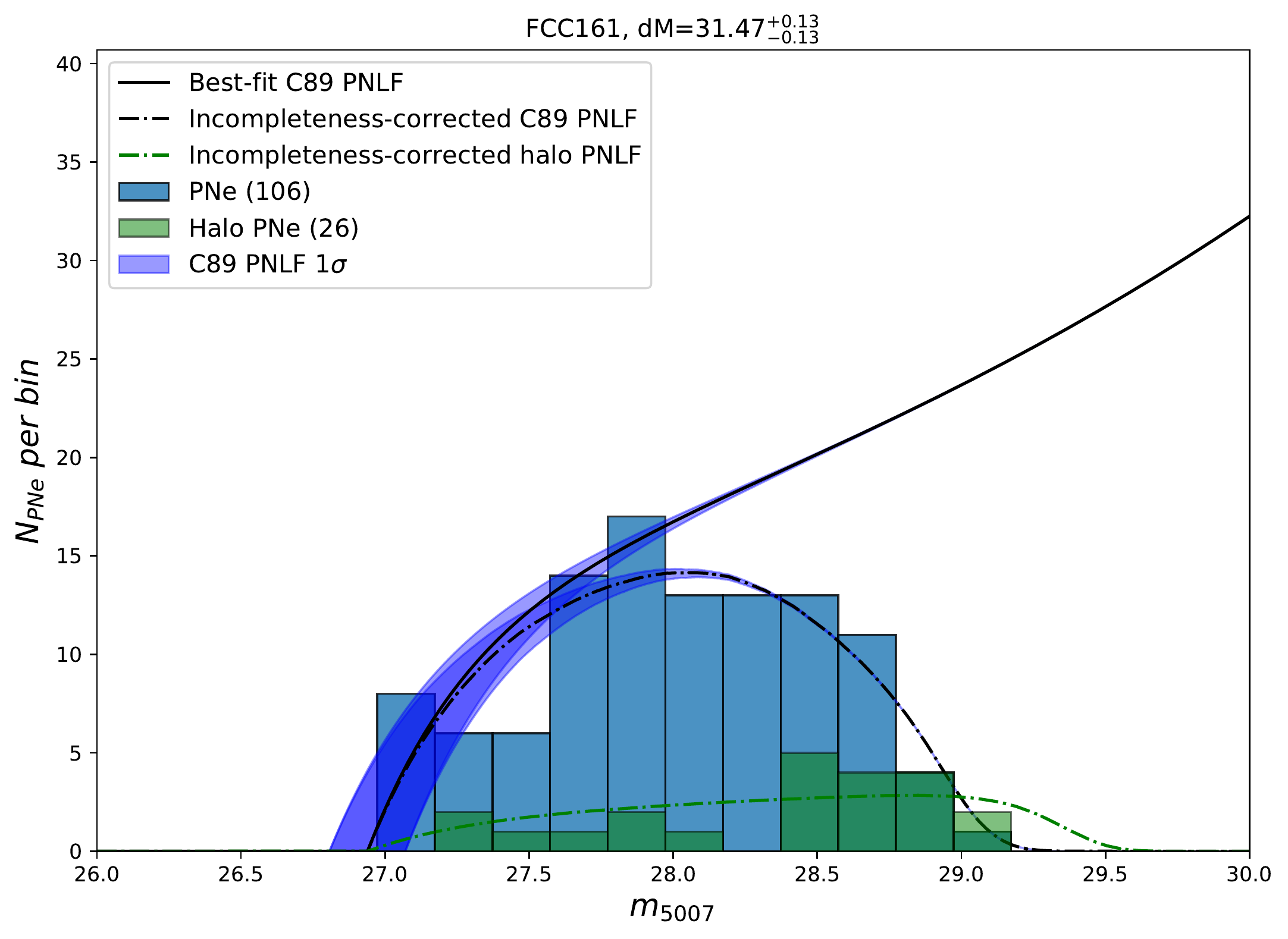} & \includegraphics[width=5.9cm]{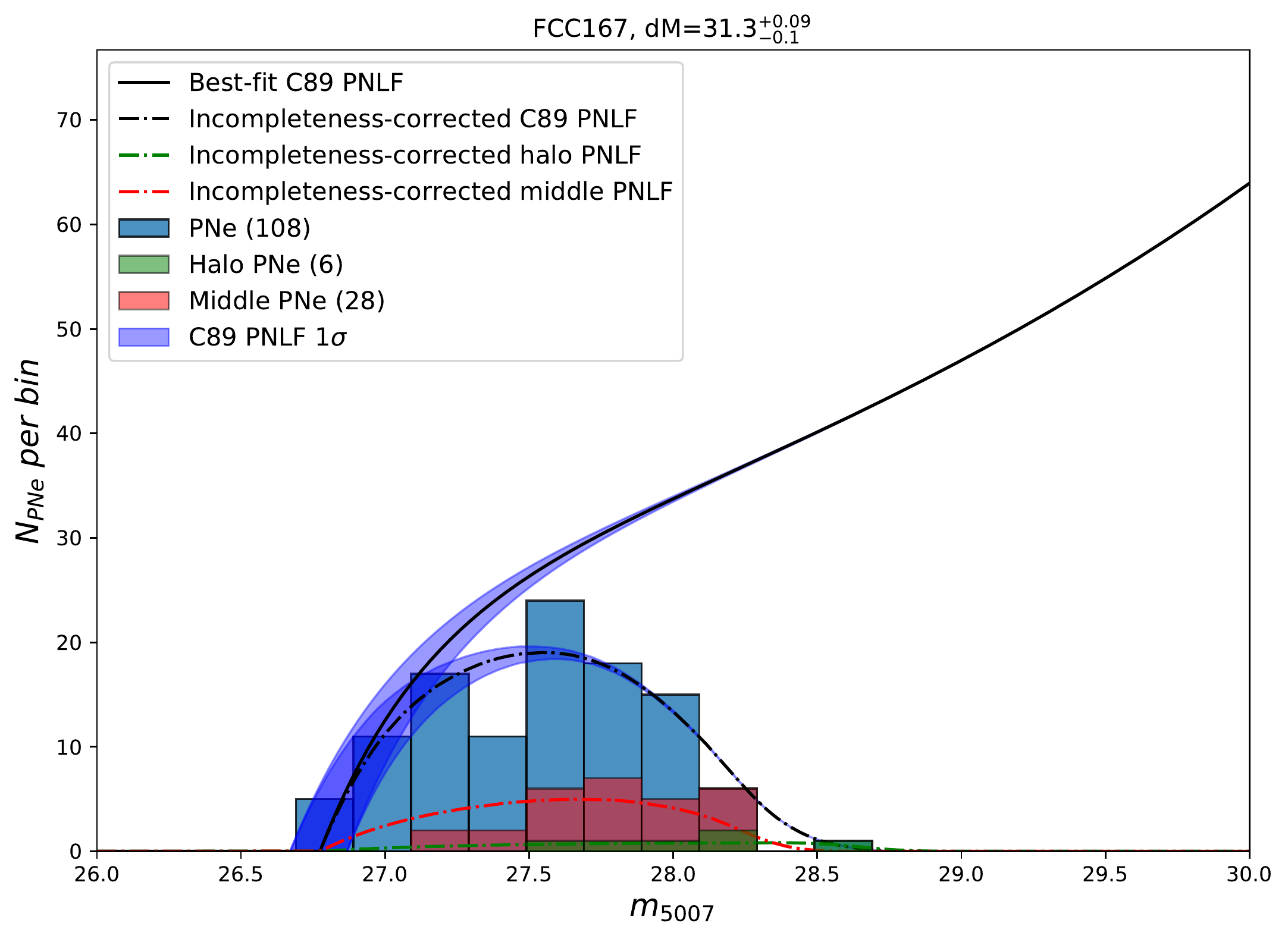} &
    \includegraphics[width=5.9cm]{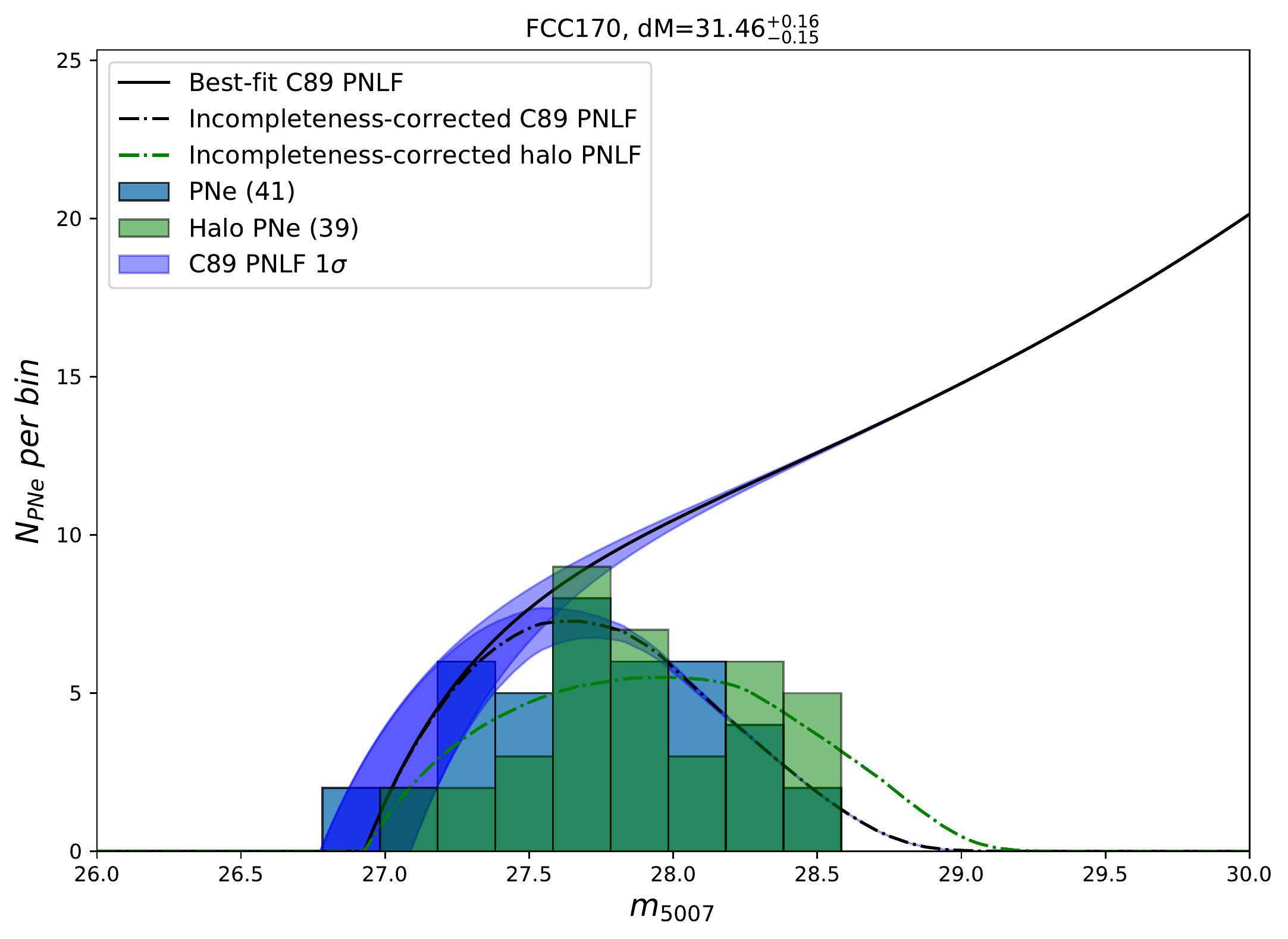} \\ 
    \tiny FCC161 & \tiny FCC167 & \tiny FCC170 \\
    \includegraphics[width=5.9cm]{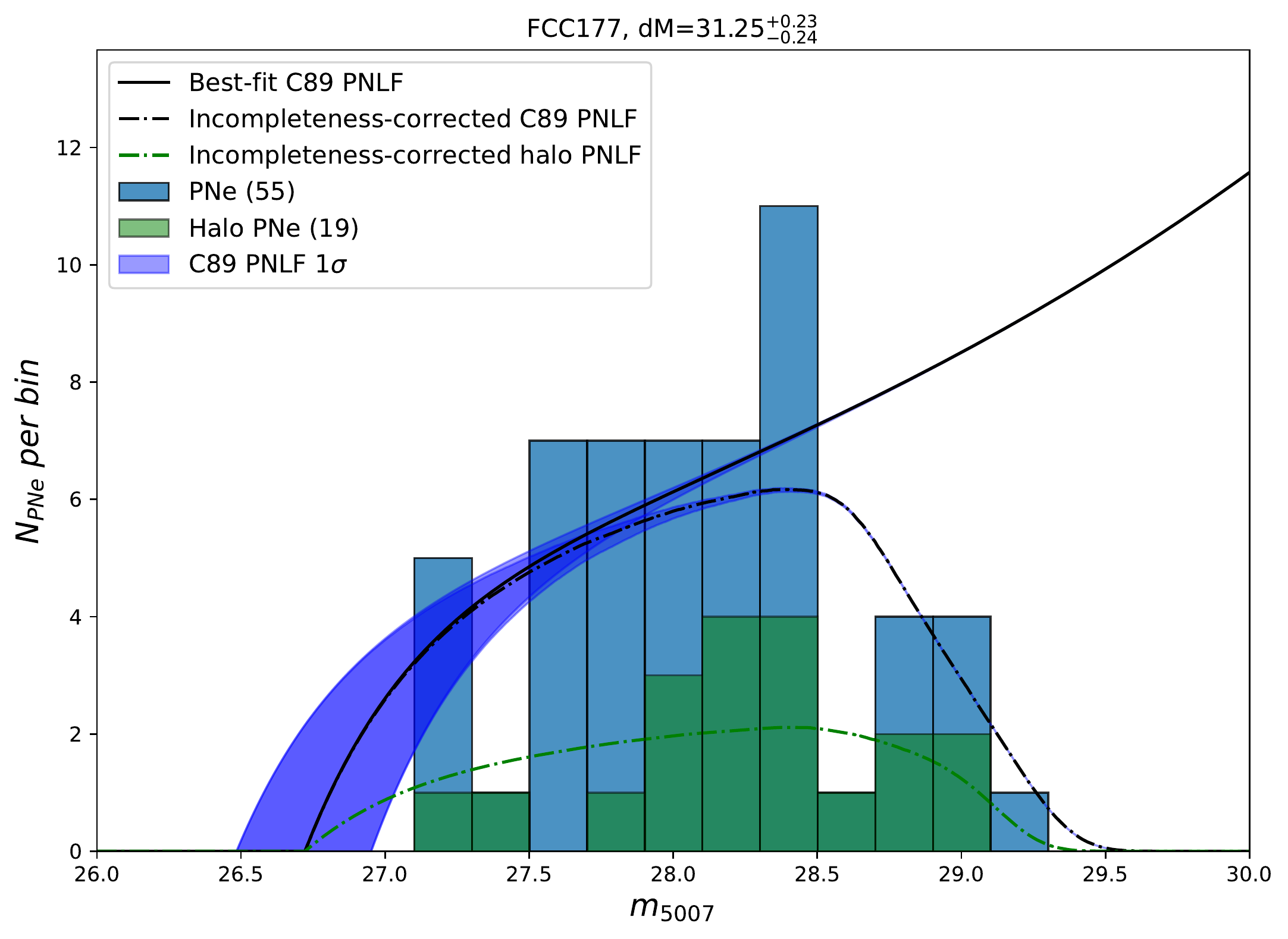} & \includegraphics[width=5.9cm]{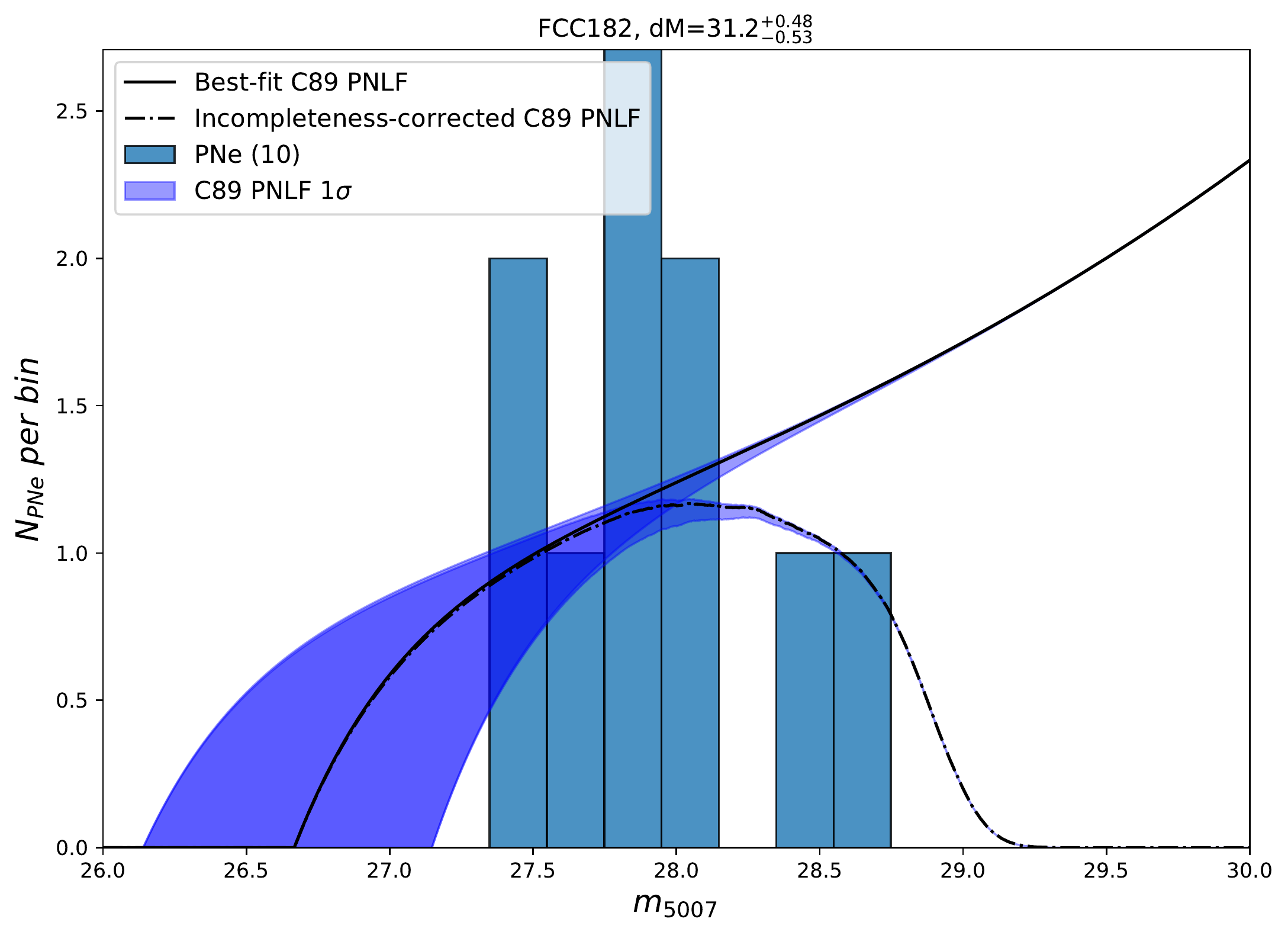} & \includegraphics[width=5.9cm]{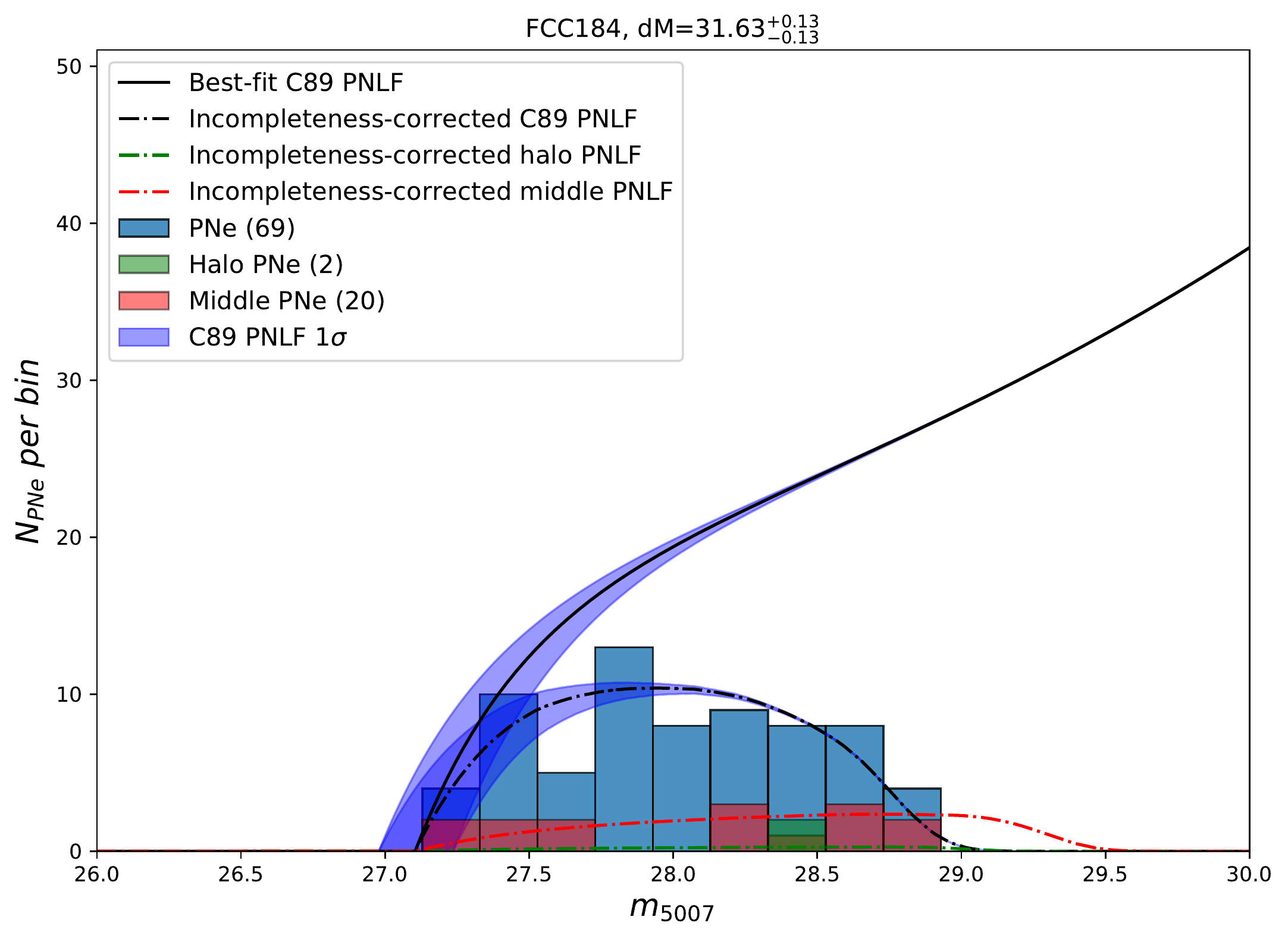} \\
    \tiny FCC177 & \tiny FCC182 & \tiny FCC184 \\
    \includegraphics[width=5.9cm]{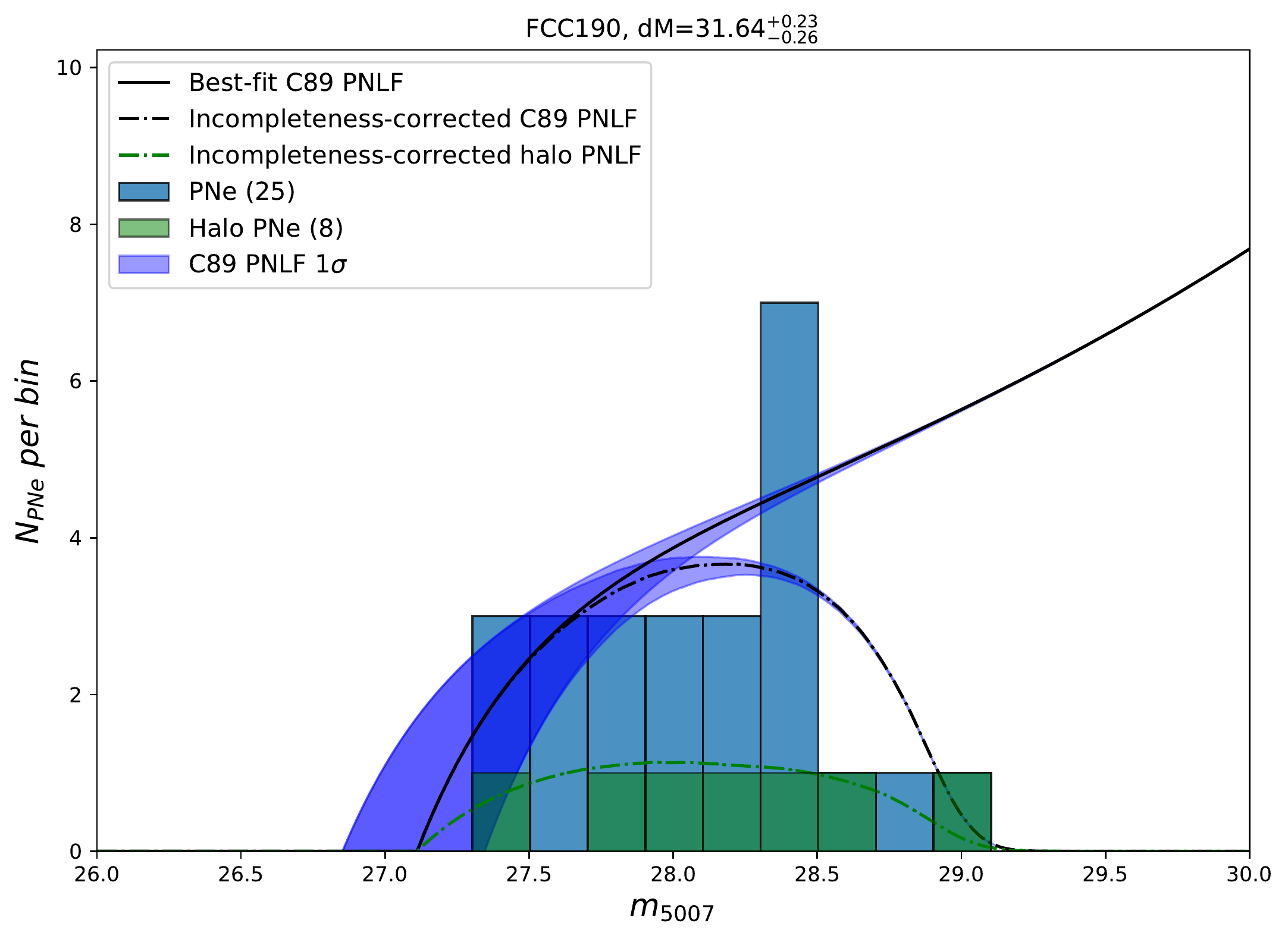} & \includegraphics[width=5.9cm]{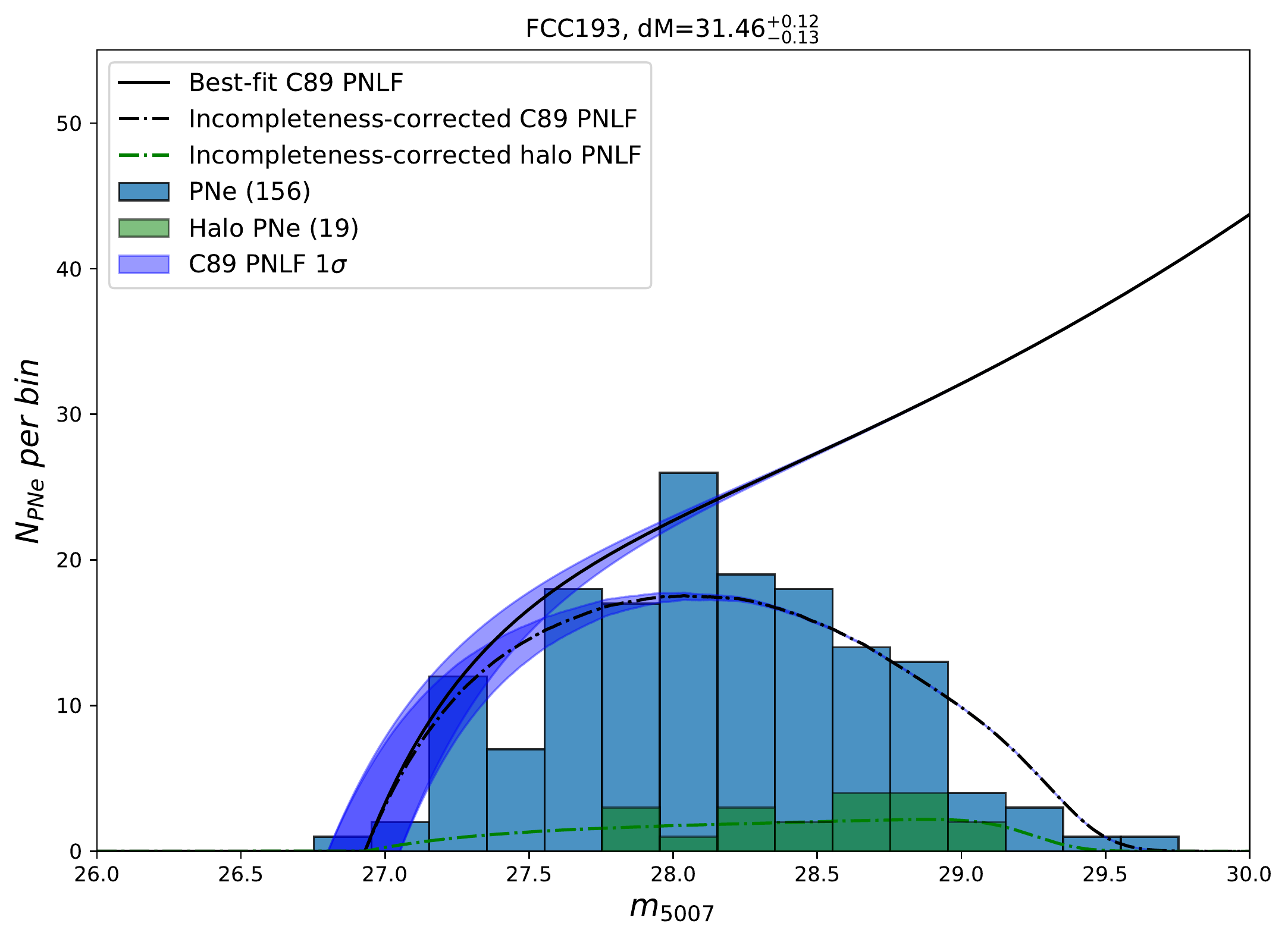} & \includegraphics[width=5.9cm]{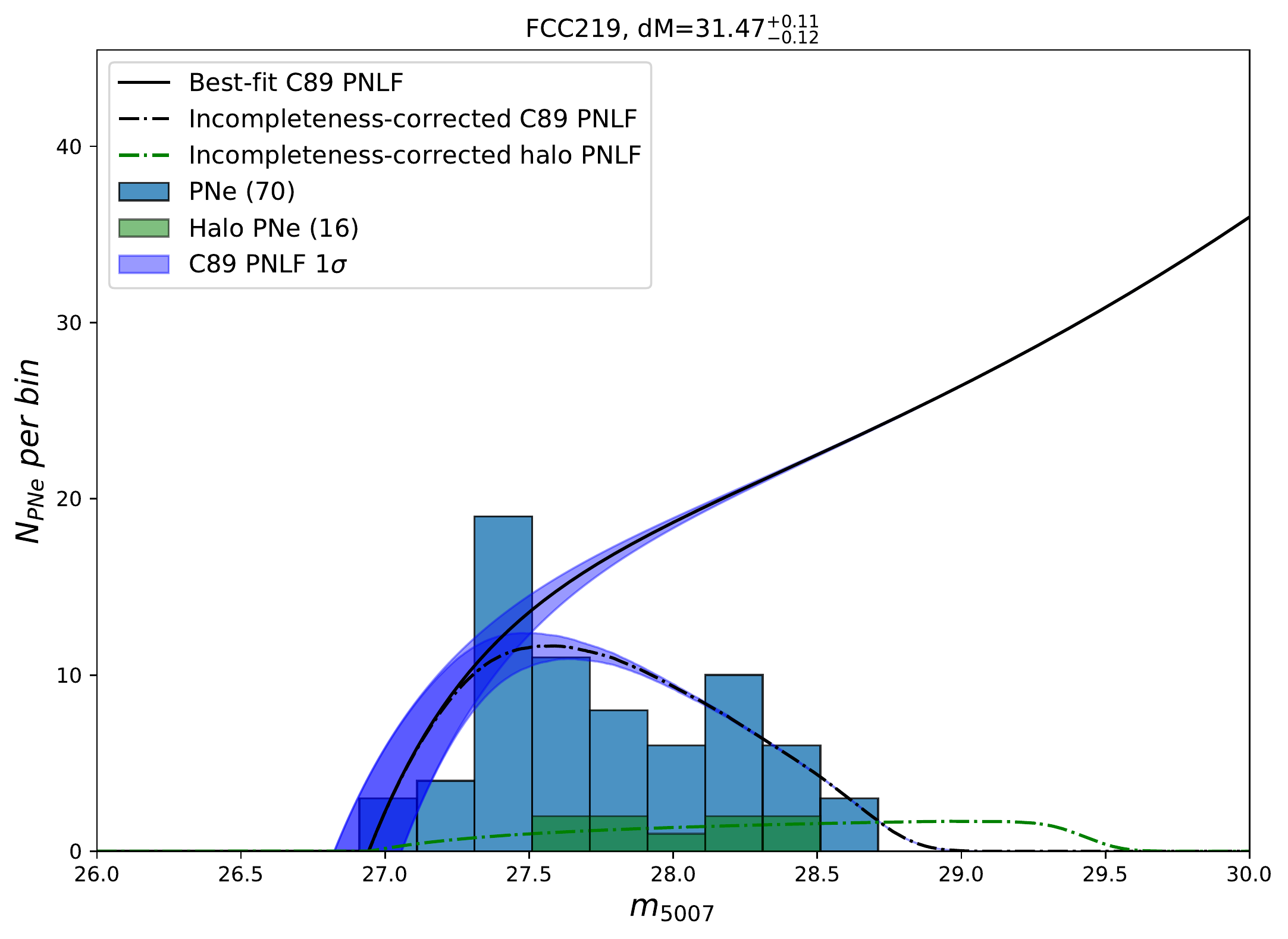} \\
    \tiny FCC190 & \tiny FCC193 & \tiny FCC219 \\
    \end{tabular}
\end{figure}

\begin{figure}
    \begin{tabular}{ccc}
        
    \includegraphics[width=5.9cm]{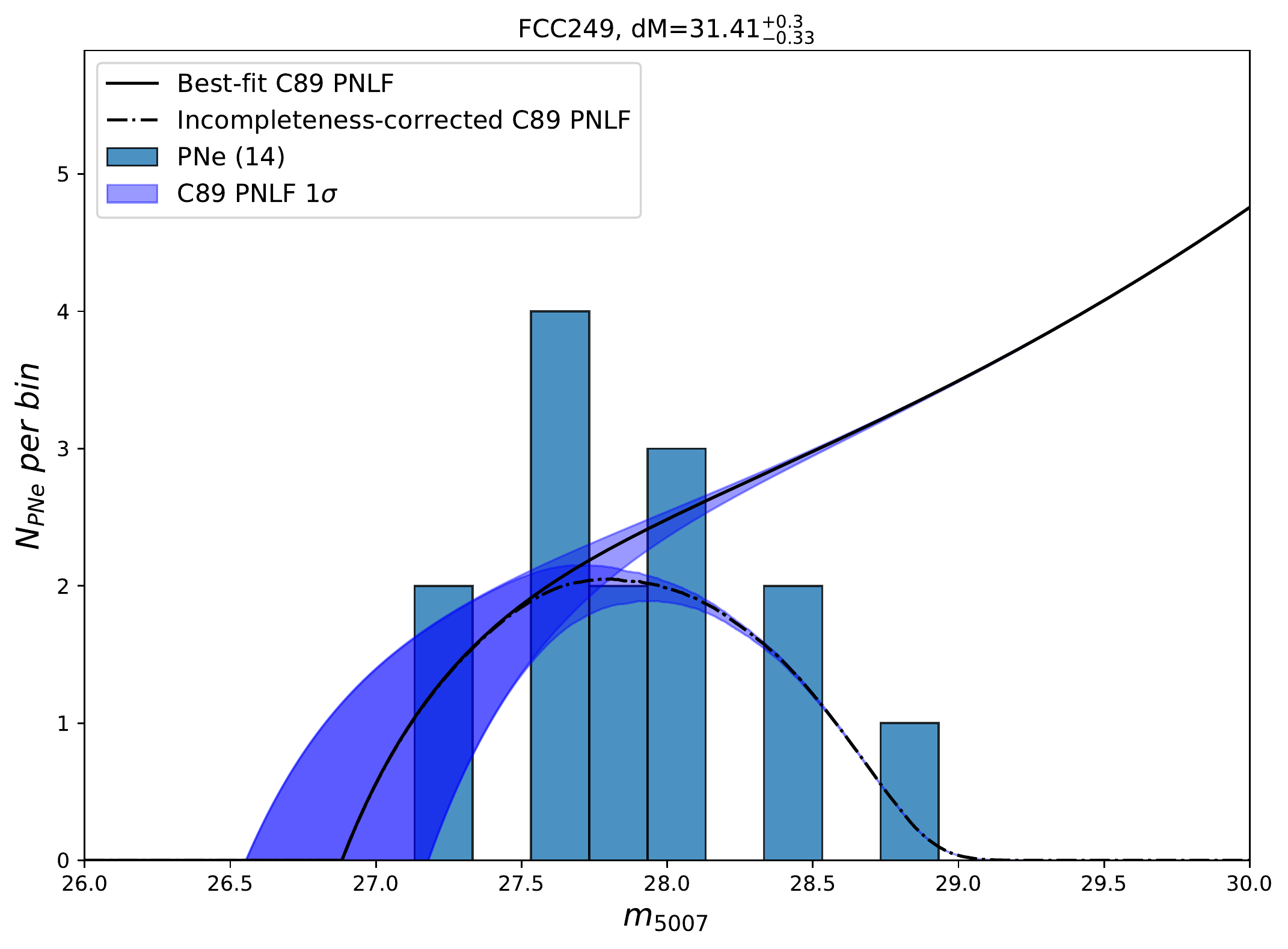} &
    \includegraphics[width=5.9cm]{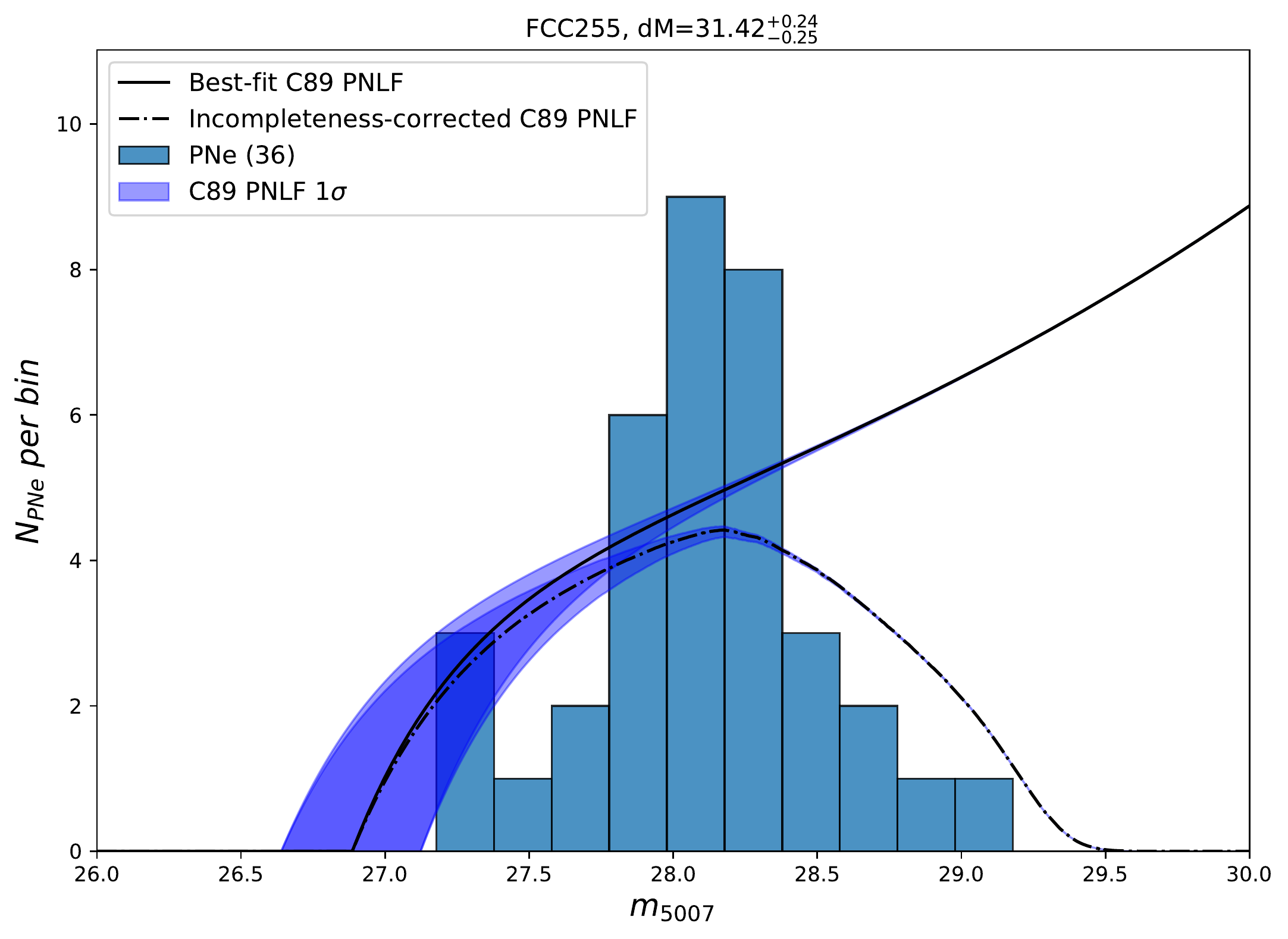} & \includegraphics[width=5.9cm]{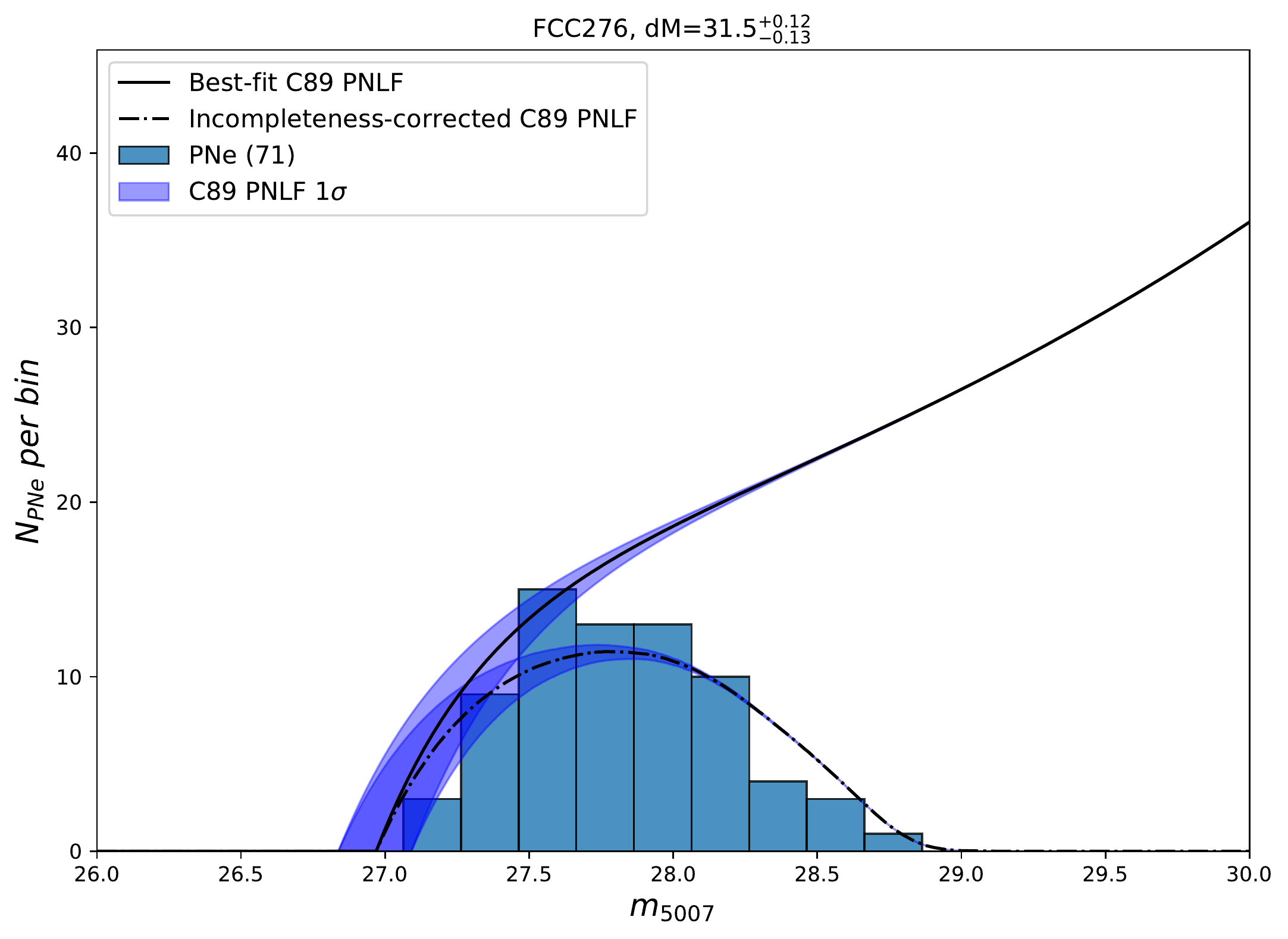} \\ 
    \tiny FCC249 & \tiny FCC255 & \tiny FCC276 \\
    \includegraphics[width=5.9cm]{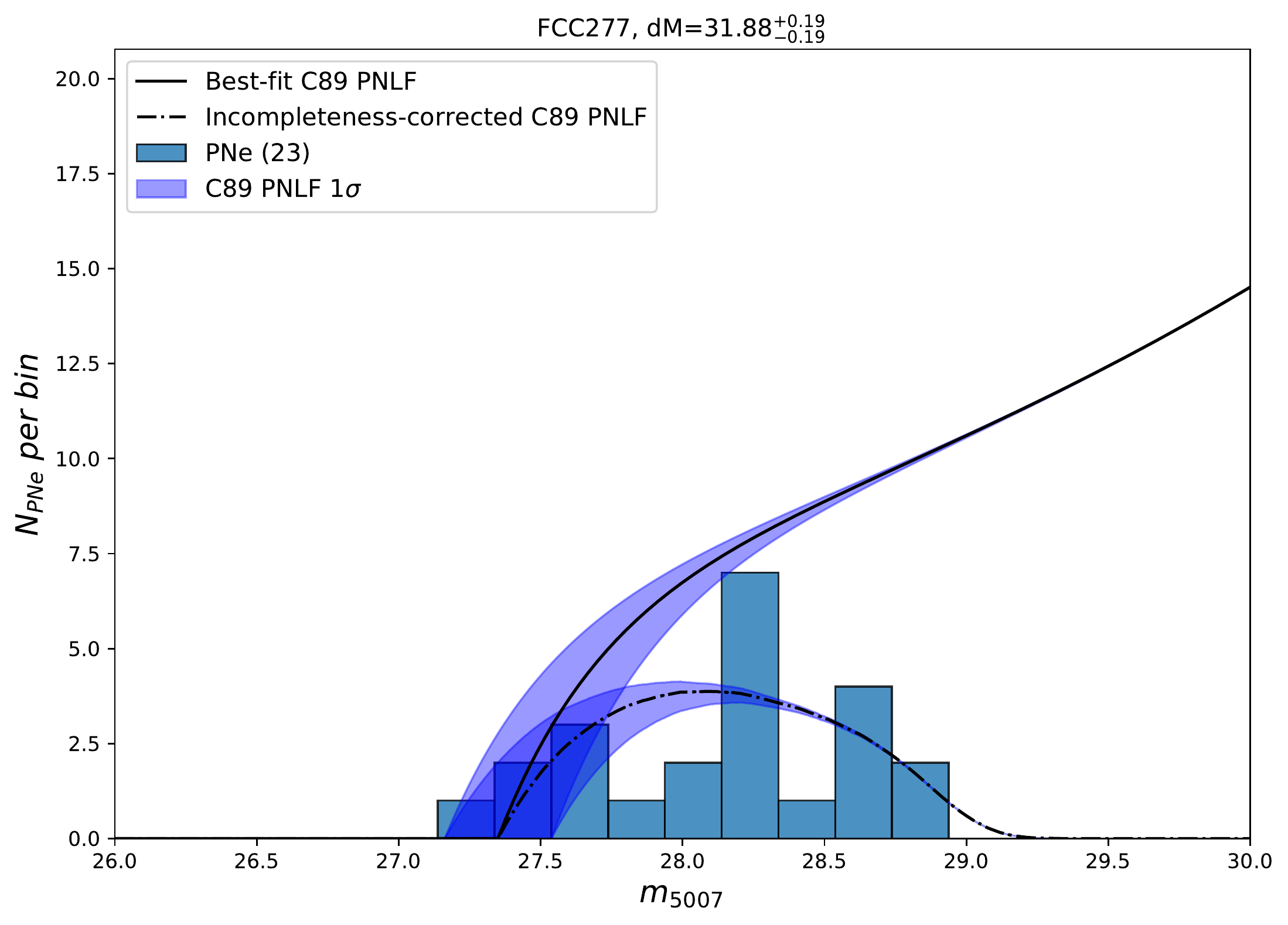} & \includegraphics[width=5.9cm]{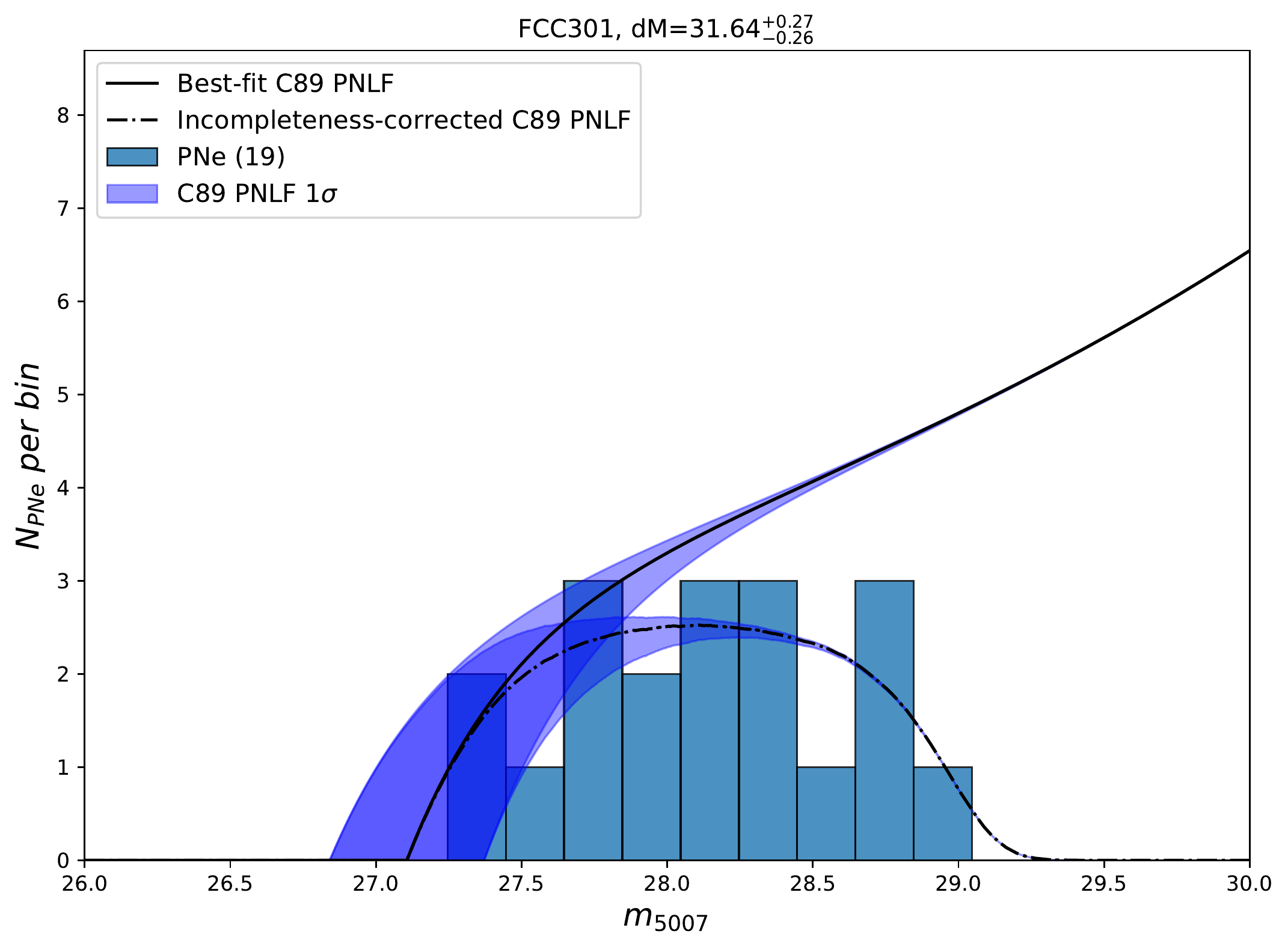} &    \includegraphics[width=5.9cm]{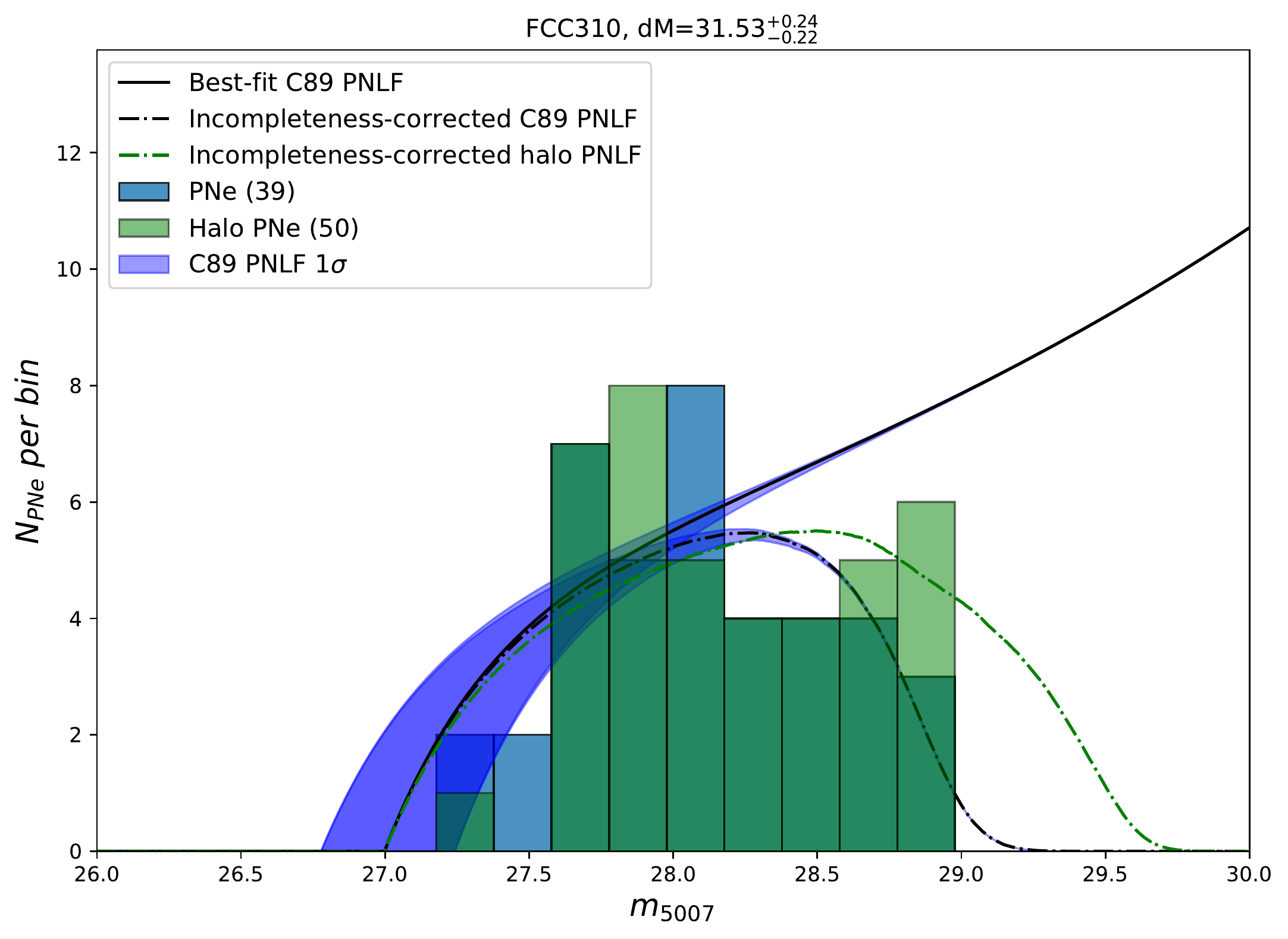} \\
    \tiny FCC277 & \tiny FCC301 & \tiny FCC310 \\
    \end{tabular}

    \caption{Observed PNLF for the PNe sources found in all the ETGs listed in \ref{tab:galaxy_info}. The entire central PNLF is fitted to derive the best distance modulus while accounting for the incompleteness of our observations. The corresponding best-fit model is shown by the dot-dashed blue line, with corresponding confidence intervals, whereas the original model PNLF - \citep[from][]{ciardullo_planetary_1989} is shown by the filled blue line. Blue bars show central PNe, red bars indicate disk (or middle) PNe, and green bars show halo PNe. The green dot-dashed lines show the PNLF model at the best-fit distance modulus, corrected for incompleteness according to the depth of the halo pointing and re-scale to match its integral to the observed number of PNe in this pointing.}    
    \label{fig:all_PNLF}

\end{figure} 
}

\newcommand{\placeFOVplotsgrid}{
\begin{figure}
    \centering
    \begin{tabular}{cc}
    \includegraphics[width=9cm]{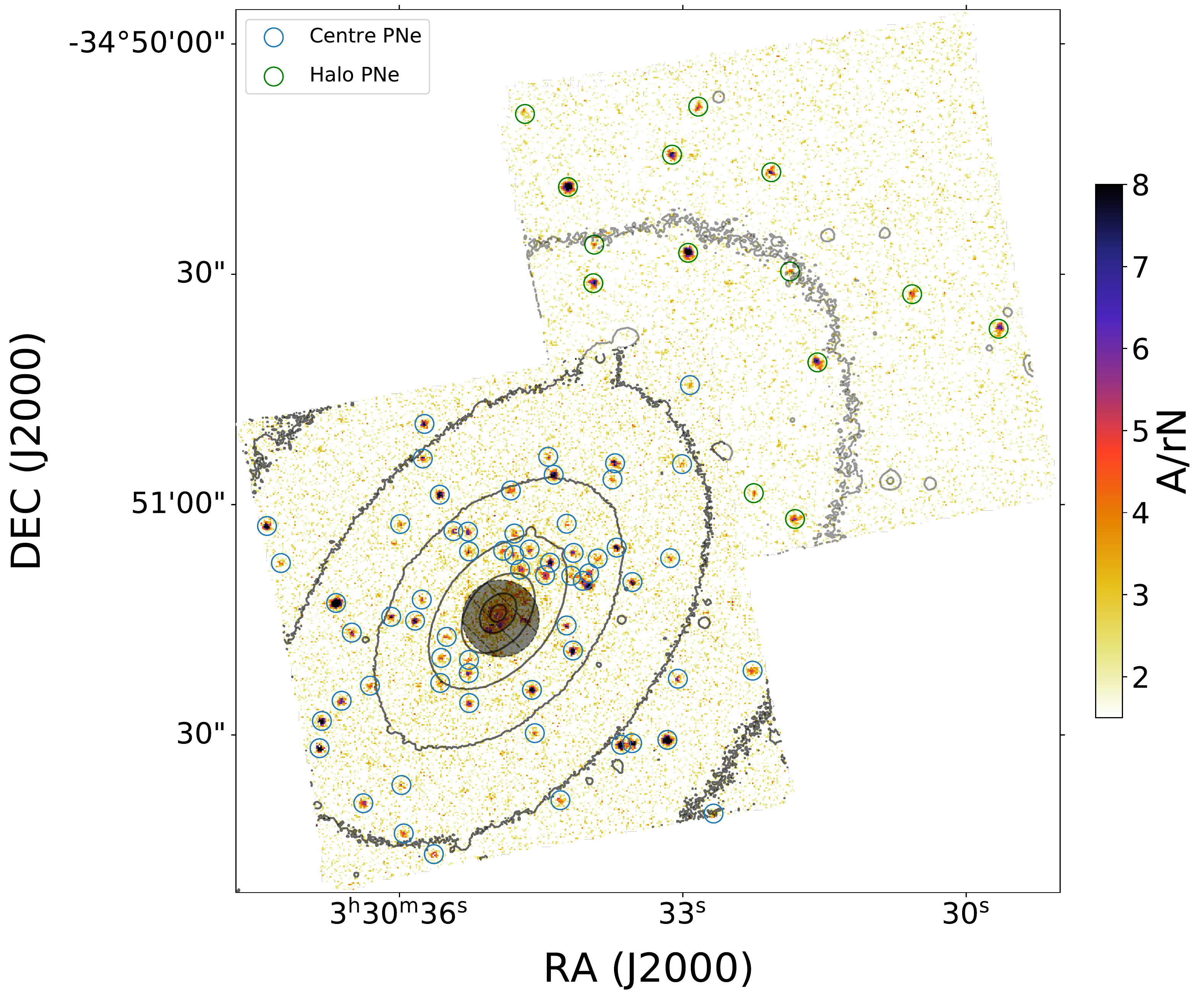} & \includegraphics[width=9cm]{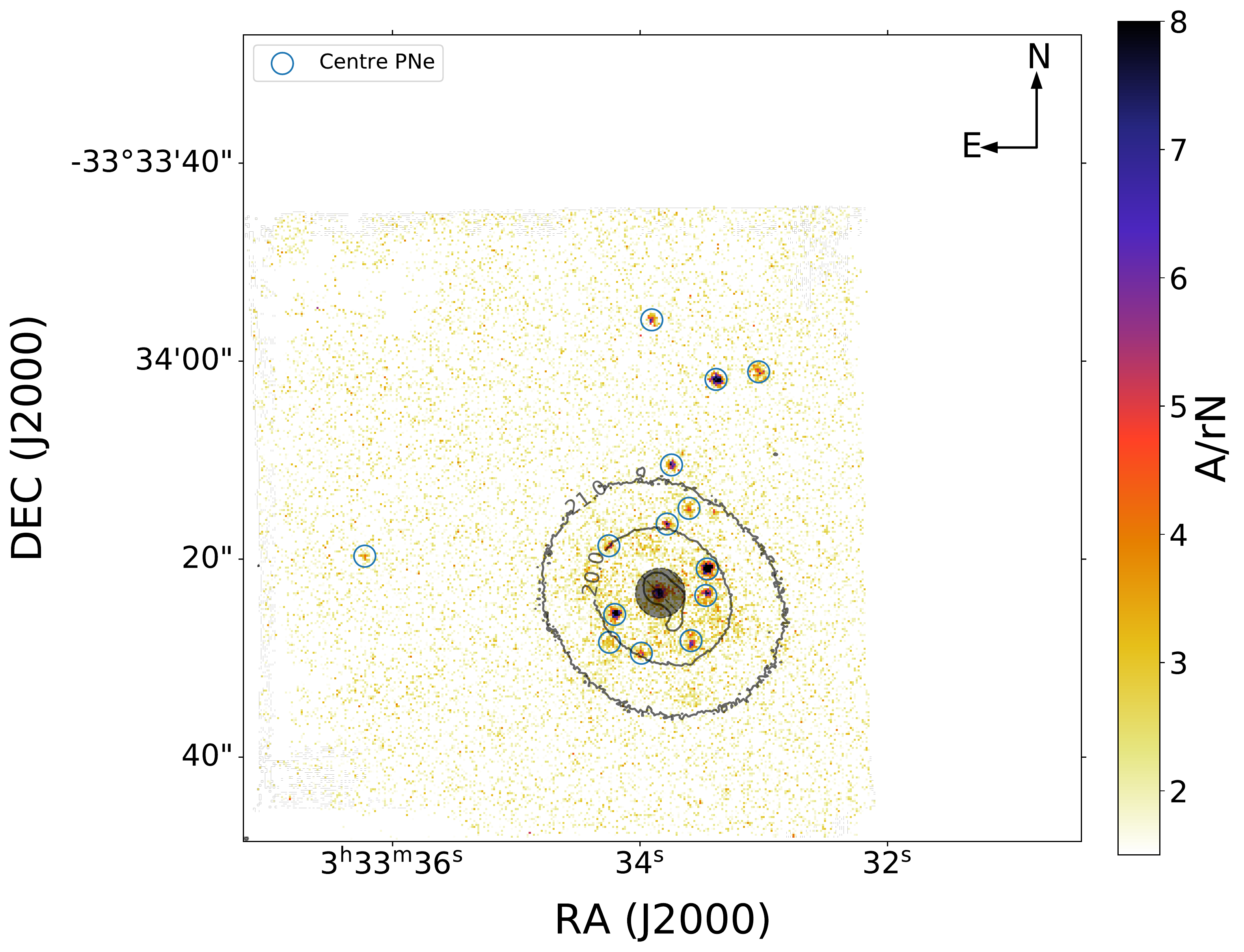} \\
    \tiny FCC083 & \tiny FCC119 \\
    \includegraphics[width=9cm]{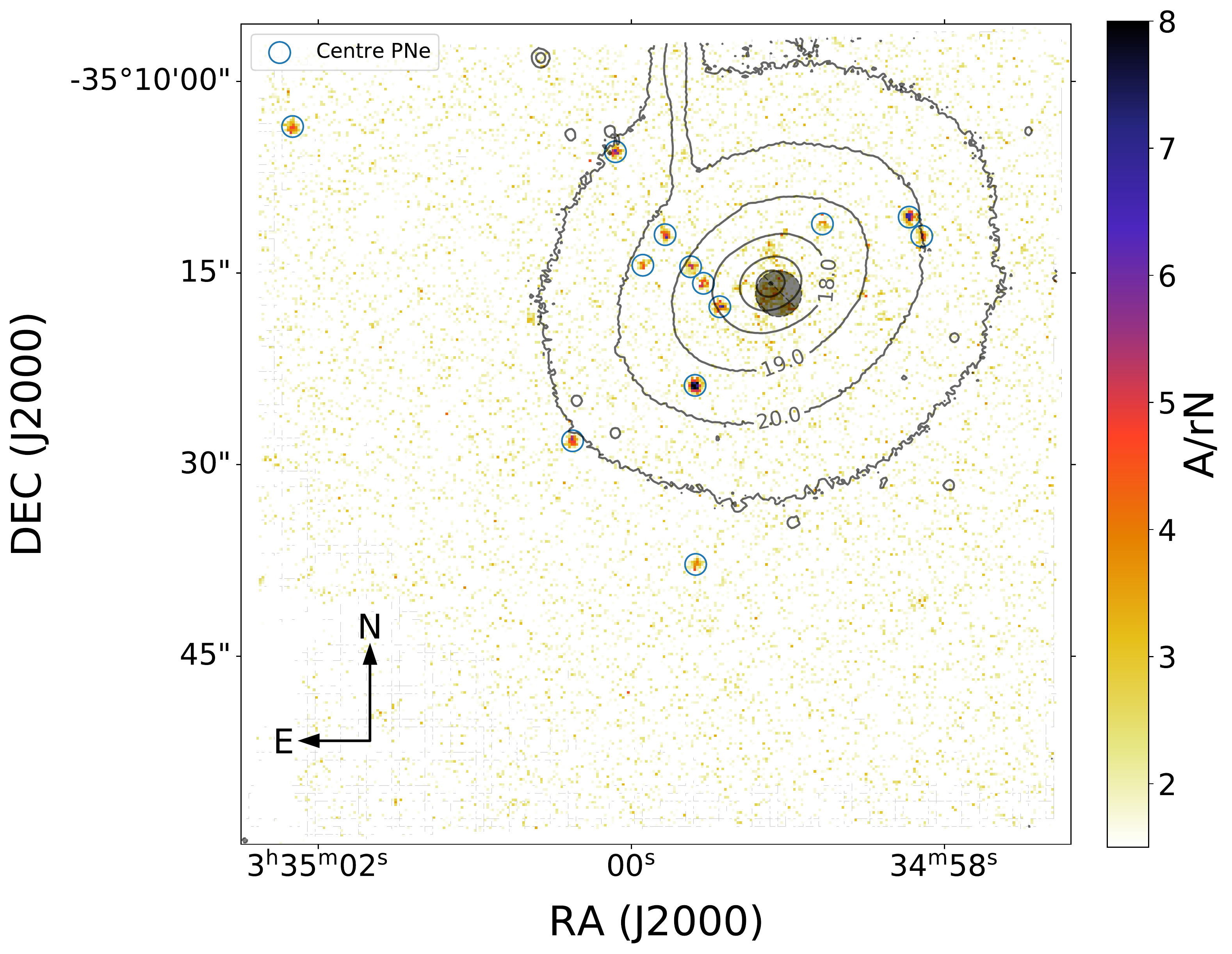} &
    \includegraphics[width=9cm]{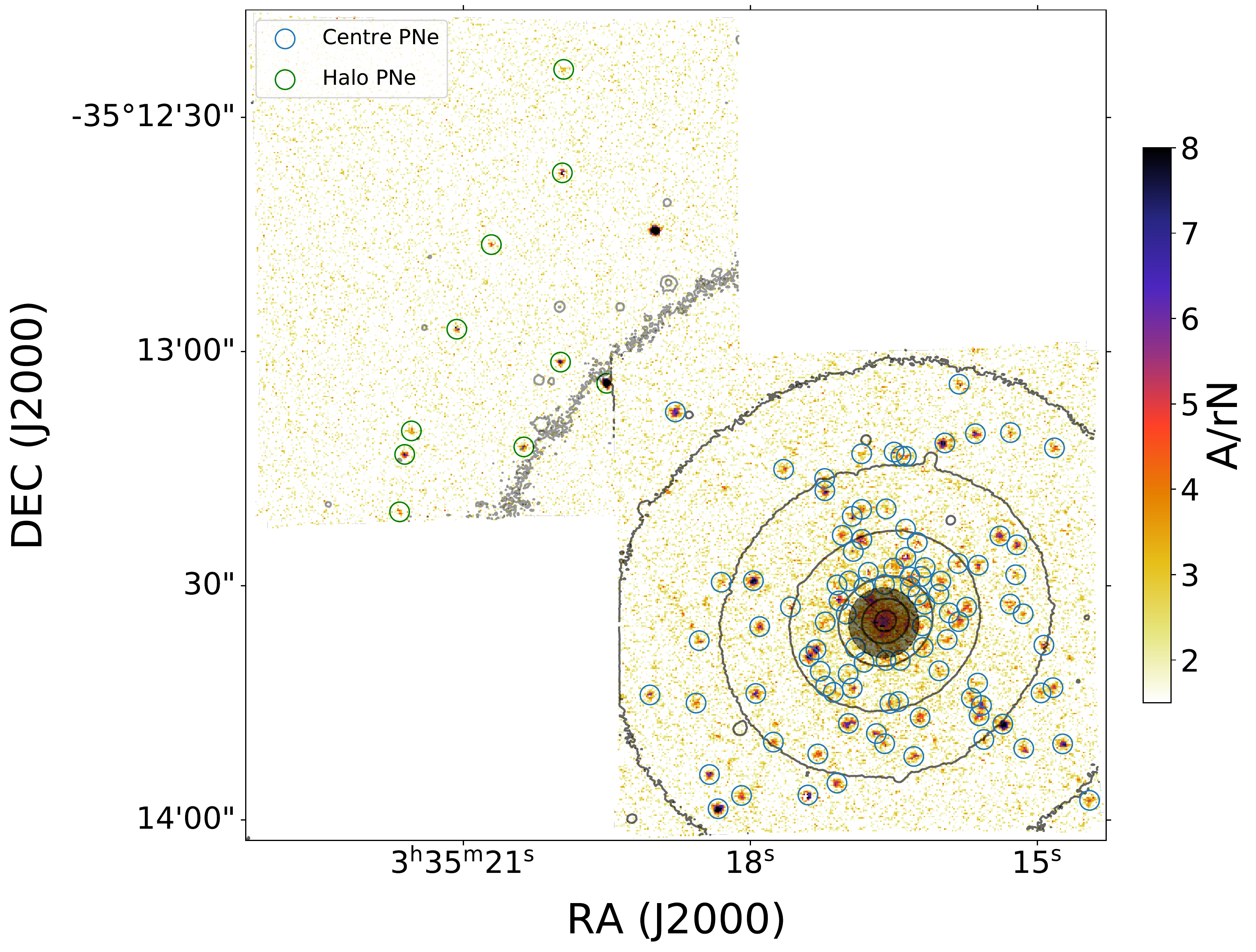} \\
    \tiny FCC143 & \tiny FCC147 \\
    \multicolumn{2}{c}{\includegraphics[width=12cm]{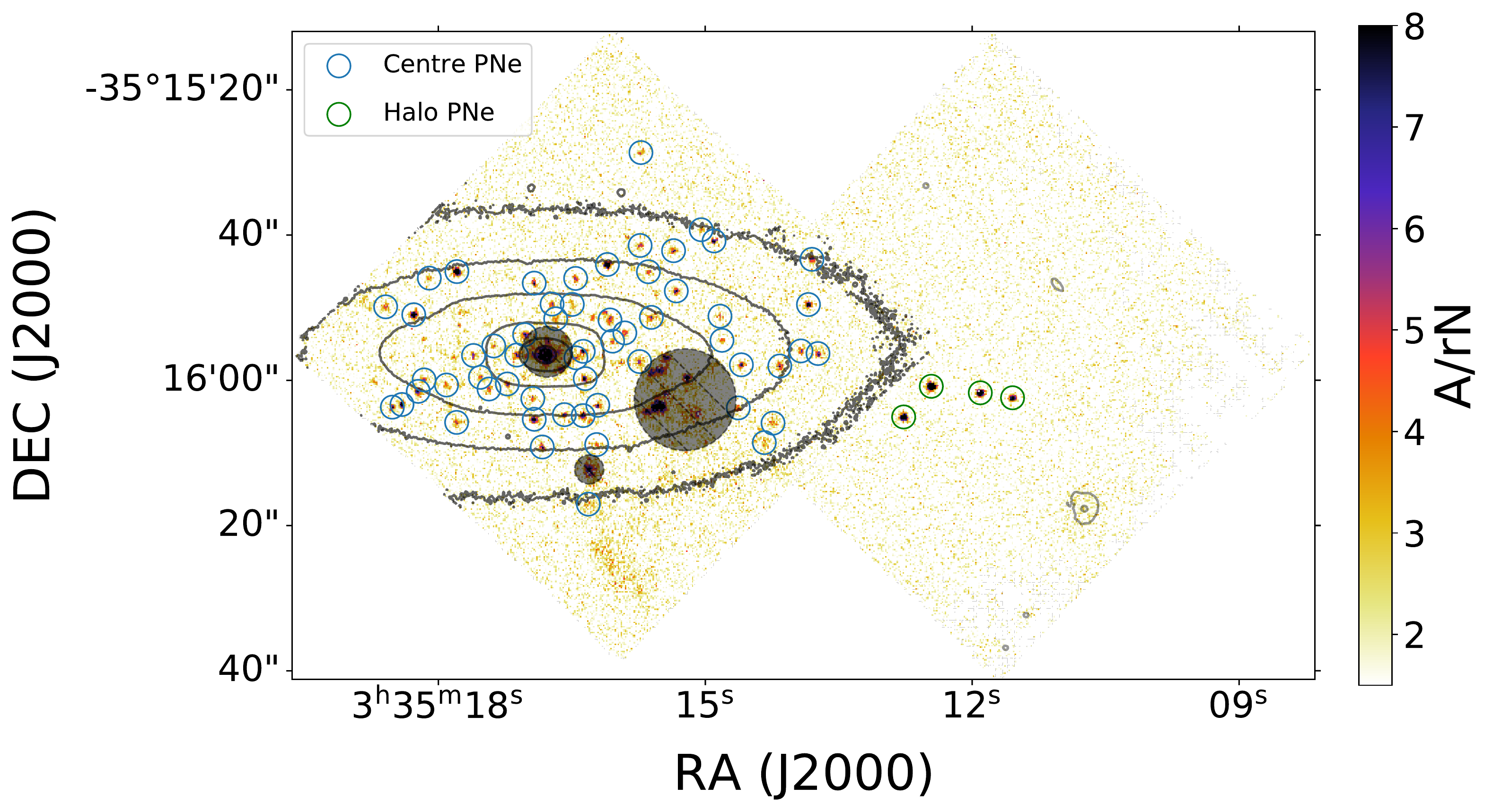} } \\
    \multicolumn{2}{c}{\tiny FCC148 }
    \end{tabular}
\end{figure}

\begin{figure}
    \centering
    \begin{tabular}{cc}
    \includegraphics[width=8.5cm]{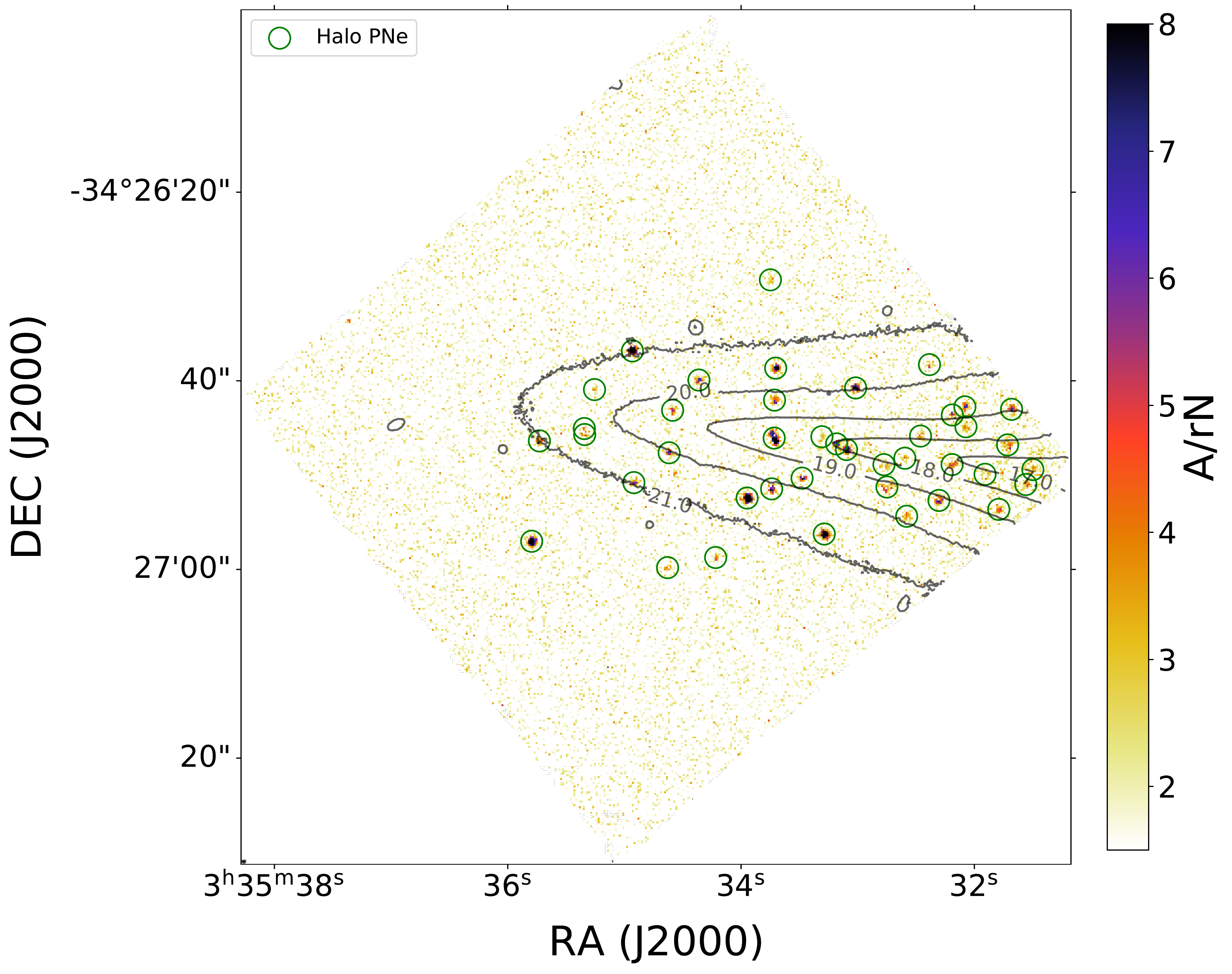} &
    \includegraphics[width=8.5cm]{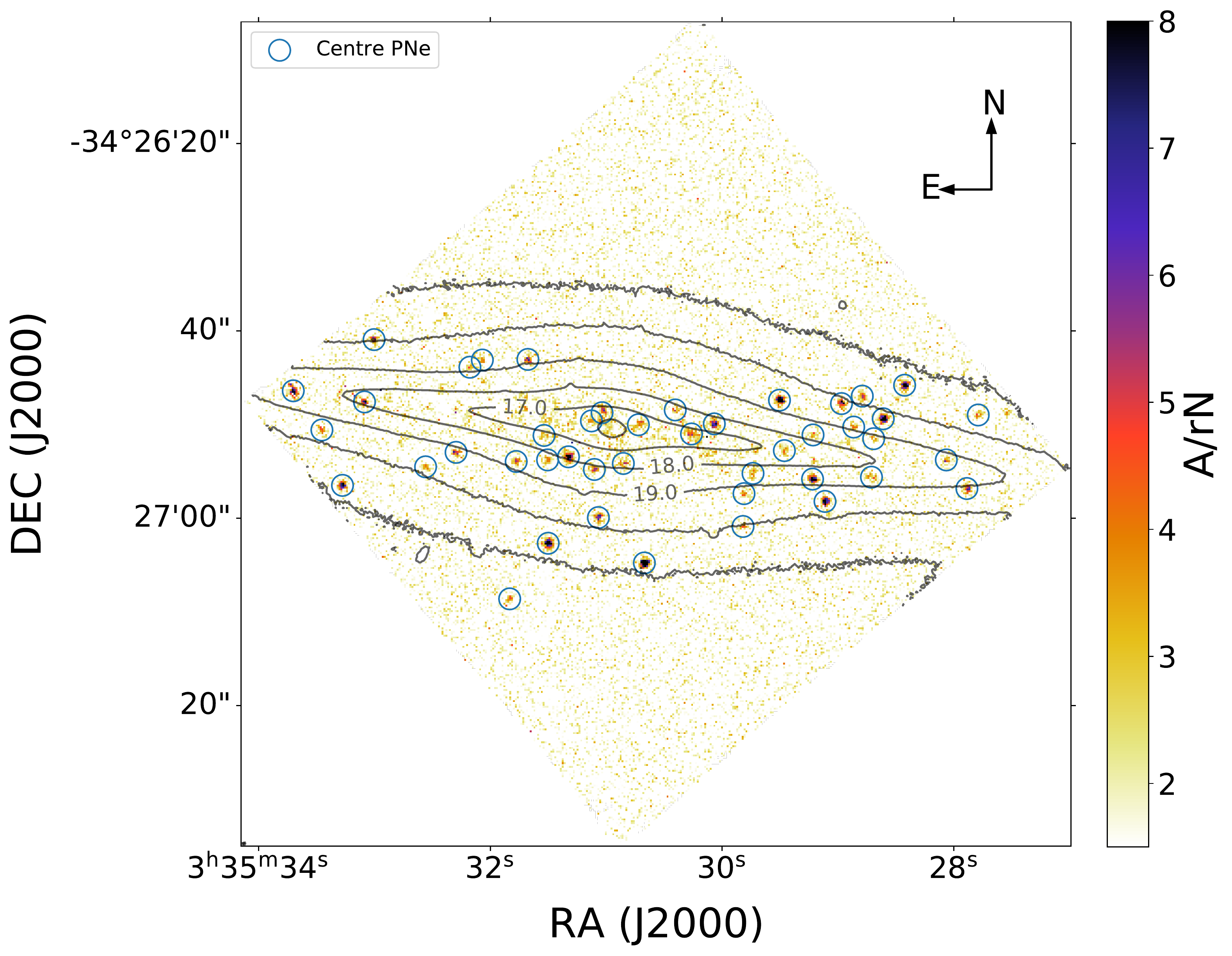} \\
    \tiny FCC153 - Halo & \tiny FCC153 - centre \\
    \multicolumn{2}{c}{\includegraphics[width=10cm]{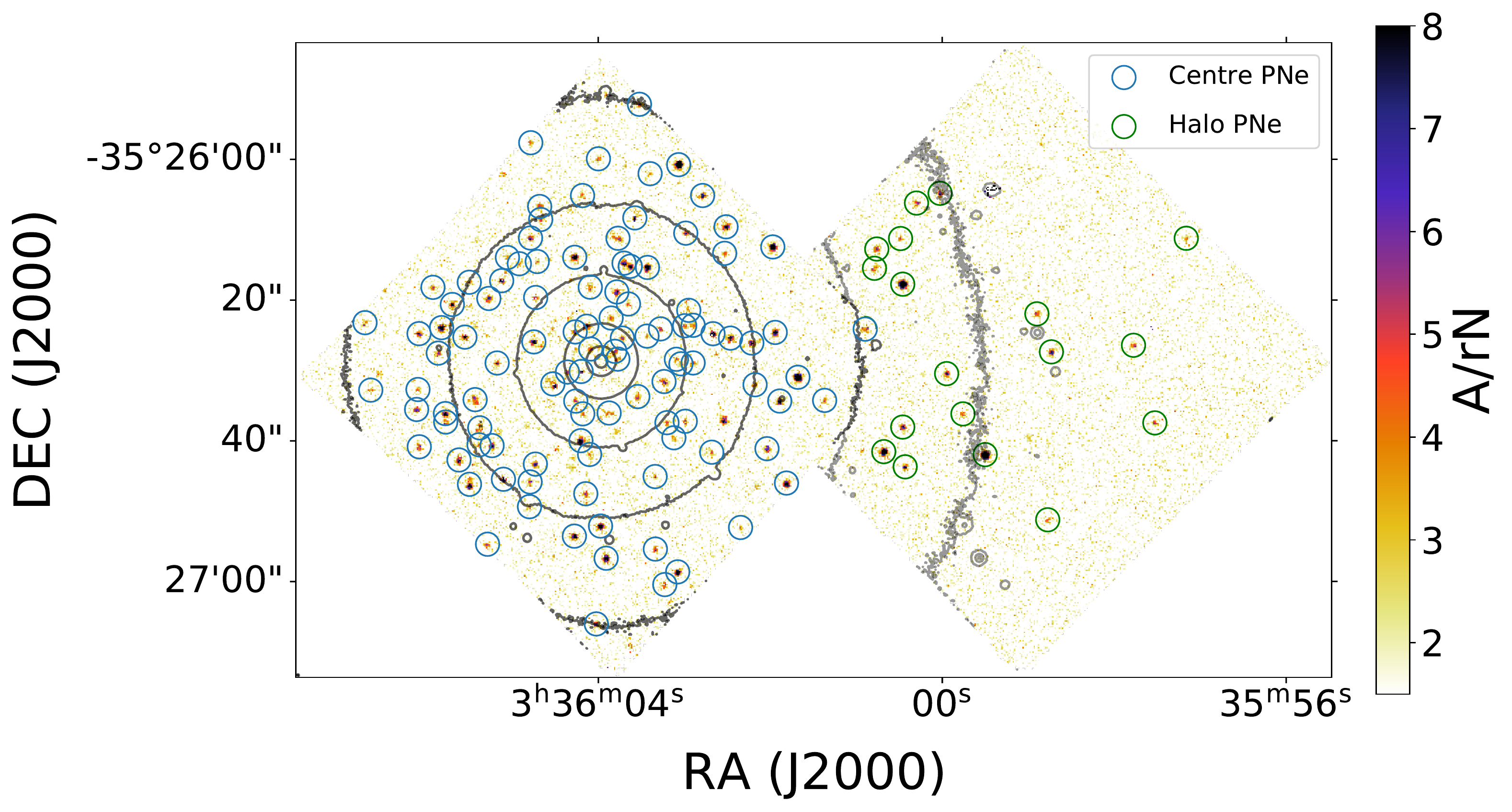} } \\
    \multicolumn{2}{c}{\tiny FCC161}\\
    \label{fig:my_label}
    \end{tabular}
    
    \begin{tabular}{cc}
    \includegraphics[width=7cm]{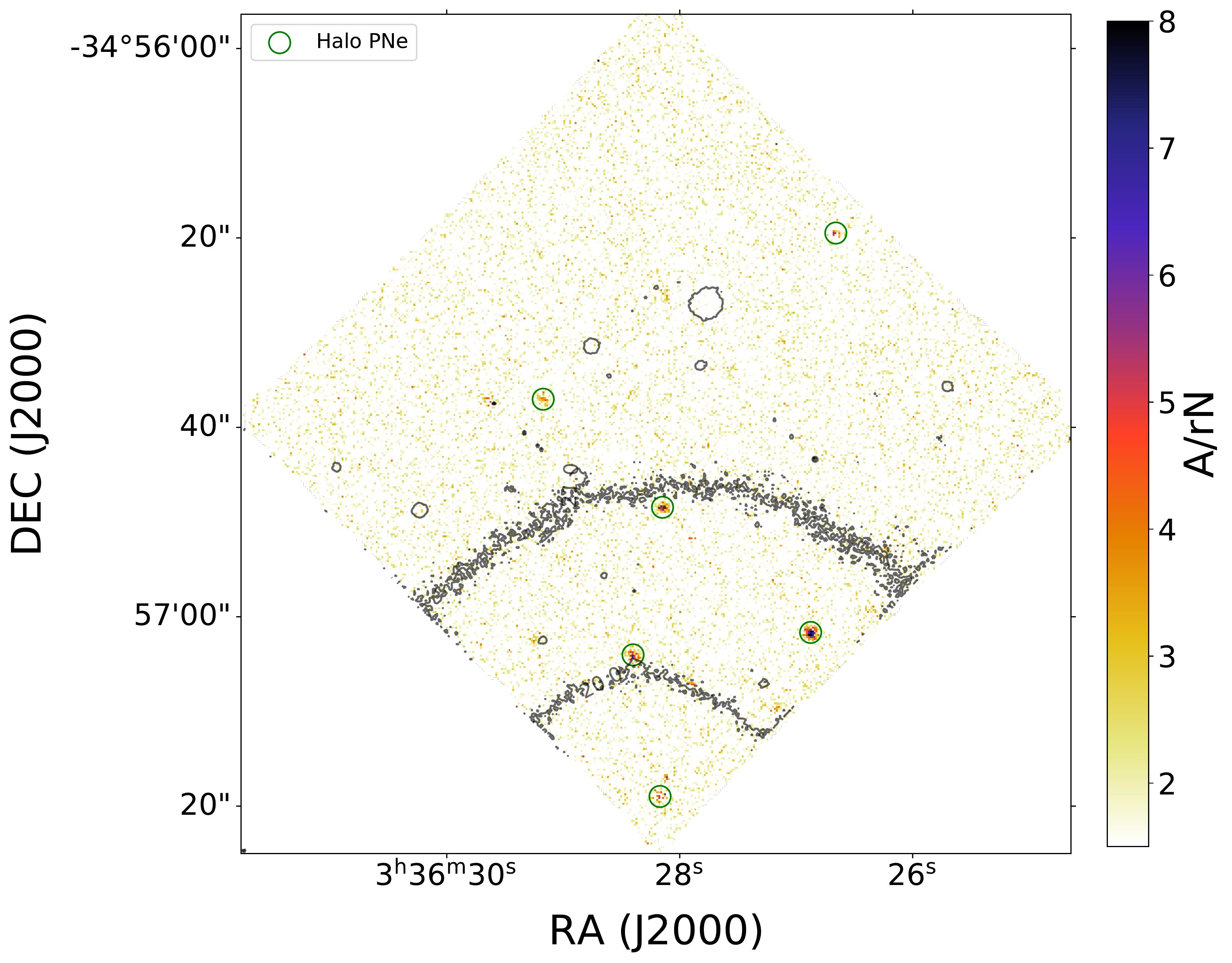} &
    \includegraphics[width=7cm]{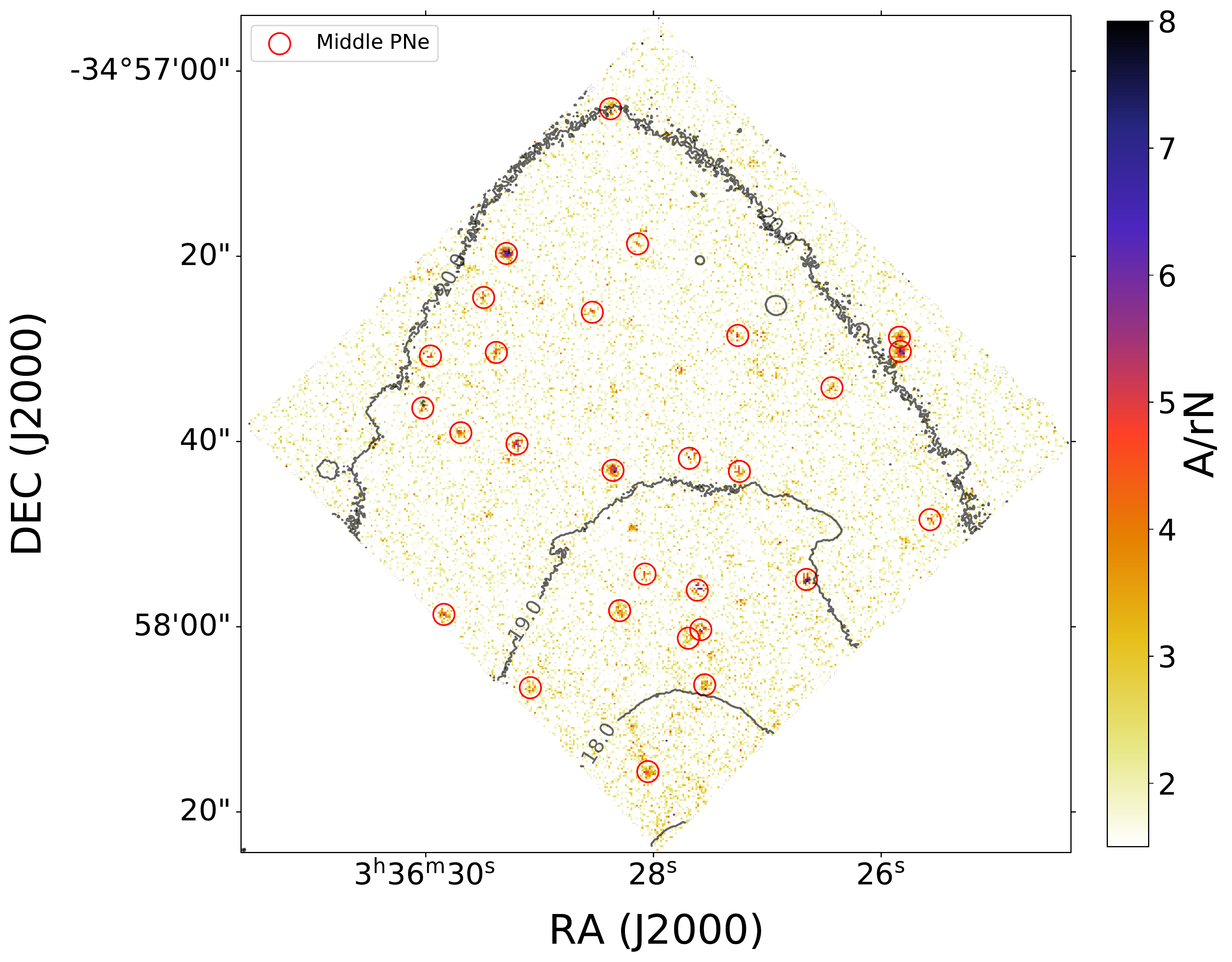} \\ 
    \tiny FCC167 - Halo & \tiny FCC167- Disk \\
    \multicolumn{2}{c}{\includegraphics[width=7cm]{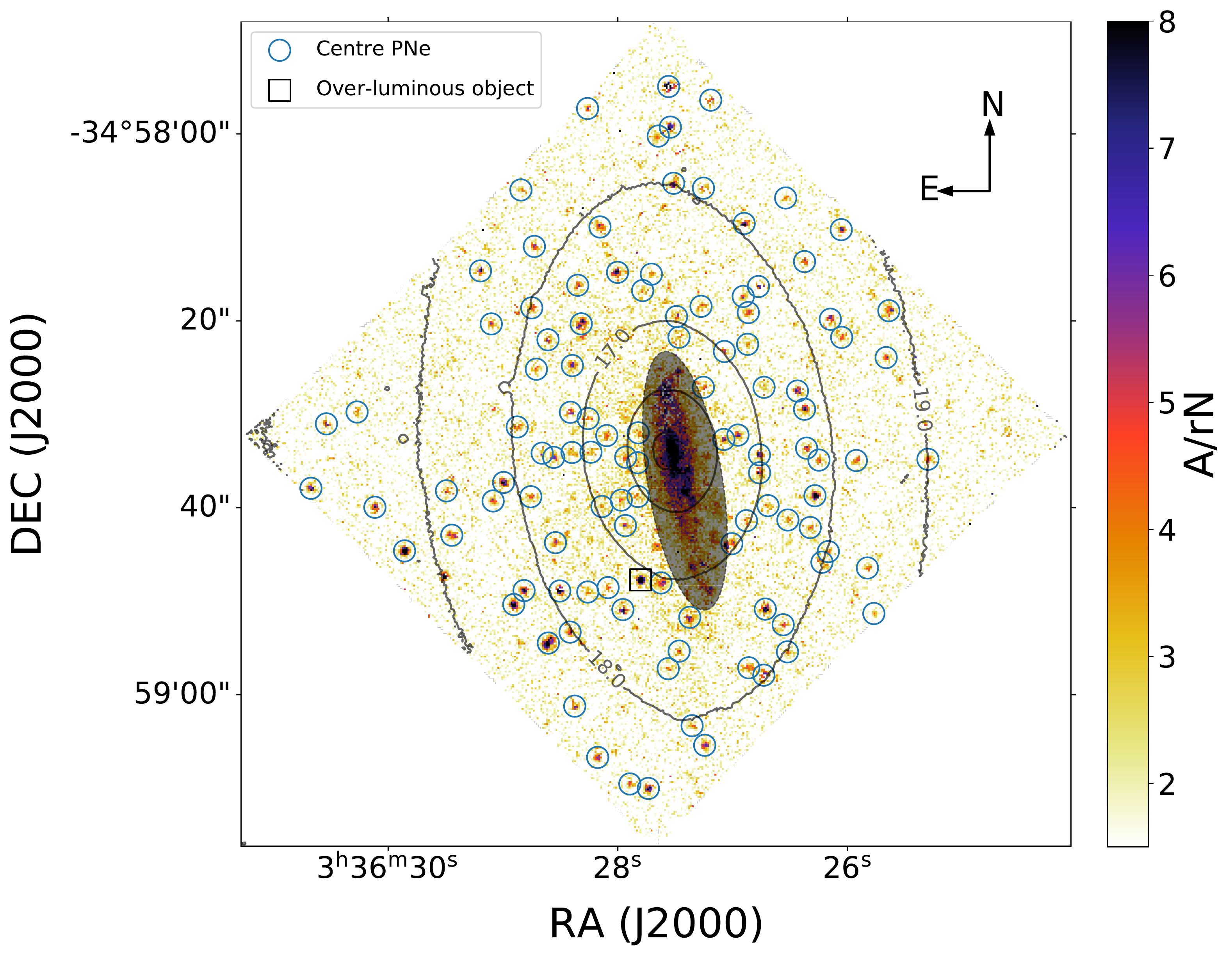} }\\
    \multicolumn{2}{c}{\tiny FCC167 - Centre} \\
    \end{tabular}
\end{figure}

\begin{figure}
    \centering
    \begin{tabular}{cc}
    \includegraphics[width=8.5cm]{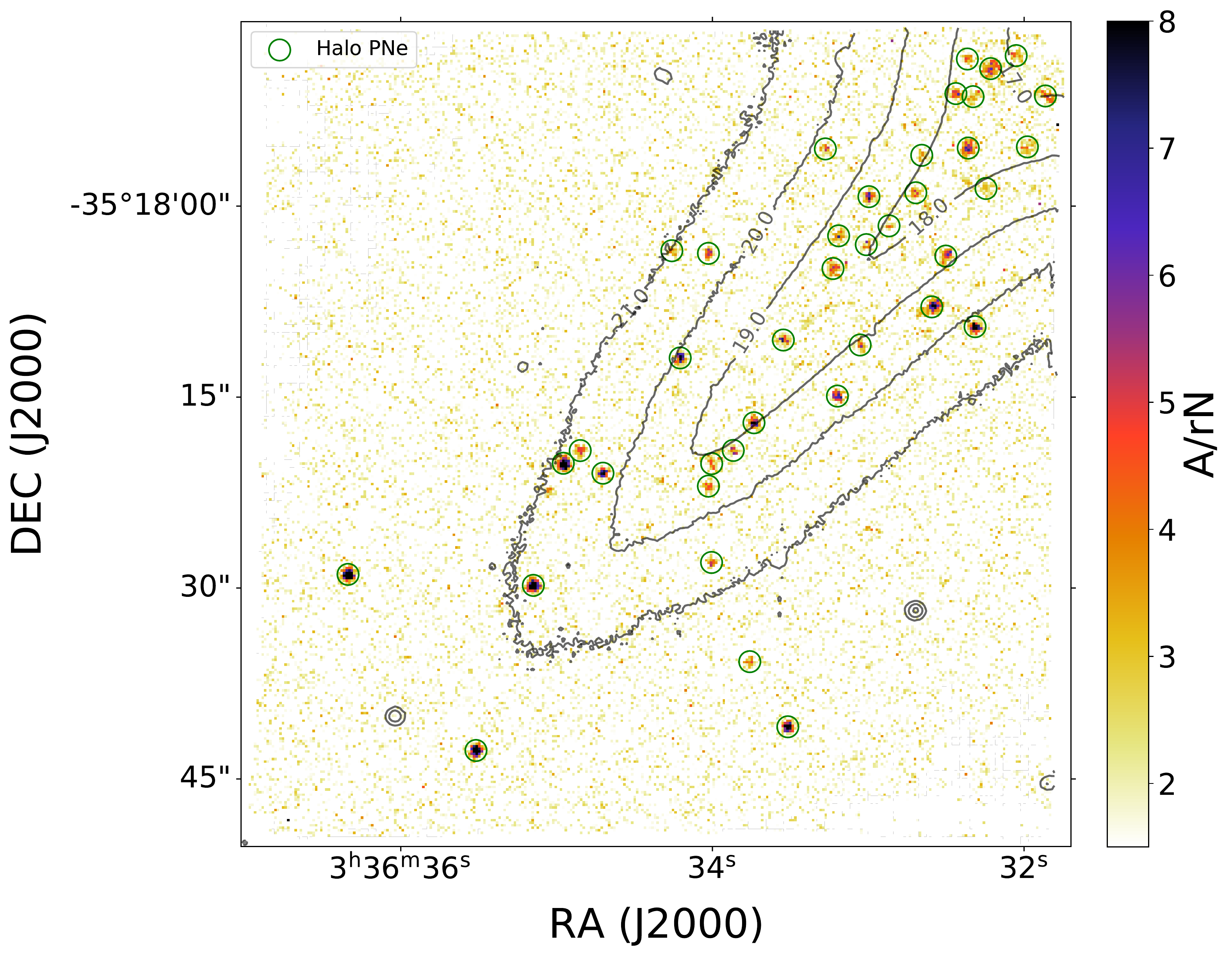} &
    \includegraphics[width=8.5cm]{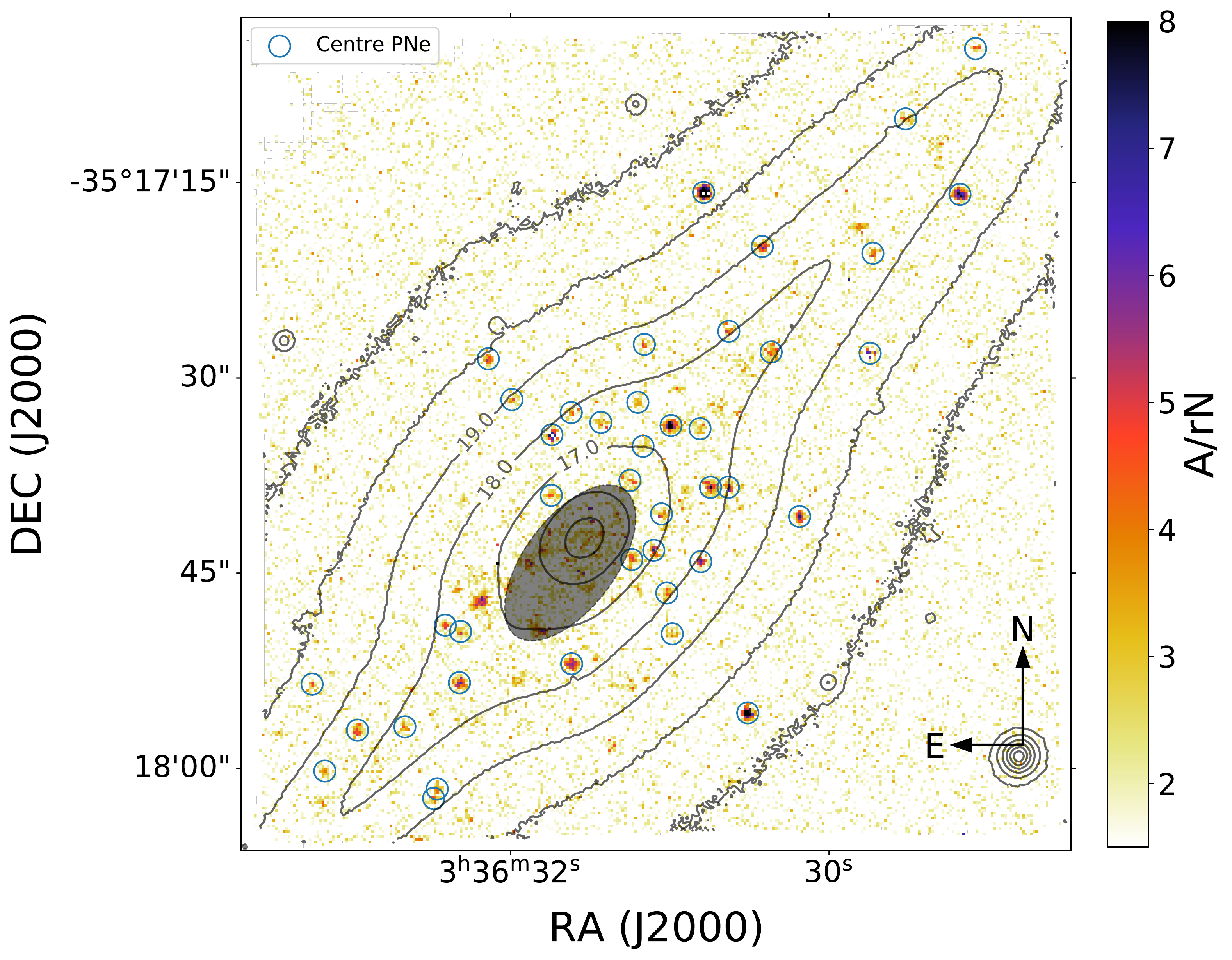} \\ 
    \tiny FCC170 - Halo & \tiny FCC170 - Centre \\
    \includegraphics[width=8.5cm]{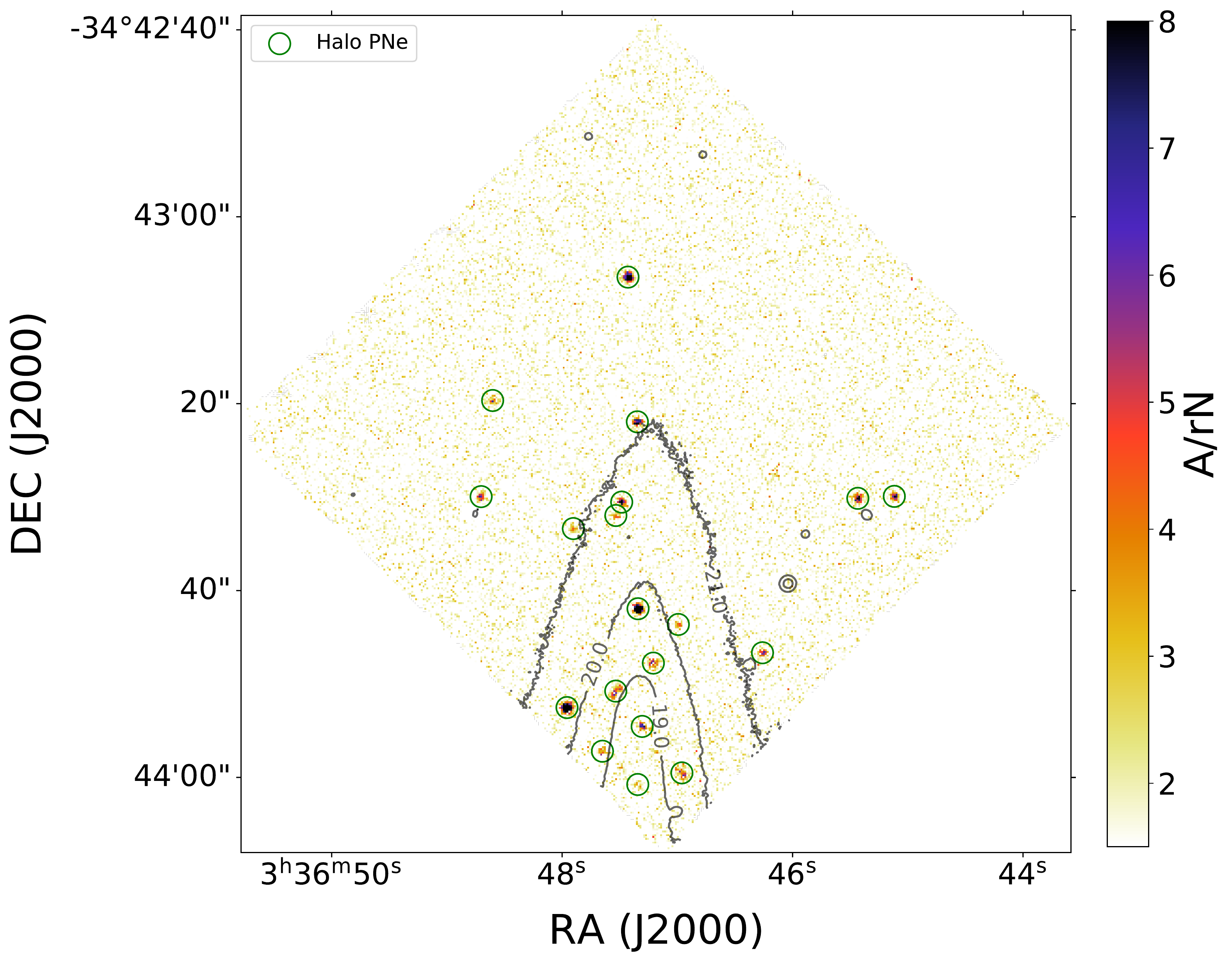} &
    \includegraphics[width=8.5cm]{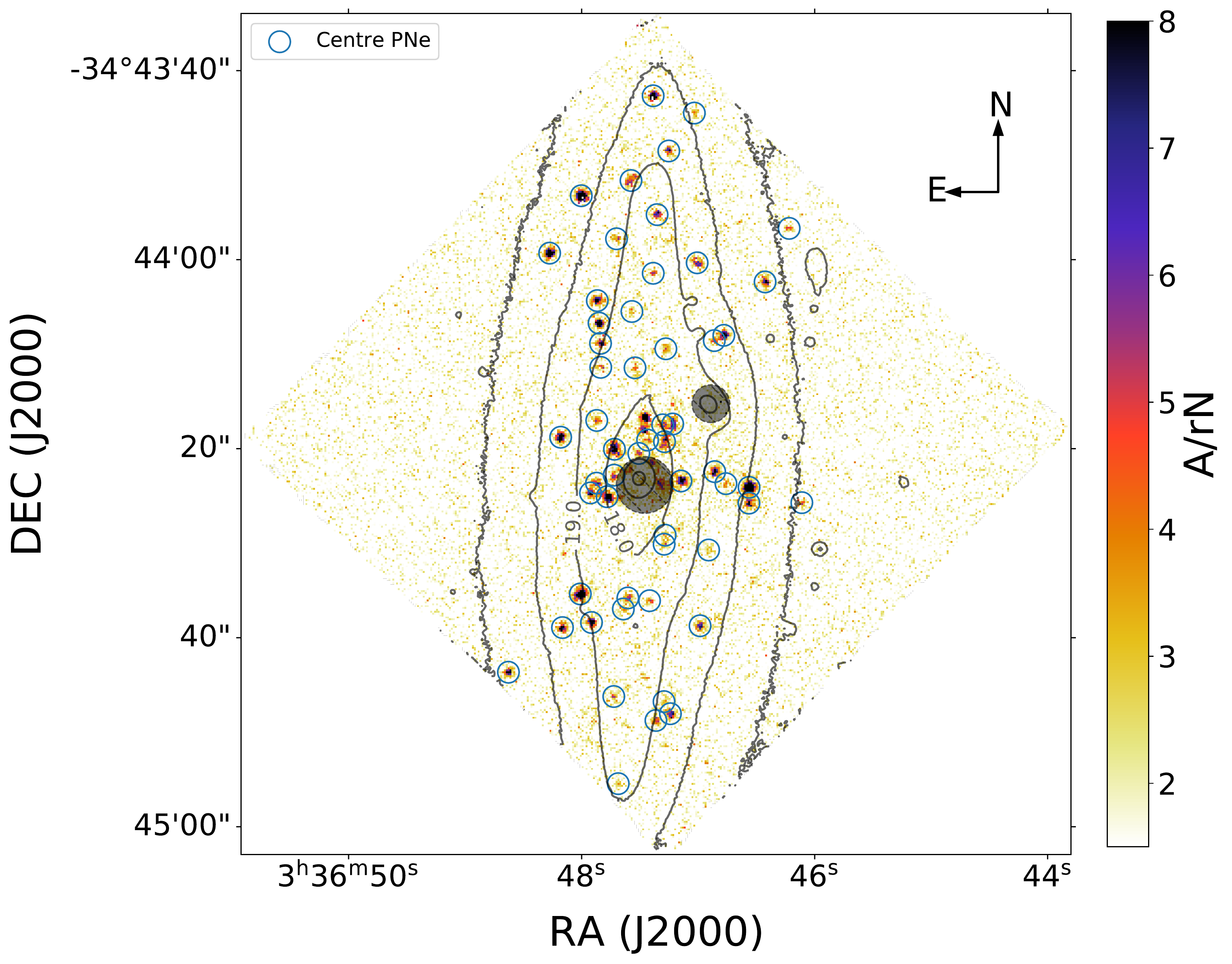} \\ 
    \tiny FCC177 - Halo & \tiny FCC177 - Centre \\
    \includegraphics[width=8.5cm]{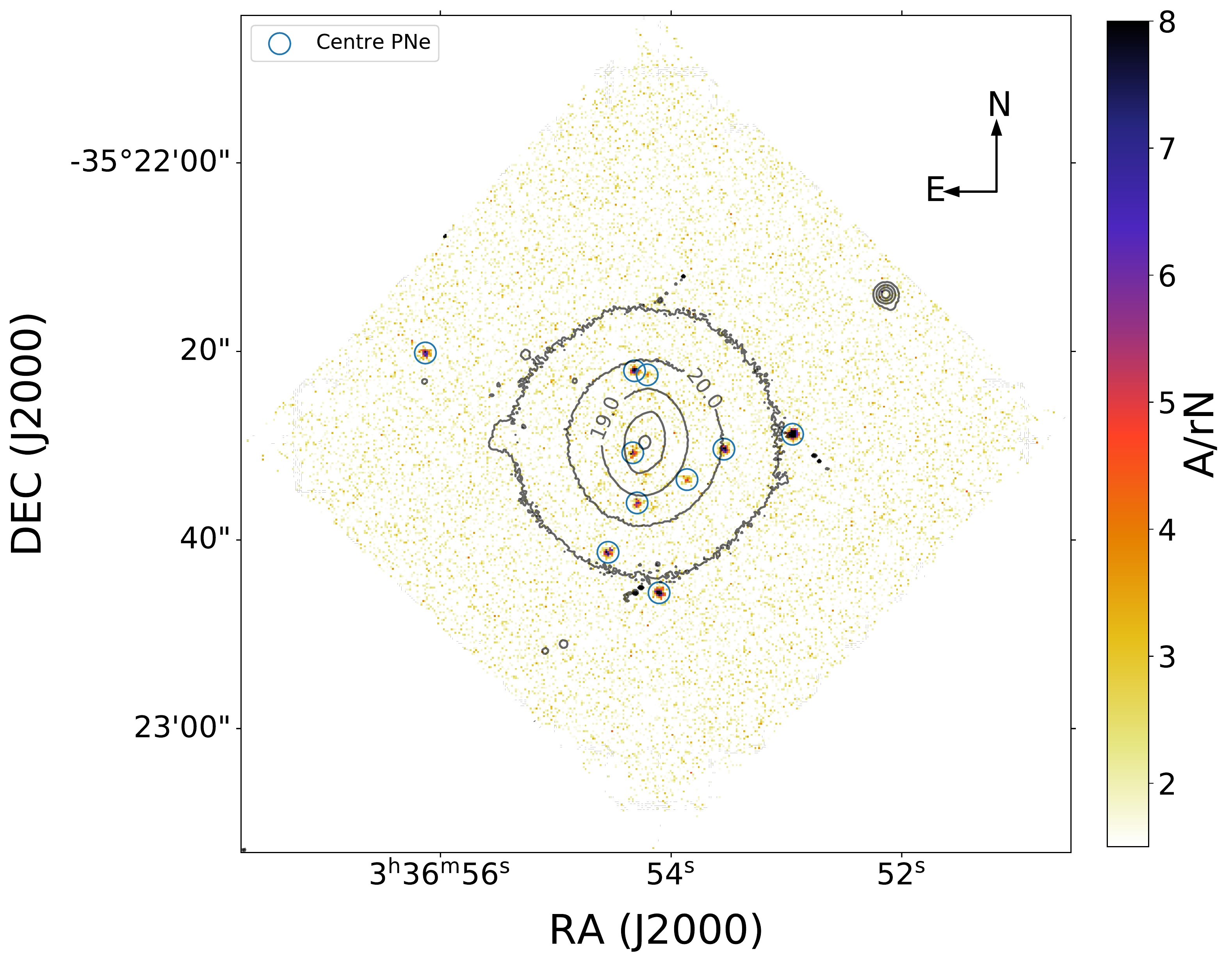} & \includegraphics[width=10cm]{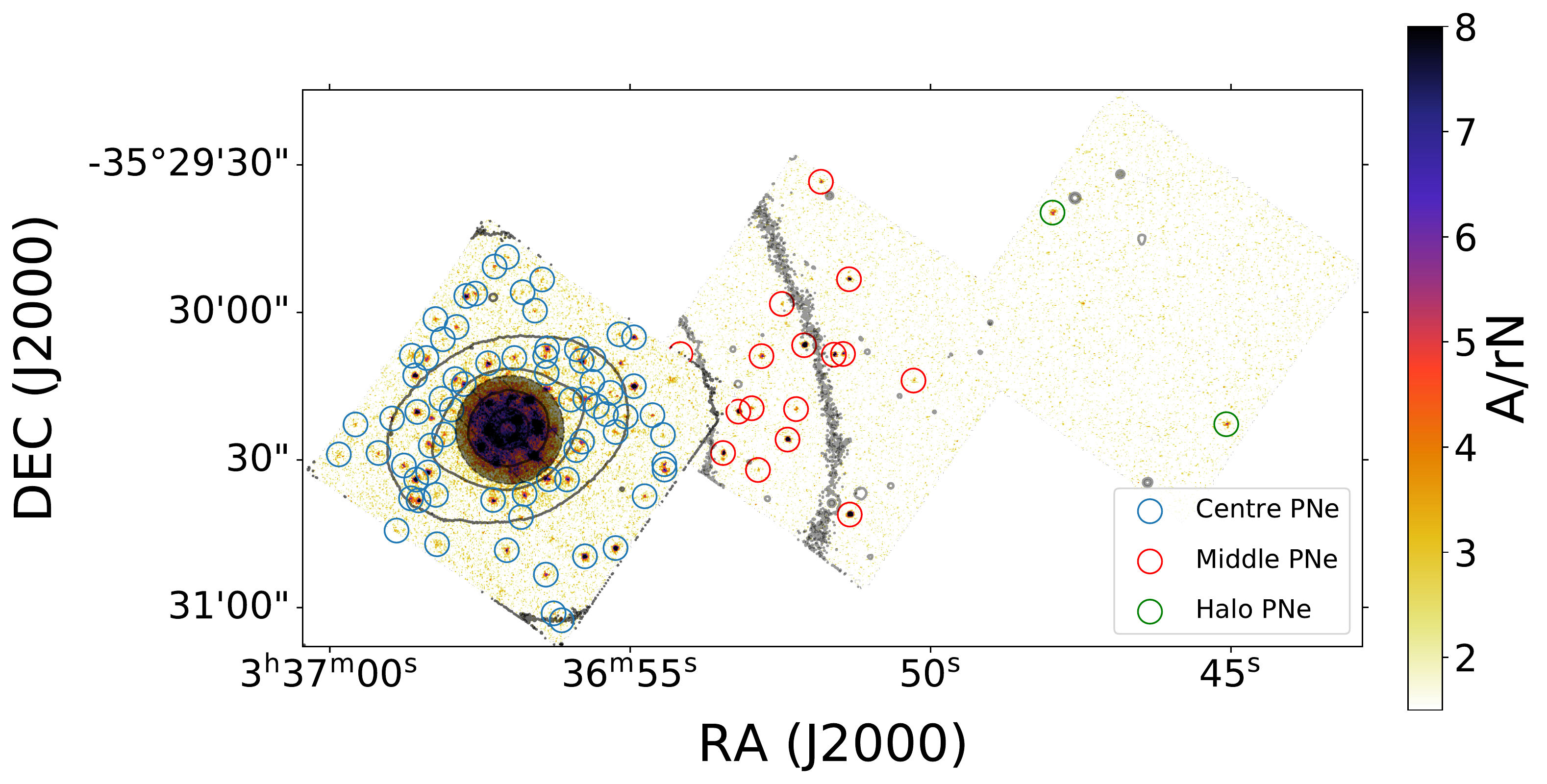} \\
    \tiny FCC182 & \tiny FCC184 \\
    \end{tabular}
\end{figure}

\begin{figure}
    \centering
    \begin{tabular}{cc}
    \includegraphics[width=8cm]{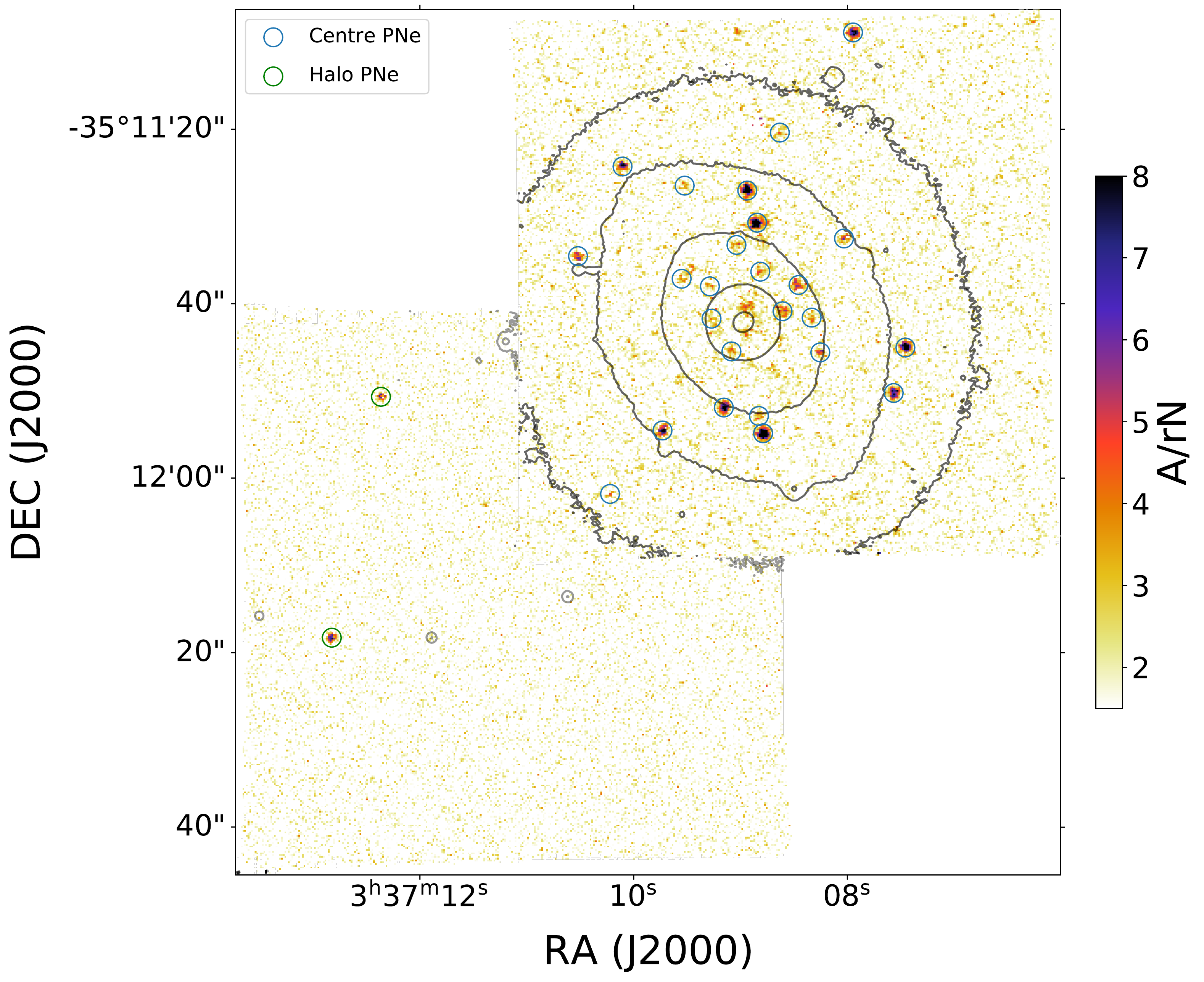} & \includegraphics[width=8cm]{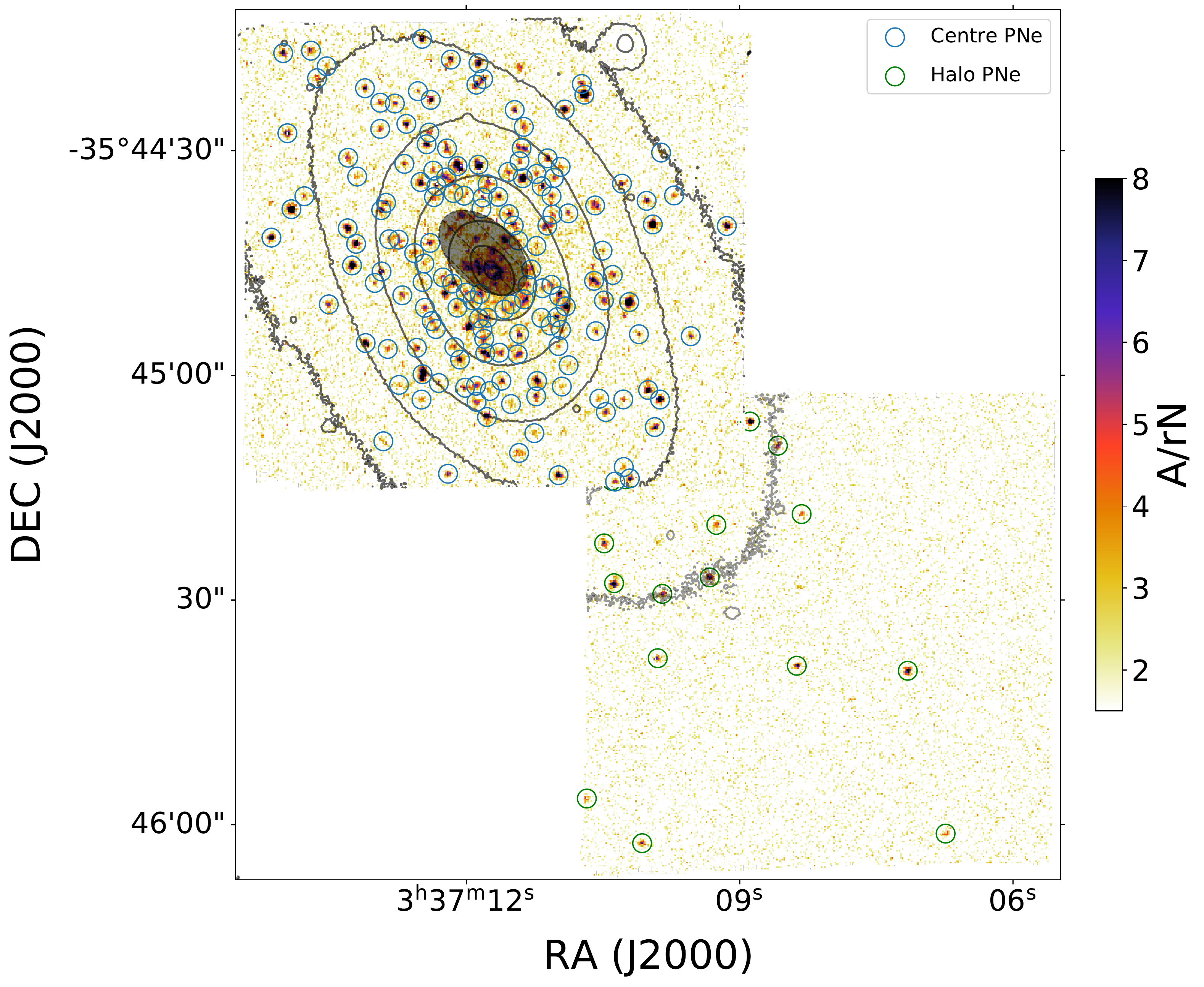} \\
    \tiny FCC190 & \tiny FCC193 \\
    \includegraphics[width=8cm]{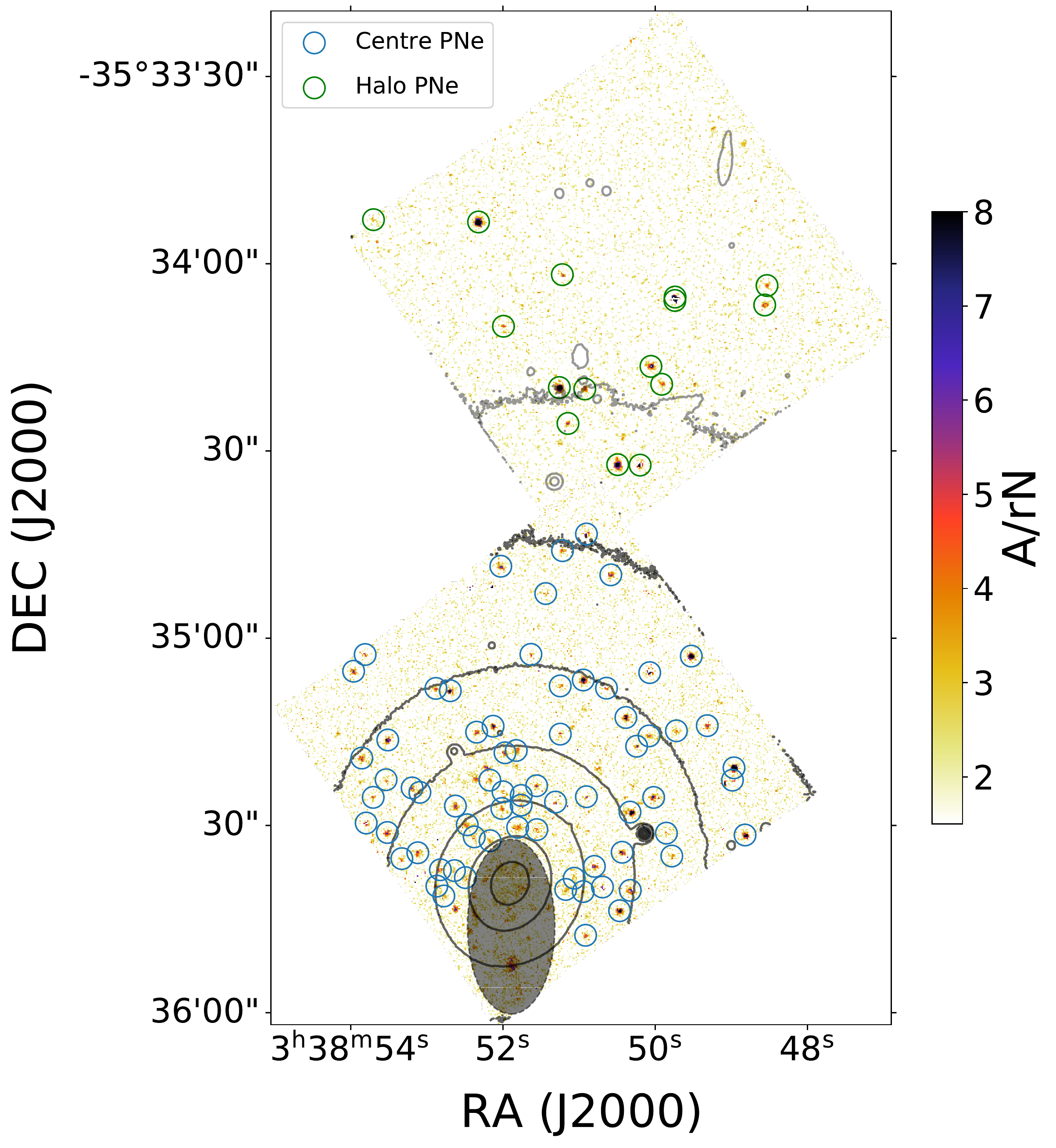} &
    \includegraphics[width=8cm]{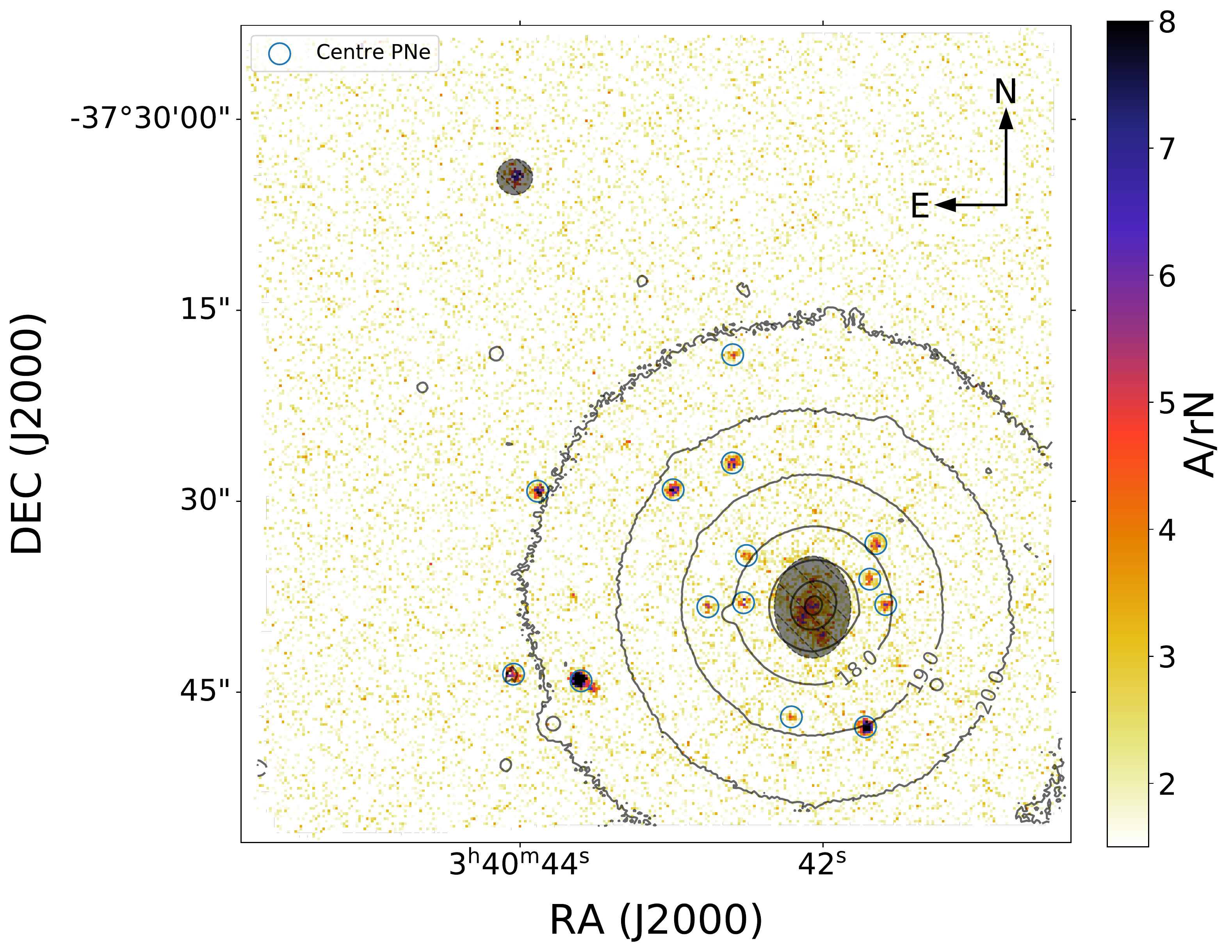} \\
    \tiny FCC219 & \tiny FCC249 \\
    \includegraphics[width=8cm]{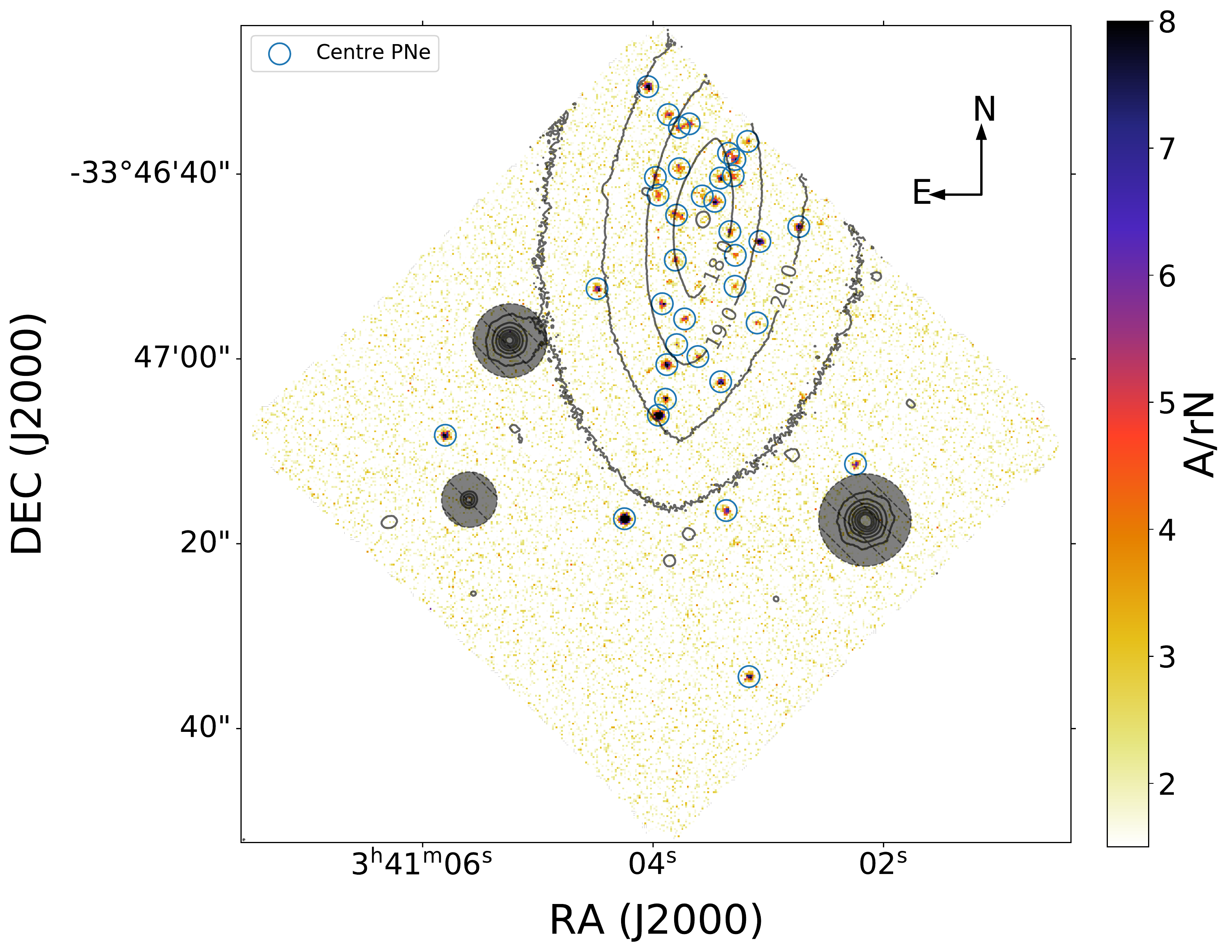} &
    \includegraphics[width=8cm]{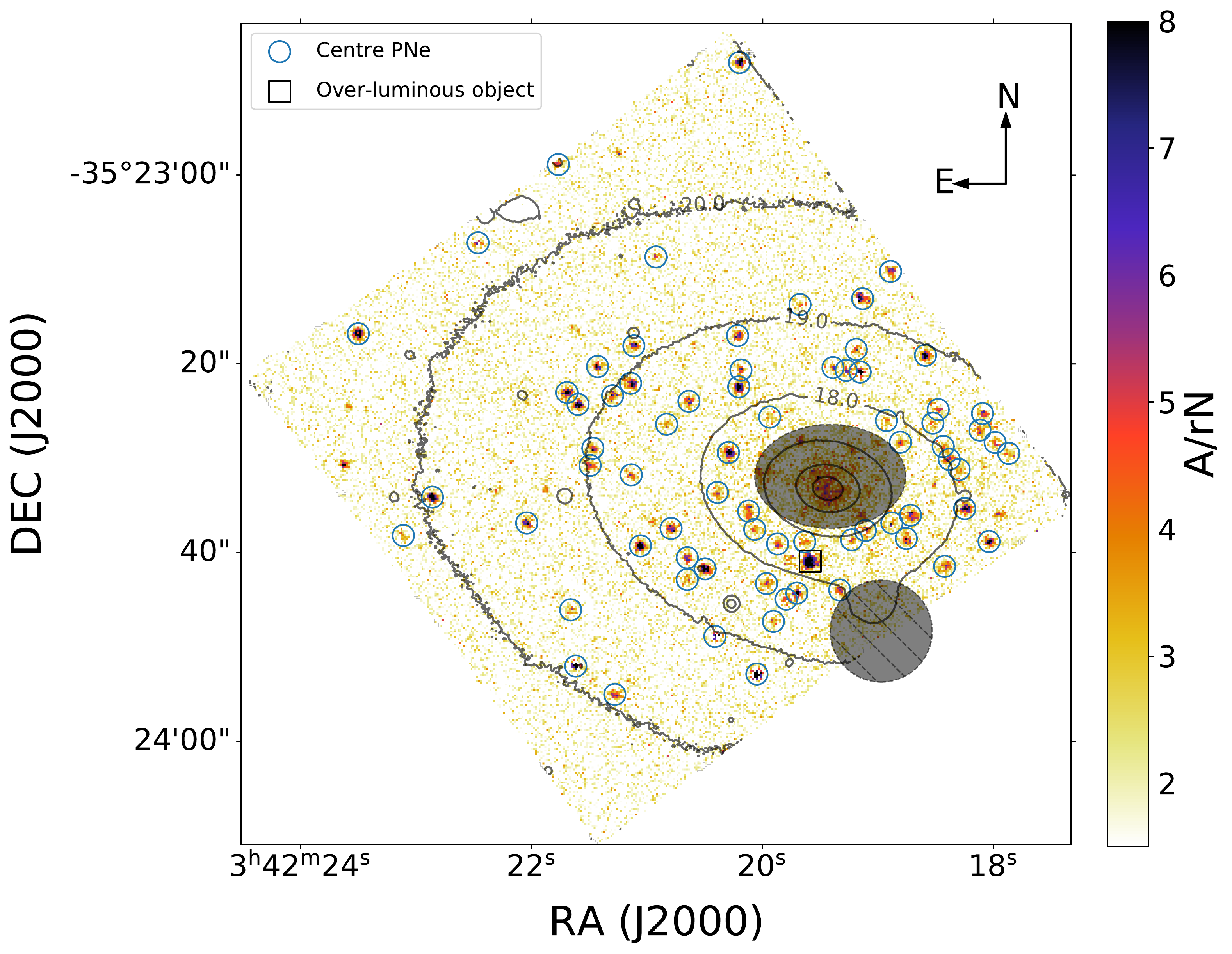} \\
    \tiny FCC255 & \tiny FCC276 \\
    \end{tabular}
\end{figure}

\begin{figure}
    \centering
    \begin{tabular}{cc}
    \includegraphics[width=8cm]{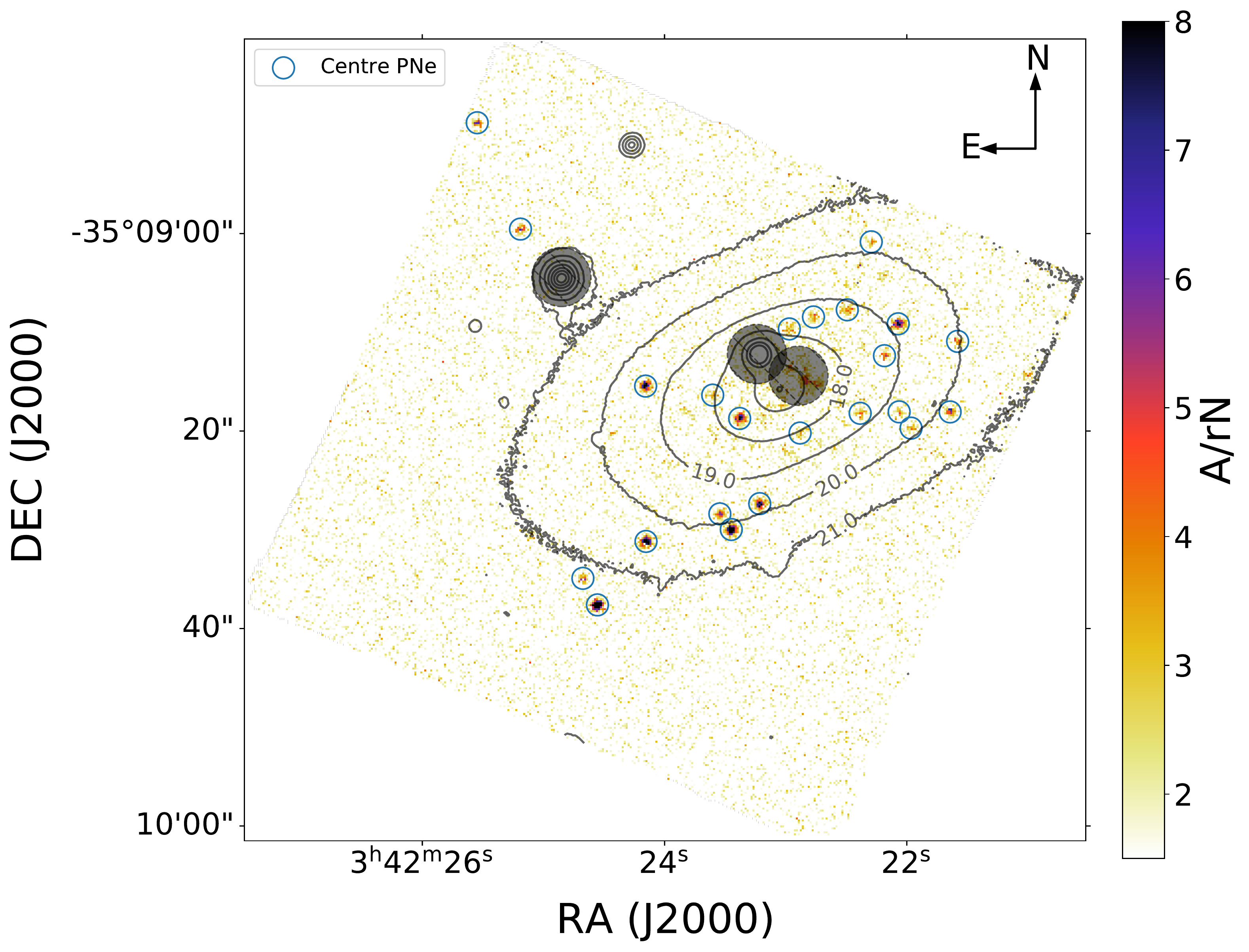} &
    \includegraphics[width=8cm]{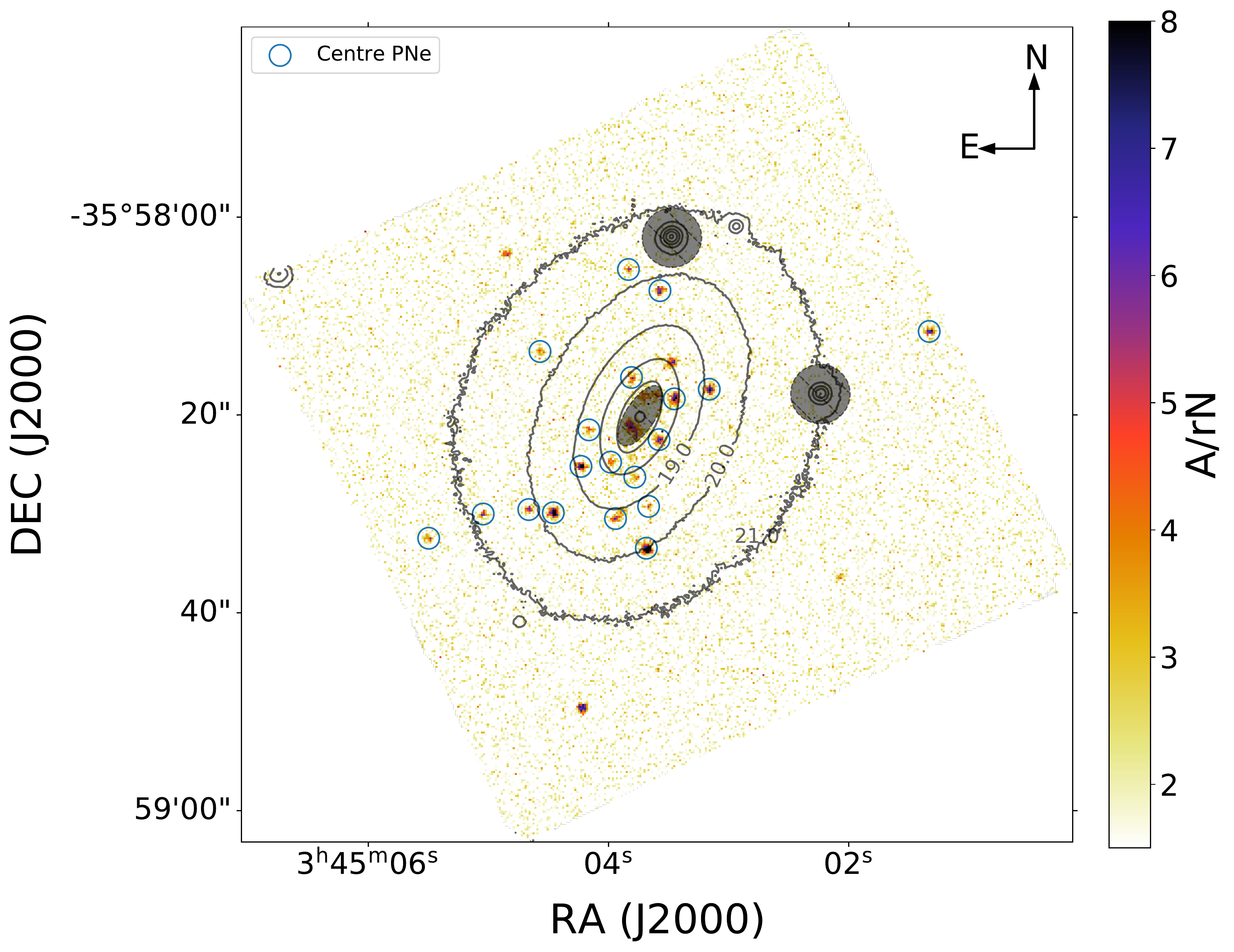} \\
    \tiny FCC277 & \tiny FCC301 \\
    \multicolumn{2}{c}{\includegraphics[width=8cm]{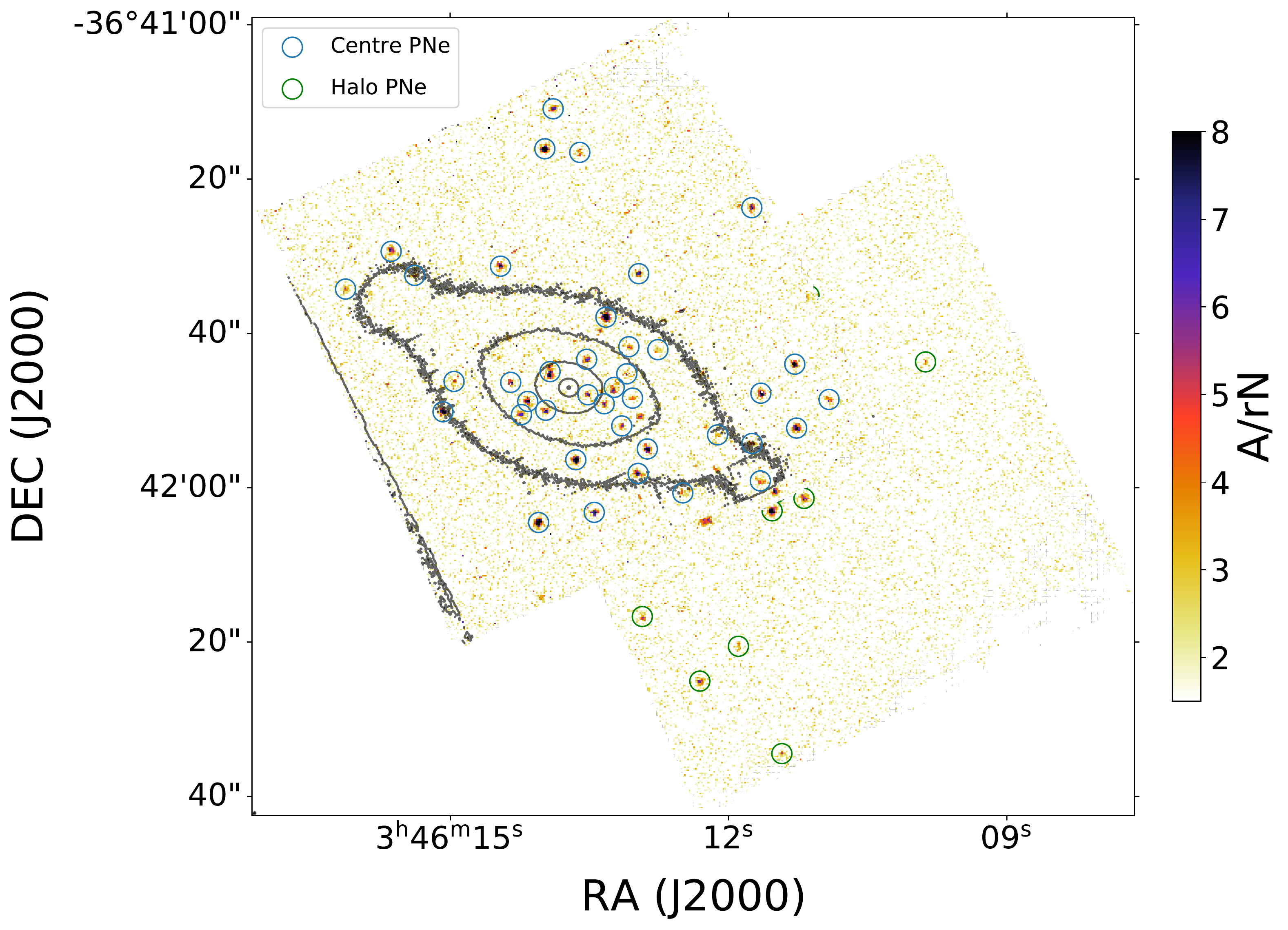} } \\
    \multicolumn{2}{c}{\tiny FCC310}
    \end{tabular}
    \caption{Maps for the A/rN ratio between the [\ion{O}{iii}] 5007 \AA{} line amplitude and residual-noise level derived from our spaxel-by-spaxel spectral fit to the MUSE spectra. Both central and halo pointing are shown, with detected [\ion{O}{iii}] sources identified by coloured circles: Blue for central PNe, red for Disk (or Middle) PNe and green for Halo PNe. Isophotes are spaced at one mag arcsec$^{-1}$ intervals. The grey dashed regions indicates the area excluded from our PNe analysis.}
    \label{fig:F3D_FOV}
\end{figure}
}

\newcommand{\placegalaxytable}{
    \begin{table*}
        \caption{PNe numbers and distances to each ETG.}
        \label{tab:galaxy_info}
        \centering
        \renewcommand{\arraystretch}{1.5}
        \begin{tabular}{c c c c c c | c c | c}
            \hline
            Galaxy & Pointing & $N_\mathrm{{PNe}}$ & $N_{2.5}$ & $\mu_\mathrm{{PNLF}}$ (mag) & $D_\mathrm{{PNLF}}$ (Mpc) & $\mu_\mathrm{SBF}$ (mag) & $\mu_\mathrm{CF2}$ (mag) & Alt. Name\\
            (1) & (2) & (3) & (4) & (5) & (6) & (7) & (8) & (9)\\
            \hline
            FCC083 & center & 62 & $150^{+22}_{-19}$ & $31.34^{+0.14}_{ -0.15 }$ & $18.53^{+1.23}_{-1.30}$ & $31.42 \pm 0.07$ & $31.42 \pm 0.24$ & NGC 1351 \\
            FCC083 & halo & 19 & $26^{+7}_{-6}$ & -- & -- & -- & -- & -- \\
            FCC119 & center & 14 & $23^{+8}_{-6}$ & $31.71^{+0.40}_{ -0.42 }$ & $21.97^{+4.10}_{-4.27}$ & $31.54 \pm 0.10$ & $31.52 \pm 0.24$ & -- \\
            FCC143 & center & 13 & $45^{+16}_{-12}$ & $31.91^{+0.28}_{ -0.27 }$ & $24.11^{+3.14}_{-2.95}$ & $31.43 \pm 0.09$ & $31.22 \pm 0.24$ & NGC 1373 \\
            FCC147 & center & 98 & $270^{+30}_{-27}$ & $31.43^{+0.11}_{ -0.11 }$ & $19.36^{+1.02}_{-0.99}$ & $31.46 \pm 0.07$ & $31.56 \pm 0.24$ & NGC 1374 \\
            FCC147 & halo & 13 & $14^{+5}_{-4}$ & -- & -- & -- & -- & -- \\
            FCC148 & center & 57 & $128^{+19}_{-17}$ & $31.56^{+0.17}_{ -0.16 }$ & $20.54^{+1.60}_{-1.55}$ & $31.50 \pm 0.07$ & $31.48 \pm 0.14$ & NGC 1375 \\
            FCC148 & halo & 16 & $23^{+7}_{-6}$ & -- & -- & -- & -- & -- \\
            FCC153 & center & 44 & $128^{+22}_{-19}$ & $31.68^{+0.14}_{ -0.17 }$ & $21.73^{+1.44}_{-1.66}$ & $31.59 \pm 0.07$ & $31.32 \pm 0.24$ & IC 1963 \\
            FCC153 & halo & 41 & $75^{+14}_{-12}$ & -- & -- & -- & -- & -- \\
            FCC161 & center & 106 & $214^{+23}_{-21}$ & $31.47^{+0.13}_{ -0.13 }$ & $19.67^{+1.19}_{-1.20}$ & -- & $31.24 \pm 0.24$ & NGC 1379 \\
            FCC161 & halo & 26 & $31^{+7}_{-6}$ & -- & -- & -- & -- & -- \\
            FCC167 & center & 108 & $402^{+43}_{-39}$ & $31.30^{+0.09}_{ -0.10 }$ & $18.23^{+0.79}_{-0.87}$ & $31.57 \pm 0.07$ & $31.35 \pm 0.15$ & NGC 1380 \\
            FCC167 & middle & 28 & $82^{+19}_{-15}$ & -- & -- & -- & -- & -- \\
            FCC167 & halo & 6 & $10^{+6}_{-4}$ & -- & -- & -- & -- & -- \\
            FCC170 & center & 41 & $133^{+24}_{-21}$ & $31.46^{+0.16}_{ -0.15 }$ & $19.57^{+1.45}_{-1.38}$ & $31.70 \pm 0.08$ & $31.69 \pm 0.28$ & NGC 1381 \\
            FCC170 & halo & 39 & $94^{+18}_{-15}$ & -- & -- & -- & -- & -- \\
            FCC177 & center & 55 & $72^{+11}_{-10}$ & $31.25^{+0.23}_{ -0.24 }$ & $17.80^{+1.88}_{-1.95}$ & $31.51 \pm 0.07$ & $31.49 \pm 0.28$ & NGC 1380A \\
            FCC177 & halo & 19 & $24^{+7}_{-6}$ & -- & -- & -- & -- & -- \\
            FCC182 & center & 10 & $14^{+6}_{-4}$ & $31.20^{+0.48}_{ -0.53 }$ & $17.35^{+3.84}_{-4.21}$ & $31.46 \pm 0.09$ & $31.44 \pm 0.28$ & -- \\
            FCC184 & center & 69 & $268^{+36}_{-32}$ & $31.63^{+0.13}_{ -0.13 }$ & $21.22^{+1.29}_{-1.23}$ & $31.43 \pm 0.09$ & $31.41 \pm 0.28$ & NGC 1387 \\
            FCC184 & middle & 20 & $27^{+8}_{-6}$ & -- & -- & -- & -- & -- \\
            FCC184 & halo & 2 & $3^{+4}_{-2}$ & -- & -- & -- & -- & -- \\
            FCC190 & center & 25 & $54^{+13}_{-11}$ & $31.64^{+0.23}_{ -0.26 }$ & $21.29^{+2.30}_{-2.54}$ & $31.54 \pm 0.07$ & $31.52 \pm 0.28$ & NGC 1380B \\
            FCC190 & halo & 8 & $22^{+11}_{-8}$ & -- & -- & -- & -- & -- \\
            FCC193 & center & 156 & $289^{+25}_{-23}$ & $31.46^{+0.12}_{ -0.13 }$ & $19.59^{+1.10}_{-1.15}$ & $31.63 \pm 0.07$ & $31.42 \pm 0.22$ & NGC 1389 \\
            FCC193 & halo & 19 & $23^{+7}_{-5}$ & -- & -- & -- & -- & -- \\
            FCC219 & center & 70 & $238^{+32}_{-28}$ & $31.47^{+0.11}_{ -0.12 }$ & $19.71^{+1.04}_{-1.08}$ & $31.54 \pm 0.07$ & $31.37 \pm 0.22$ & NGC 1404 \\
            FCC219 & halo & 16 & $18^{+6}_{-4}$ & -- & -- & -- & -- & -- \\
            FCC249 & center & 14 & $31^{+11}_{-8}$ & $31.41^{+0.30}_{ -0.33 }$ & $19.17^{+2.62}_{-2.91}$ & $31.80 \pm 0.08$ & $31.82 \pm 0.24$ & NGC 1419 \\
            FCC255 & center & 36 & $58^{+11}_{-10}$ & $31.42^{+0.24}_{ -0.25 }$ & $19.20^{+2.10}_{-2.18}$ & $31.50 \pm 0.07$ & $31.48 \pm 0.28$ & ESO 358-G050 \\
            FCC276 & center & 71 & $235^{+31}_{-28}$ & $31.48^{+0.12}_{ -0.13 }$ & $19.76^{+1.13}_{-1.21}$ & $31.46 \pm 0.07$ & $31.50 \pm 0.22$ & NGC 1427 \\
            FCC277 & center & 23 & $109^{+28}_{-23}$ & $31.88^{+0.19}_{ -0.19 }$ & $23.76^{+2.07}_{-2.05}$ & $31.58 \pm 0.08$ & $31.56 \pm 0.28$ & NGC 1428 \\
            FCC301 & center & 19 & $46^{+13}_{-10}$ & $31.64^{+0.27}_{ -0.26 }$ & $21.24^{+2.60}_{-2.59}$ & $31.47 \pm 0.08$ & $31.06 \pm 0.24$ & ESO 358-G059 \\
            FCC310 & center & 39 & $72^{+14}_{-12}$ & $31.53^{+0.24}_{ -0.22 }$ & $20.21^{+2.27}_{-2.05}$ & $31.50 \pm 0.07$ & $31.48 \pm 0.28$ & NGC 1460 \\
            FCC310 & halo & 50 & $68^{+11}_{-10}$ & -- & -- & -- & -- & -- \\
            \hline
        \end{tabular}
        \tablefoot{Galaxies covered in this PNe catalogue, including entries for central, middle, and halo regions where covered. Galaxy name (1) and location (2), number of detected PNe (3), expected number of PNe (from PNLF) within 2.5 magnitudes of the bright-end cut-off (4), PNLF derived distance modulus (5) and corresponding distance in Mpc (6). We also include the latest distance modulus values of \citet{blakeslee_acs_2009,blakeslee_surface_2010} (7) and from the CosmicFlows-3 catalogue \citep{tully_cosmicflows-3_2016} (8). Finally, the alternative galaxy name is provided for simpler identification (9).}
        
    \end{table*}
    }
    
\begin{document} 
    
    \title{The Fornax3D project: Planetary nebulae catalogue and independent distance measurements to Fornax cluster galaxies\thanks{Tables A.1 to A.36 are only available at the CDS via anonymous ftp to ... or via ...}}
    
    \author{T. W. Spriggs\inst{1} 
    \and
    M. Sarzi \inst{2}
    \and
    P. M. Gal\'an-de Anta \inst{2, 3}
    \and
    R. Napiwotzki \inst{1}
    \and
    S. Viaene \inst{4}
    \and
    B. Nedelchev \inst{2}
    \and
    L. Coccato \inst{5}
    \and
    E.\,M.\,Corsini \inst{6, 7}
    \and
    K. Fahrion \inst{5}
    \and
    J. Falcón-Barroso \inst{8, 9}
    \and
    D. A. Gadotti \inst{5}
    \and
    E. Iodice \inst{10, 5}
    \and
    M. Lyubenova \inst{5}
    \and
    I.\,Martín-Navarro \inst{11, 8, 9}
    \and
    R. M. McDermid \inst{12, 13}
    \and
    L. Morelli \inst{18}
    \and
    F. Pinna \inst{11}
    \and 
    G. van de Ven \inst{14}
    \and
    P. T. de Zeeuw \inst{15, 16}
    \and
    L. Zhu \inst{17}
    }
    \institute{Centre for Astrophysics Research, University of Hertfordshire, College Lane, Hatfield AL10 9AB, UK 
    \newline e-mail: tspriggs@outlook.com
    \and
    Armagh Observatory and Planetarium, College  Hill, Armagh BT61 9DG, Northern Ireland, UK 
    \and
    Astrophysics Research centre, School of Mathematics and Physics, Queen's University Belfast, Belfast BT7 INN, UK 
    \and
    Sterrenkundig Observatorium, Universiteit Gent, Krijgslaan 281, 9000 Gent, Belgium 
    \and
    European Southern Observatory, Karl-Schwarzschild-Stra{\ss}e 2, 85748 Garching bei Muenchen, Germany 
    \and
    Dipartimento di Fisica e Astronomia ‘G. Galilei’, Università di Padova, vicolo dell’Osservatorio 3, 35122 Padova, Italy 
    \and
    INAF–Osservatorio Astronomico di Padova, vicolo dell’Osservatorio 5, 35122 Padova, Italy 
    \and
    Instituto de Astrof\'isica de Canarias, Calle V\'ia L\'actea s/n, 38200 La Laguna, Tenerife, Spain 
    \and
    Depto. Astrof\'isica, Universidad de La Laguna, Calle Astrof\'isico Francisco S\'anchez s/n, 38206 La Laguna, Tenerife, Spain 
    \and
    INAF–Osservatorio Astronomico di Capodimonte, Salita Moiariello 16, 80131, Naples, Italy 
    \and
    Max-Planck-Institut f\"ur Astronomie, K\"onigstuhl 17, 69117 Heidelberg, Germany 
    \and 
    Astronomy, Astrophysics and Astrophotonics Research Centre, Department of Physics and Astronomy, Macquarie University, Sydney, NSW 2109, Australia 
    \and 
    ARC Centre of Excellence for All Sky Astrophysics in 3 Dimensions (ASTRO 3D), Australia 
    \and
    Department of Astrophysics, University of Vienna, T\"urkenschanzstra{\ss}e 17, 1180 Vienna, Austria 
    \and
    Sterrewacht Leiden, Leiden University, Postbus 9513, 2300 RA Leiden, The Netherlands 
    \and
    Max-Planck-Institut f\"ur extraterrestrische Physik, Gie{\ss}enbachstra{\ss}e, 85741 Garching bei M\"unchen, Germany 
    \and
    Shanghai Astronomical Observatory, Chinese Academy of Sciences, 80 Nandan Road, Shanghai 200030, China 
    \and
    Instituto de Astronom\'ia y Ciencias Planetarias, Universidad de Atacama, Avenida Copayapu 485, Copiap\'o, Chile 
    }
    
    \date{Received ; accepted }
    \titlerunning{Fornax3D: PNe Census of ETG's}
    
    \abstract
     {Extragalactic planetary nebulae (PNe) offer a way to determine the distance to their host galaxies thanks to the nearly universal shape of the planetary nebulae luminosity function (PNLF). Accurate PNe distance measurements rely on obtaining well-sampled PNLFs and the number of observed PNe scales with the encompassed stellar mass. This means either disposing of wide-field observations or focusing on the bright central regions of galaxies.
     In this work we take this second approach and conduct a census of the PNe population in the central regions of galaxies in the Fornax cluster, using VLT/MUSE data for the early-type galaxies observed over the course of the Fornax3D survey.
     Using such integral-field spectroscopic observations to carefully separate the nebular emission from the stellar continuum, we isolated [\ion{O}{iii}] 5007 \AA{} sources of interest, filtered out unresolved impostor sources or kinematic outliers, and present a catalogue of 1350 unique PNe sources across 21 early-type galaxies, which includes their positions, [\ion{O}{iii}] 5007 \AA{} line magnitudes, and line-of-sight velocities.
     Using the PNe catalogued within each galaxy, we present independently derived distance estimates based on the fit to the entire observed PNLF observed while carefully accounting for the PNe detection incompleteness.
     With these individual measurements, we arrive at an average distance to the Fornax cluster itself of 19.86 $\pm$ 0.32 Mpc ($\mu_\mathrm{PNLF}$= 31.49 $\pm$ 0.04 mag).
     Our PNLF distance measurements agree well with previous distances based on surface brightness fluctuations, finding no significant systematic offsets between the two methods as otherwise reported in previous studies.
    }
    
    \maketitle
    %

\section{Introduction}
Extragalactic planetary nebulae (PNe) are detectable at great distances, via their strong [\ion{O}{iii}] 5007 \AA{} emission, with surveys extending out as far as the Coma Cluster, $\sim$ 100 Mpc, where PNe were detected within the intracluster medium \citep{gerhard_detection_2005}. Their ionised [\ion{O}{iii}] 5007 \AA{} emission is unresolved at such distances, and when embedded within the bright stellar continuum of the galaxy only a careful modelling of the stellar continuum allows one to measure the PNe [\ion{O}{iii}] flux.
%
This is why traditionally the study of extragalactic PNe is focused in the halo region of galaxies, where the stellar background is minimal.
Previous studies have applied a range of observation techniques for the task of detecting extragalactic PNe: photometric observations using either 'on-off' band imaging or narrow-band filters of regions outside the galaxy central core \citep{mcmillan_planetary_1993, feldmeier_calibrating_2007, herrmann_planetary_2008}, while counter-dispersed slit-less spectroscopy was employed for the Planetary Nebulae Spectrograph (PN.S) instrument, used in \citet{douglas_pns_2007} and \citet{coccato_kinematic_2009}, and later in the extended PN.S (ePN.S) in \citet{pulsoni_extended_2018}.
However, these techniques fall short when exploring the central regions of galaxy, where spectroscopy is instead needed to uncover the presence of PNe amidst the stellar background and diffuse ionised-gas emission.
Integral-field spectroscopy is ideal in this respect, in particular when using instruments such as MUSE \citep[Multi Unit Spectroscopic Explorer][]{bacon_muse_2010} which, with its large field-of-view allows one to efficiently map the PNe populations of nearby galaxies \citep[e.g. ][]{kreckel_revised_2017}

An example of the growing spatial coverage that single observations can now achieve, across a single galaxy, would be M31. Thanks to its relatively close proximity, the PNe are detectable across the entire galaxy \citep{ciardullo_planetary_1989, merrett_deep_2006, pastorello_planetary_2013, jacoby_survey_2013, martin_sitelle_2018, bhattacharya_survey_2019-1, bhattacharya_survey_2019}.

Over the past 32 years, one of the chief motivations for detecting extragalactic PNe has been to use them as tracer for the stellar kinematics in the faint outskirts of galaxies, where direct spectroscopic measurement are prohibitive. Indeed, numerous studies have shown how PNe can be used in this way to probe the structure of galaxy stellar halos and to measure their dark matter content\citep{romanowsky_dearth_2003,merrett_deep_2006, douglas_pns_2007, coccato_kinematic_2009, kafle_need_2018, martin_sitelle_2018, longobardi_kinematics_2018, pulsoni_extended_2018, bhattacharya_survey_2019-1}. 
At the same time, research has also shown that the distribution of PNe absolute magnitudes in [\ion{O}{iii}] 5007 \AA{}, $\mathrm{M_{5007}}$, appears to be nearly invariant between galaxies, with a noticeable cut-off at the bright end \citep{jacoby_planetary_1989-1, ciardullo_planetary_1989, jacoby_planetary_1989, ciardullo_planetary_1989-1}. Notable discrepancies around the invariant nature of the PNLF arise when observing beyond 2 magnitudes from the cut-off, with evidence of dips appearing towards the faint end of the PNLF within the Large and Small Magellanic Clouds (LMC and SMC, respectively), as found by \citet{jacoby_survey_2002} and \citet{hernandez-martinez_emission_2009}. As such, a generalised analytical planetary nebulae luminosity function (PNLF)  was first proposed by \citet{ciardullo_planetary_1989}, who used the PNe observed in M31 as Cepheid Variable distance estimates to calibrate the PNLF bright-end absolute magnitude [\ion{O}{iii}] 5007 \AA{}, $\mathrm{M_{5007}^{*}}$ at a value of $-4.48$ mag.

More recently, a value of $\mathrm{M_{5007}^{*}} = -4.53 \pm 0.06$ mag has been decided on for galaxies with stellar populations that are more metal rich than the LMC \citep{ciardullo_planetary_2012}, with the stated census that $\mathrm{M_{5007}^{*}}$ does fade in smaller, more metal-poor galaxies.


In regards to extra-galactic distance determinations, the PNLF method has proven itself a reliable and unique tool, in particular given that PNe are expected to reside within all types of galaxies. 
Supernovae type Ia (SNIa) measurements are indeed serendipitous by nature, whereas using surface brightness fluctuations (SBF) and Cepheid variable stars is restricted mostly only to early- and late-type galaxies, respectively.
Previous works found evidence of a systematic difference between distances derived from the PNLF and SBF \citep{ciardullo_comparison_1993}, although more recently both  \citet{kreckel_revised_2017} and \citet{hartke_halo_2017} 
found that their PNe-based distances, to NGC628 and M49 respectively, are in good agreement with SBF measurements.
%
As for the application of the PNLF method to the Fornax cluster, beyond our previous work \citep[][hereafter S20]{spriggs_fornax_2020}, there are recorded efforts using the aforementioned photometry and slit-less spectroscopy approaches \citep{mcmillan_planetary_1993, teodorescu_planetary_2005, feldmeier_calibrating_2007}. Each one reporting good agreement to the respective distances measurements available to their dates. Now though, with more detailed SBF, SNIa, and other distance estimators, we see that the Fornax cluster may be further away than the initial estimates ranging between 16 Mpc and 19 Mpc. 

Here, we continue the detection work of S20, and carry out a census of PNe populations within the central, middle and halo regions of the remaining early-type galaxies (ETGs) of the Fornax3D (F3D) survey \citep{sarzi_fornax3d_2018}. We introduce our PNLF modelling methodology, improving upon the method of S20. We then report independent distance estimates for each galaxy, and compare our findings with other results and techniques from the literature. 

This paper is organised as follows. In Sect.~2, we introduce the F3D data set, along with reduction steps. Sect.~3 reviews the process of PNe detection, filtering, and cataloguing, as introduced in S20, then develops on these practises through the introduction of our PNLF modelling approach and how we tested the robustness and reliability of this method. The PNe catalogue is introduced in Sect.~4, detailing the contents, with notes on objects of unique interest. This is accompanied by the PNLF distances and how they stack up against distance measurements from other methods. We also briefly discuss the 3D structure of the Fornax cluster, as prescribed by our PNLF distances. Finally, we wrap up our findings and draw some future perspectives in Sect.~5

\section{Data}
Our work uses the MUSE data of the F3D survey, in the Wide Field Mode (WFM) so as to reach a spatial sampling of 0.2\arcsec $\times$ 0.2\arcsec\ across a $1\arcmin \times 1\arcmin$ field of view (FOV). The wavelength coverage that MUSE spans 4650--9300 \AA{}, with a spectral sampling of 1.25 \AA{} $\textrm{pixel}^{-1}$. The Fornax3D survey included multiple MUSE pointings (between two and three) to cover extended objects, and to ensure in particular that observations for ETGs reached down to the same limiting surface brightness of $\mu_B=25\ \textrm{mag} \ \textrm{arcsec}^{-2}$. The survey ensured that the central pointings would be observed under good seeing conditions (for a FWHM $\leq$ 0.8\arcsec) precisely to enable a study of point sources such as PNe and globular clusters \citep[e.g.\ ][]{fahrion_constraining_2019} as well as of galactic nuclei. In the outer regions the main interest was to characterise the properties of stellar population in the faint outskirts of galaxies, which is why these regions were observed under more lenient seeing conditions but over longer exposure times.

The MUSE data was reduced using the MUSE pipeline \citep{weilbacher_design_2012, weilbacher_muse-drp_2016}, within the ESOREFLEX \citep{freudling_automated_2013} environment, as reported in \citet{sarzi_fornax3d_2018} and \citet{iodice_fornax_2019}. The reduction phase deals with key steps such as telluric correction, sky-subtraction and absolute flux calibration. All datacubes have been flux calibrated and this flux calibration has been further verified using HST images, as in \citet{sarzi_fornax3d_2018} and S20. 
The Fornax3D data taken from different pointings were aligned and presented as a single mosaic in \citet{iodice_fornax_2019}. For the purposes of detecting PNe it is important to account for the different image quality of these different observations, and therefore here we analyse separately the combined observations from each individual pointing. 

\subsection{Residual cube preparation}
Our PNe detection method is based on residual data, that is, reduced MUSE spectra that have first been analysed by the Galaxy IFU Spectroscopy Tool pipeline \footnote{\href{https://abittner.gitlab.io/thegistpipeline}{https://abittner.gitlab.io/thegistpipeline}} \citep[GIST,][]{bittner_gist_2019} and finalised by removing the stellar continuum. The GIST pipeline employs both pPXF \citep{cappellari_parametric_2004,cappellari_improving_2017} and PyGandALF (see GIST documentation): the first handles spaxel-by-spaxel modelling of the stellar continuum, using the full MILES stars library of templates \citep{vazdekis_miuscat_2012,vazdekis_uv-extended_2016} while the second focuses on the nebular emissions. The results of the GIST pipeline efforts are then utilised to produce a residual data cube after subtracting the modelled stellar continuum, which therefore still contains the emission lines. We then later re-fit the residual cube for [\ion{O}{iii}] emission only, imposing a fixed line-width on the emission lines. This was found in S20 to produce a better map in [\ion{O}{iii}] 5007 \AA{} emissions, which led to a greater number of PNe detected.

\section{Methods}
\subsection{Compiling the PNe catalogue}

In order to detect and characterise PNe in our sample galaxies using the MUSE data we adopt the methodology that is described in full in S20\footnote{Code hosted on Github: \href{https://github.com/tspriggs}{https://github.com/tspriggs} \citep{thomas_spriggs_tspriggsmuse_pne_fitting_2021}} and which consists of the following steps:

Firstly, we carry out a spaxel-by-spaxel fit to the [\ion{O}{iii}] 4959,5007 \AA{} emission line doublet in the residual cube, imposing a fixed intrinsic line-width of $30\,\rm km\,s^{-1}$ typical of unresolved PNe in order to better highlight the presence of such unresolved [\ion{O}{iii}] sources. Previously identified regions with diffuse nebular emission or with significant and systematic stellar-continuum fit residuals are masked.

Secondly, we extract MUSE residual minicubes around each candidate PNe source 1.8\arcsec$\times$1.8\arcsec wide and over a 100 \AA\ wavelength range. This is followed by the evaluation of the imaging point-spread function (PSF). This is achieved through the simultaneous fit of the residual minicubes for the four best-detected PNe candidate sources, in terms of their central value for the A/rN ratio, using the 3D cube-fitting process described in S20.

Thirdly, we fit for the PNe candidate total [\ion{O}{iii}] flux and kinematics using the 3D cube-fitting process of S20 while now holding the spatial distribution of the [\ion{O}{iii}] flux to that of the previously derived PSF. following this, we conduct an evaluation of the minicube fit quality and point-source detection. Candidate PNe with central A/rN ratio less than three are deemed undetected and are discarded, as are sources whose minicube fits returned $\chi^2$ values above the 3$\sigma$ limit for the $\chi^2$ statistics (for the given number of degree of freedom).

Finally, we conduct a spectral characterisation and PNe impostor identification process. After extracting PSF-weighted aperture spectra around each PNe candidate sources (from the original MUSE cube) we used GIST's GandALF fits along to extract  H$\beta$, H$\alpha$, [\ion{N}{ii}]\,6583  and [\ion{S}{ii}]\,6716,6731 doublet line measurements, subsequently identifying PNe impostor sources from supernovae remnants or compact \ion{H}{ii} regions using diagnostic diagrams as seen in S20. This is accompanied by the identification of any kinematic interlopers among the remaining PNe candidate. PNe sources with exceedingly high or low velocities (beyond a 3 $\sigma$ limit) compared to the stellar line-of-sight velocity distribution (LOSVD) along the PNe direction are not considered to belong to the galaxy under consideration.

Following these steps, we finally arrive at a catalogue of 1350 unique PNe across 21 ETGs in the Fornax cluster, located either within the central, middle or halo pointings of the F3D survey. 
For these objects, we list their spatial coordinates, line-of-sight velocities, as well as their measured fluxes, and magnitudes with appropriate errors, in the measured [\ion{O}{iii}] 5007\AA{} emission lines.
The catalogue also includes PNe impostor sources and PNe interlopers, as well as a handful of PNe objects that were considered to be over-luminous (see, e.g. the case of FCC~167 in S20) and which will also be excluded by our following PNLF analysis. 

\subsection{PNLF fitting and distance estimation}

In S20 we followed the simplest approach for estimating the distance to a galaxy based on the brightest observed PNe. In this case the PNLF distance modulus is simply  $\mu_\mathrm{PNLF} = \mathrm{m_{5007,\rm min}} - \mathrm{M_{5007}^{*}}$ where $\mathrm{m_{5007,\rm min}}$ is the [\ion{O}{iii}] 5007 \AA{} magnitude of the brightest PN and $\mathrm{M_{5007}^{*}}$ is the M31 Cepheid distance calibrated absolute magnitude of the PNLF bright cut off. 
This approach is relatively sound in the case of galaxies presenting a significant number of PNe (as in FCC~167 and FCC~219, the objects studied in S20) but will clearly lead to biased results when dealing with few PNe. Indeed in this case we are more likely to miss the rarest objects at the bright cut-off of the PNLF, thus deducing an overestimated distance.

Due to a large range in the number of PNe observed across our 21 galaxies, we decided to implement a distance estimation technique based on the entire observed PNLF, taking full advantage of our knowledge for the PNe-detection completeness function in each object in order to exploit the whole range of observed $\mathrm{m_{5007}}$ values. We follow the completeness correcting method of S20, assessing the level of incompleteness of our samples over the observed FOV. In short though, we calculate the completeness of detection for a given apparent magnitude ($m_{5007}$) from the fraction of galaxy stellar light within the MUSE FOV, where a PNe of that particular magnitude can be detected. This procedure is also informed by the PSF of the observation, along with the signal to noise cut of 3 $\times$ A/rN. This is also the method applied, and described, within \citet{galan-de_anta_fornax_2021}.

Then, for each galaxy we explore the parameter space of $\mu_\mathrm{PNLF}$, evaluating for a value that would lead to a completeness-corrected PNLF model through simply minimising the Kolmogorov-Smirnov statistic D. To achieve this, we utilise the Python based Scipy package's 'KS\_1samp' function, passing the supremum value (D) to the scalar minimiser of LMfit \citep{newville_lmfit_2014, newville_lmfitlmfit-py_2019}, applying the Nelder-Mead optimisation method \citep{nelder_simplex_1965}.
In terms of model PNLF, we used the analytic form provided by \citet{longobardi_planetary_2013}
\begin{equation}
    N(M) = c_{1} \ e^{ \ c_2 \ M_{5007}} \ \big[1 - e^{ \ 3( M^{*}_{5007} - \ M_{5007} )}\big],
    \label{eq:PNLF}
\end{equation}
where $c_{1}$ is a normalisation factor, $c_{2}$ defines the functional form of the PNLF, $\mathrm{M_{5007}^{*}}$ is the bright-end cutoff value first  and finally, $\mathrm{M_{5007}}$ is the absolute magnitude of the PNe in [\ion{O}{iii}] 5007\AA{}. In that follows we adopt the most recently calibrated values of $\mathrm{M_{5007}^{*}} = -4.53 \pm 0.06$ for the bright-end cut-off \citep{ciardullo_planetary_2012} and hold $c_{2}$ to the standard value of 0.307 derived by \citet{ciardullo_planetary_1989}. 
We note that systematic variations in $\mathrm{M_{5007}^{*}}$ have been reported at very low stellar metallicities, below [M/H] < -0.5 \citep{bhattacharya_survey_2021}, but since only one of our objects (FCC\,119, see Tab. 1C in \citet{iodice_fornax3d_2019}) would appear to fall in this regime, such potential systematic effect would not greatly impact our overall results.
Appendix A shows that in principle our methodology can estimate the $c_{2}$ parameter but also that our current data do not provide strong evidence for departing from the assumed $c_{2} = 0.307$ value. Since the KS-statistics is independent of normalisation, we effectively ignore $c_{1}$ while looking for the best $\mu_\mathrm{PNLF}$ value through our KS-statistic minimisation. Instead, we finally adjust the normalisation of our best completeness-corrected model for the observed PNLF by simply integrating it and matching the result to observed number of PNe.
In this case, availing of the entire PNe sample allowsfor tighter constraints on any estimates for the total number of expected PNe by integrating the PNLF down to some magnitude limit, as this depends on the original Poisson uncertainty on the actual number of PNe used to constrain the PNLF model in first place. 

\placefigsimsDMtwoobjects
\placefigsimsDM
\placefigerrorcomparison

To evaluate our uncertainties around our best $\mu_\mathrm{PNLF}$ values we resort to simulations. Namely, starting from the best-model PNLF (at the best $\mu_\mathrm{PNLF}$ and accounting for the completeness), we sample such a parent distribution $N$ times to produce a $N$ synthetic observed PNLF, which we then re-fit through our KS-statistic minimisation routine to obtain $N$ best $\mu_\mathrm{PNLF}$  values. The $16^{\rm th}$ and $84^{\rm th}$ percentile of such a $\mu_\mathrm{PNLF}$ distribution provides our $\pm 1\sigma$ uncertainties on the measured $\mu_\mathrm{PNLF}$. Typically, $N=100$ deliver stable results. 
In sampling the best-model PNLF we account for Poisson statistics. For this, each time we sample the parent PNLF model we use a varying number drawn from a Poisson distribution with expectation value equal to the observed number of PNe. Accounting for Poisson statistics is necessary considering how the formation of PNe is itself a stochastic process. Put it simply, the observed number of PNe is likely to be different even if we were hypothetically to observe in the same way two identical galaxies at the same distance, which in turn would lead to slightly different PNLF distance estimates.
To further account for observational errors in our magnitude measurements, we perturb our original data according to our measured errors and add in quadrature to the error budget the standard deviation for the $\mu_\mathrm{PNLF}$ distribution that is obtained in this way.
Both accounting for Poisson statistics and observational errors contribute very little to the error budget, around 3\% an 2\%, respectively. Indeed, as will be shown in the next section, how well $\mu_\mathrm{PNLF}$ can be estimated depends primarily on the number of PNe and on depth of our detection completeness.
%
%
Finally, we note that \cite{ciardullo_planetary_2012} report a formal error of 0.06 mag for $\mathrm{M_{5007}^{*}}$, which we also add in quadrature to our PNLF errors.

\subsection{PNLF simulations}

Simulations act as a useful diagnostic tool to understand how our PNLF modelling handles different PNe samples sizes ($N_\mathrm{{PNe \, sim}}$), especially when dealing with small PNe numbers where we wish to understand whether and to which extent our solution may be biased.
The simulations are based on a finely sampled intrinsic PNLF model that assumes the standard $c_\mathrm{2 \, in}$ = 0.307 value, which is then shifted taking a distance modulus of $\mu_\mathrm{PNLF \, in}$ = 31.45 mag and finally corrected for incompleteness. From such a parent PNLF profile we then draw $N_\mathrm{{PNe \, sim}}$ apparent $m_\mathrm{{5007\,sim}}$ values to create a large set of synthetic PNLF, with  $m_\mathrm{{5007\,sim}}$ varying between 5 and 200 in order to cover the range of recorded $N_\mathrm{{PNe}}$ seen in our central pointings (Table \ref{tab:galaxy_info}). Each synthetic PNLF is then passed to our PNLF-fitting algorithm for parameter estimation. 

Assuming the PNLF given in Eq.~\ref{eq:PNLF} and while only fitting for $\mu_\mathrm{PNLF}$, Fig.~\ref{fig:simulations_dM_FCC193_FCC147} shows how this parameter is recovered with increasing precision with increasing PNe sample size, without incurring in any significant bias even at low $N_\mathrm{{PNe \, sim}}$ numbers. The top panel of Fig.~\ref{fig:simulations_dM_FCC193_FCC147} is based on simulations that adopt the completeness profile observed in FCC~193, where our data allow one to detect PNe some 2.5 magnitudes below the PNLF bright cut-off, whereas the lower panel shows the case of FCC~147 where only probe the PNLF 1.5 magnitudes below this limit. In both instances the scatter in the recovered $\mu_\mathrm{PNLF}$ scales as $1/\sqrt{N_\mathrm{{PNe \, sim}}}$, but at a given number of PNe the accuracy in recovering $\mu_\mathrm{PNLF}$ would actually decrease for deeper the observations.
Accounting within our simulation for the entire range of completeness functions observed across our sample (Fig.~\ref{fig:simulations_dM}) produces an average trend with $N_\mathrm{{PNe \, sim}}$ for the scatter in the recovered $\mu_\mathrm{PNLF}$ that captures well the errors in our actual $\mu_\mathrm{PNLF}$ measurements for our 21 objects (Fig.~\ref{fig:sim_data_error_comparison}).
Overall, this exercise gives us confidence in our $\mu_\mathrm{PNLF}$ estimates and errors, even in objects with less than 20 PNe.


\section{Results and discussion}

In this section, we first review the catalogue of the PNe found within the ETG population of the F3D survey. Then, moving onto the results of applying our PNLF fitting method (see Sect.~3) to each galaxy. We then discuss the accuracy of the PNLF as a distance indicator, when compared to SBF measurements.

\subsection{The F3D PNe catalogue}
We present a catalogue of PNe detected across 21 galaxies, reporting a total of 1350 unique PNe. Though the PNe of each galaxy are our main interest, each galaxy's catalogue contains sources highlighted as impostors, following the criteria of S20. Such impostors are labelled, as described in Section~3.1, following the criteria of S20: supernova remnants (SNR), HII regions, and over-luminous sources (of which we find two in total; one each in FCC\,167 and FCC\,276). Only the sources in our catalogues that have the ID of 'PN' are used for the PNLF modelling and distance estimate efforts.

A sample of our catalogued PNe, from FCC\,083, can be found in appendix A, where the two regions covered in the F3D survey are labelled as: 'center' and 'halo'. The rest of the catalogue is stored as digital records, uploaded to the VizieR On-line Data Catalog. The table contains a 'Source ID' column that follows the same naming convention introduced in S20, using the F3D label used at the start. The RA and Dec are included for future source cross-matching. '$m_{5007}$' is the apparent magnitude of the [\ion{O}{iii}] 5007 \AA{} emission. 'A/rN' is the amplitude-to-residual noise ratio, essentially a signal-to-noise measure. 'LOSV (km s$^{-1}$)' is the reported observed line-of-sight velocity of the source, derived from the wavelength position of the [\ion{O}{iii}] 5007 \AA{} emission line. The 'label' column states each source's assigned identity: PN, SNR (supernova remnant), HII (compact HII region), or OvLu (over luminous source). Finally, an index column is included for easier source comparison, highlighting sources that appear in the overlap of pointings: center C-00, middle M-00 and halo H-00.

All the galaxies in Table~\ref{tab:galaxy_info}, are shown in Figs.~\ref{fig:all_PNLF} and \ref{fig:F3D_FOV}. Every galaxy's FOV is shown via their signal-to-noise map in [\ion{O}{iii}] 5007 \AA{}, with catalogued PNe circled. Alongside the FOV, we present the modelled PNLF of the catalogued sources. If there is halo or middle data present, they are also presented in different colours, as explained in the legends of each plot. The blue histograms show the distribution of detected PNe $m_{5007}$ values, while the black line indicates a scaled form of the \citet{jacoby_planetary_1989} form of the PNLF. We then scale this PNLF to form the incompleteness-corrected PNLF, shown by dashed lines (colour dependant on pointing location). The blue shaded regions surrounding the C89 PNLF line, as well as the dashed incompleteness-corrected profile, show the 1 $\sigma$ confidence limit. After reviewing the literature for works that may contain matching PNe, the only other surveys that contained such matches (for FCC\,167 and FCC219 by \citet{feldmeier_calibrating_2007} and \citet{mcmillan_planetary_1993} respectively) have already been discussed and compared with our samples in S20. We are then confident that we are cataloguing sources never before detected.

\placefigvelocitycomparison

\subsection{Dynamical tracers}
The PNe observed within the central regions of a galaxy can be assumed to be travelling with the same systemic velocity of the galaxy. To this end, we can verify this assumption by taking the median of the LOSV values, as measured from the 3D fitting code, using the mean wavelength position to measure each PNe's individual LOSV. Fig.~\ref{fig:velocity_comparison} shows the comparison of our median LOSV values, to those given by \citet{iodice_fornax3d_2019}, who present the heliocentric systemic velocity of each galaxy, as measured from the stars in the central regions. In agreement with our original assumption, we see that the PNe median LOSV values and \citet{iodice_fornax3d_2019} systemic velocities are in very good agreement.

\placegalaxytable

\subsection{PNLF distances}

Our deep MUSE data allow one to evaluate the accuracy and reliability of the standard candle nature of the PNLF, as an independent distance measurement technique. Using the PNe catalogue that we have produced we can now derive independent PNLF distance estimates for each of the 21 ETGs targeted by the F3D project as well as an overall average distance to the Fornax cluster, which is estimated to reside $\sim$ 20 Mpc away \citep{tonry_sbf_2001,blakeslee_acs_2009, tully_cosmicflows-2_2013, de_grijs_clustering_2020-1}. 

Table \ref{tab:galaxy_info} contains the result from each galaxy, along with the halo and middle regions, where available. It lists the number of detected PNe ($N_\mathrm{PNe}$), as well as the number of expected PNe within 2.5 magnitudes of the PNLF cut-off ($N_{2.5}$). Errors on $N_{2.5}$ are derived on the basis of the Poisson uncertainties associated to the actual observed number of PNe $N_\mathrm{PNe}$, after rescaling. Here, we also present our best-fit values of $\mu_\mathrm{{PNLF}}$, and distance (in Mpc), as measured from the PNe populations. 
Table~\ref{tab:galaxy_info} also includes revised distance estimates for the FCC~167 and FCC~219 ($\mu_\mathrm{PNLF}$ = 31.30 $\pm$ 0.07 mag and 31.48 $\pm$ 0.10 mag, respectively) that are consistent with the values reported in S20 (31.24 $\pm$ 0.11 mag and 31.42 $\pm$ 0.10 mag), owing to the fact that in first place these objects have a fair number of PNe. For FCC~219, we also note the agreement between PNLF and TRGB method, where \citet{hoyt_carnegie_2021} report a distance of 31.36 $\pm$ 0.04 $\pm$ 0.05 mag, which is within the uncertainty of our proposed 31.46$^{+0.11}_{-0.12}$. 
In the case of FCC\,167, \citet{roth_towards_2021} have recently used the same MUSE data set to benchmark their refined 'on-off' band imaging PNe detection methodology, finding a smaller distance for FCC~167, with $\mu_\mathrm{PNLF} = 31.10 \, \pm \, 0.04$ mag. 
Although this comparison highlights a radial trend for the difference between our respective $m_{5007}$ measurements, possibly due to some unaccounted stellar-population effect (which we limit through a detailed spectral fitting), the origin for the distance offset appears to lie with the brightest PNe in the central pointing FCC~167, which all reside in low-stellar background regions.
In this regime the simulations of S20 show that our $m_{5007}$ measurements are rather robust, leaving room only for an offset in absolute flux calibration to explain the reported difference in the inferred distance. In this respect we avail of HST images to check the calibration of the MUSE data \citep[see][for details]{sarzi_fornax3d_2018}, which in the case of FCC~167 only required a 3\% re-scaling towards lower fluxes (adding 0.03 mag). For what follows, we note that across our sample such a re-scaling averages out, indicating no systematic offset. 

\placefigSBFminPNLF
\placefigTensionSBFminPNLF

To validate our PNLF distance estimates, we started by comparing our individual to measurements to the SBF distances, which exist for all but one of our sample galaxies (FCC~161). Using the latest SBF values from either \citet{blakeslee_acs_2009} or \citet{blakeslee_surface_2010} (based on HST-ACS in the F850LP and F814W pass-band, respectively), Fig.~\ref{fig:PNLF_minus_SBF} shows how overall our PNLF distances agree well with their SBF counterparts, with a median difference of $\Delta \mu = \mu_\mathrm{{PNLF}} - \mu_\mathrm{{SBF}} = 0.003$ mag and a standard deviation of 0.21 mag. 
In comparison, \citet{ciardullo_planetary_2012} reported a median $\Delta \mu = -0.33$ mag corresponding to SBF distances being on average larger than PNLF estimates.
Taking a step further, in Fig.~\ref{fig:tension_PNLF_minus_SBF} we show the distribution for the statistical tension between the PNLF and SBF distance measurements, dividing the $\Delta \mu$ values by the quadratic sum of their respective errors. Figure~\ref{fig:tension_PNLF_minus_SBF} further shows how our PNLF values agree on average with the SBF distances, with a large majority of cases showing less than a 1$\sigma$ tension.
How our PNLF distance measurements compare to the SBF estimates of \citet{blakeslee_acs_2009} can be further appreciated in Fig.~\ref{fig:Dist_comp_Bl}, where 15 of the 20 common objects indeed are consistent within the errors, as would be expected when adopting 1$\sigma$ errors.

Of the galaxies covered here, we note the presence of three edge-on galaxies: FCC\,153, FCC\,170 and FCC\,177. Both FCC\,170, and FCC\,177 measured $\mu_{\mathrm{PNLF}}$ show distinguished offsets from their SBF distance counterparts (Fig.~\ref{fig:Dist_comp_Bl}). This would agree with both \citet{blakeslee_distances_1999} and \citet{mei_acs_2007}, where the former reported differences between SBF distances and other distance methods for these objects, whereas the latter noted that SBF measurements for edge-on galaxies poses a greater challenge, compared to measurements with face-on galaxies.
In fact, \citet{ciardullo_planetary_2012} also remarks that the two galaxies that show the largest disagreement between SBF and PNLF distance measurements in their sample are edge-on. 
The disagreement between these two distance methods is believed, in part, to arise from the steeper gradient in stellar light along the line of sight of an edge-on galaxy, compared to the smoother stellar profile of a face-on galaxy.

\citet{iodice_fornax3d_2019} find evidence of a potentially ongoing interaction between FCC\,143 and FCC\,147, from the Fornax Deep Survey \citep[FDS,][]{iodice_fornax_2019} images. The \citet{blakeslee_acs_2009} SBF distances indicate a separation of $\sim$ 0.3 Mpc. We are at least able to confirm the distance of FCC\,147 with a excellent agreement with the SBF measurement. However, the presence of only a handful of PNe within FCC\,143 means that our distance estimate is no where near as reliable compared to that of FCC\,147.

\placefigDistBl
\placefigDistCF

\subsection{The distance to the Fornax cluster}
 
Taking the weighted average of the PNLF derived distances, we estimate a distance to the Fornax cluster of 19.86 $\pm$ 0.32 Mpc. 
This is in remarkably good agreement, with similar uncertainties, with both 's SBF mean distance of 20.0 $\pm$ 0.3 $\pm$ 1.4 Mpc \citep{blakeslee_acs_2009} and recommended mean distance of 19.1$^{+1.4}_{-1.2}$ Mpc, from Cepheid distances, SBF, and Tip of the Red Giant Branch (TRGB) methods \citep{de_grijs_clustering_2020-1}. 
Our reported average distance corresponds to a distance modulus of  $\mu_\mathrm{PNLF}$ = 31.49 $\pm$ 0.04 mag and is based on all 21 ETGs listed in Table~\ref{tab:galaxy_info}, whereas \citet{blakeslee_acs_2009} report a median distance modulus to Fornax of $\mu_\mathrm{SBF}$ = 31.51 $\pm$ 0.09 mag based on over 43 objects they imaged with HST.

To further gauge the reliability of our PNLF derived distances, Fig.~\ref{fig:Dist_comp_CF3} compares our values with those in the CosmicFlows-3 catalogue of \citet{tully_cosmicflows-3_2016}. This catalogue contains statistically derived mean values obtained combining published distances derived using Supernova Type Ia, the Fundamental Plane (FP), Globular Cluster Luminosity Function (GCLF), and the TRGB. As such measurements come with their own errors and potential biases, the values of CosmicFlows-3 catalogue show larger overall uncertainties compared to the SBF measurements of \citet{blakeslee_acs_2009}. Our PNLF values agree within the errors with the CosmicFlows-3 distances for all but two objects, providing further confidence also in the case of FCC~161 that was not targeted by \citet{blakeslee_acs_2009}. This is reflected also by the agreement between our PNLF average distance to Fornax and the value inferred using the CosmicFlows-3 value for our sample galaxies, which stands at 19.4 $\pm$ 0.4 Mpc ($\mu_\mathrm{PNLF}$= 31.44 $\pm$ 0.04 mag).

\placefigtangRmag
\placefigstructure

\subsection{Fornax cluster spatial structure}

Our PNLF distance measurements allows us to map out the ETG population of the Fornax cluster, comparing our independently derived distances within the cluster, and more specifically, to its brightest cluster galaxy (BCG); namely NGC\,1399 (FCC\,213). Here, we treat NGC\,1399 as the central point of the cluster, with our ETGs situated around it. Before mapping the spatial distribution of our sample galaxies within the cluster, we first review the current distance estimates for NGC\,1399. 

\citet{blakeslee_acs_2009} reported a distance to NGC\,1399 of 20.9 $\pm$ 0.9 Mpc, though this was updated to 21.1 $\pm$ 0.7 Mpc in \citet{blakeslee_surface_2010}. A few years earlier and always based on SBF, \citet{tonry_sbf_2001} measured a distance of 19.95 $\pm$ 1.47 Mpc. The CosmicFlows-3 survey reports a statistical mean distance from multiple sources of 22.08 $\pm$ 0.12 Mpc \citep{tully_cosmicflows-2_2013}.
MUSE data for NGC\,1399 were collected both during the F3D survey, covering two external regions, and over the course of previous campaigns for the central regions (Prog. ID. 296.B-5054(A)). Upon reviewing these observations, which were taken under less than ideal seeing conditions compared to our other targets, we identified seven potential PNe in one of the F3D halo pointings. Initial estimates from converting the apparent magnitude of brightest PNe directly to the cut-off position produced an initial distance guess of $\sim$ 22 Mpc. Although this may agree with the results of \citet{tully_cosmicflows-2_2013}, the large uncertainties that would come with modelling such a sparse PNLF ($\sim$0.45 mag, see Fig.~2) would not allow us to replace the previous other estimates for the distance to NGC~1399 nor provide a robust reference to help with the present discussion. 

Though BCGs are expected to be close to the barycentre of a cluster, it is not always the case that they reside at the geometric centre of the cluster. NGC~1399 resides deep in the potential well of the cluster as traced by the X-ray emission of its hot intracluster gas, although the cluster itself comprises both of a main virialised region around NGC~1399 and of a secondary, offset group of in-falling objects around NGC~1316. 
In this respect, it is interesting to note how the \citet{blakeslee_acs_2009,blakeslee_surface_2010} distance measurements for NGC~1399 would seeming place it not only behind the majority of our sample galaxies located within the virial radius of the Fornax cluster (according both to their and our PNLF measurements) but also beyond their own average distance to the cluster based also on objects further out.
On the other hand, although the \citet{tonry_sbf_2001} distance for NGC~1399 comes with a larger error, it is closer to the average cluster distances presented in \citet{blakeslee_surface_2010}, \citet{tully_cosmicflows-2_2013}, and \citet{de_grijs_clustering_2020-1} as well as our own estimate for Fornax cluster distance.
It would therefore seem a more indicative distance to NGC~1399 and the barycentre of the Fornax cluster.

Another expectation for the structure of a galaxy cluster is mass segregation as a result of dynamical friction, in particular for less massive clusters where tidal stripping cannot act as an obstacle to mass segregation as it does in the most massive ones \citep[see, e.g.][]{kim_yzics_2020}.
To start investigating whether that happened in Fornax, in Fig.~\ref{fig:tangential_dist_RMag} we show the tangential separation of each galaxy from NGC\,1399 compared to their $r$-band absolute magnitude \citep{iodice_fornax_2019}, used as proxy for their mass. As projected on the sky, the majority of galaxies reside within 1 degree separation, with the uncertainty in their PNLF distance decreasing with brighter $r$-band magnitudes. To a first order, the more luminous and massive galaxies appear closer to the expected cluster centre (i.e. NGC\,1399), with less luminous objects being more spread out in projected sky separation. 
To more accurately depict the 3D structure of the Fornax cluster, we compare the spatial distances between each galaxy to NGC\,1399, against their absolute $r$-band magnitude. 

Considering both the distance estimate to NGC~1399 of \citet{blakeslee_surface_2010} and \citet{tonry_sbf_2001}, in Fig.~\ref{fig:fornax_structure_comp} we produce three different scenarios for the spatial distribution of the ETGs covered here. 
The first panel (left), compares the distances taken from \citet{blakeslee_acs_2009, blakeslee_surface_2010} to their own distance of 21.1 Mpc for NGC\,1399 \citep{blakeslee_surface_2010}. The galaxies are scattered between 0.5 and 2.1 Mpc, with little indication for a trend with absolute $r$-band magnitude. 
The second panel (middle) compares the SBF distances of \citet{blakeslee_acs_2009} against the NGC\,1399 distance of 19.95 Mpc \citep{tonry_sbf_2001}. In contrast to the first panel, we find that the majority of galaxies are concentrated within 1 Mpc of NGC\,1399. 
The final panel (right) compares our PNLF distances to the \citet{tonry_sbf_2001} NGC\,1399 distance. Here, we find that, again, the majority of galaxies reside within 1.5 Mpc, though we also note some galaxies showing a separation of $\sim$ 2.5 Mpc. Although in this case we note that the brightest object is closer to NGC1399, overall from these comparisons we do not find clear evidence of mass segregation within the virial radius of the Fornax Cluster, in particular when we consider the large uncertainties in projected PNLF distance, which would correspond to similarly wide errors in the 3D separation. 

\subsection{Planetary nebulae sample sizes across the cluster}
 
The accuracy of our PNLF distance measurement depends primarily  on the PNe sample size, which in turn for a simple and relatively old stellar population is expected to simply scale with the luminosity of that population \citep{renzini_global_1986}.
For our sample of ETGs this is confirmed by trend shown in Fig.~\ref{fig:N_PNe_vs_r_mag}, between the $r$-band magnitude corresponding to the stellar flux encompassed by our single F3D pointings (without accounting regions excluded in our PNe analysis) and the expected number of PNe within 2.5 magnitudes from the bright cut-off of the PNLF, as inferred from our best match to the observed PNLF in these same pointings. Much of the scatter in the observed trend can be explained by the varying imaging quality underlying our PNLF measurements, whereby a broader PSF lead to less PNe being detected and to a larger uncertainty in the inferred intrinsic number of PNe. Other factors may also be at work, however, such a dependency with stellar metallicity for the specific number of PNe \citep[][]{buzzoni_planetary_2006} leading to more PNe rich population in metal regions, as would be probed better by our middle or halo pointing in the case of large objects. 
Following the efforts of \cite{martin-navarro_fornax_2019}, Martin-Navarro et al. (2021, in prep) now presents stellar-population parameter maps for the entire, which provides a strong basis to further investigate the link between PNe and their parent populations in a future paper.

\placefigRmagNPNe


\section{Conclusions}

Extending the work of \citet{spriggs_fornax_2020} to the full sample of ETGs targeted by the Fornax3D survey we have presented a catalogue of 1350 unique PNe sources across 21 different objects. The catalogue includes the PNe positions, their [\ion{O}{iii}] 5007 \AA{} flux and LOSV, with the PNe population of each galaxy being catalogued in standardised tables, available online: Tables\,\ref{tab:FCC083_center} to A.36 .

Using the luminosity function observed in each galaxy (limited to the central pointing for larger objects) we have have derived independent distance estimates by matching such PNLFs across their entire magnitude range thanks to a careful treatment of the PNe detection incompleteness. Using simulations we have checked the behaviour of our estimated distance errors and demonstrated that our PNLF distance estimates remain unbiased even when based on a low number of PNe. Furthermore, although the present data do not allow one to set meaningful constraints on the shape of the PNLF, as characterised by the generalised form introduced by \citet{longobardi_planetary_2013}, our methods are capable of detecting possible variations for PNe sample sizes not too far from the one we presently obtained.

Our PNLF distance estimate are generally consistent within the errors with distances derived using SBF \citep[][]{blakeslee_acs_2009,blakeslee_surface_2010} or through the combination of various methods other than PNLF \citep[][]{tully_cosmicflows-2_2013,tully_cosmicflows-3_2016}. In the case of SBF, on average our PNLF distance modulii closely match the SBF values, thus appearing to largely reduce previous tensions between these two methodologies.

With PNLF distances for 21 ETGs within the virial radius of the Fornax cluster we estimated a weighted average distance to Fornax of 19.86 $\pm$ 0.32 Mpc. This is consistent within the errors with the distance to this cluster derived from SBF values for all 43 ETGs observed \citet[][20 Mpc]{blakeslee_acs_2009} across the whole cluster and the one recommended by \citet[][19.1 Mpc]{de_grijs_clustering_2020-1} combining a variety of methods.  

We also investigated the spatial distribution of the ETGs of the F3D survey, in particular in relation to the BCG NGC~1399. Though we do not produce our own distance to this BCG, we review different cluster structure scenarios that arise from the NGC\,1399 SBF distances of \citet{blakeslee_acs_2009,blakeslee_surface_2010} and \citet{tonry_sbf_2001}, corresponding to 21.1 Mpc and 19.95 Mpc, respectively.
Assuming the \citet{blakeslee_surface_2010} value, NGC~1399 would be behind most of the objects we observe and $\sim$1 Mpc further than the average SBF distance provided by \citet[][]{blakeslee_acs_2009} for the Fornax cluster.
On the other hand, that the SBF distance to NGC\,1399 of \citet{tonry_sbf_2001} agrees better with \citet[][]{blakeslee_acs_2009}  distance to Fornax and the average PNLF distance provided by our objects inside the virial radius. This is consistent with the expectation that NGC~1399 is well centred on the potential well of the Fornax cluster, as traced also by the distribution of the hot intracluster medium.
Mass segregation should be facilitated in a relatively modest cluster such as Fornax, but we found no strong
evidence for it, independent of the assumed distance to NGC~1399 and whether we used SBF of PNLF distances. More accurate distance estimates will be needed to further explore the 3D structure of Fornax and other nearby clusters.

The present catalogue will be of great values for future investigation of the central PNe populations of ETGs, in particular in relation to the properties of their parent stellar population.
The MUSE data at hand will allow us to measure stellar age, metallicity, $\alpha$-element abundance \citep{martin-navarro_fornax_2019}, and the star-formation history \citep{pinna_fornax_2019-1, pinna_fornax_2019} of the stellar population in our sample galaxies in the same regions where our PNe are detected. Using this information, we explore in particular how the luminosity specific number of PNe $\alpha_{2.5}$ derived here will vary across the sample as a function of of these stellar population properties, in addition to measurements for the near and far-UV continuum, testing the results of previous comparisons \citep[e.g.][]{buzzoni_planetary_2006}. 

To conclude, we note the exciting prospects performing adaptive-optics (AO) observations with MUSE observations for the purpose of exploring the PNe population of nearby galaxies. 
Indeed, if present seeing-limited MUSE data allow one to detect PNe and measure PNLF distances out to a distance of 20 Mpc, finding over a hundred of PNe in the most massive objects and thus measuring distances within a 5\% accuracy \citep[see also S20][]{roth_towards_2021}, AO-assisted observations have the potential to at least double the reachable distance with the PNLF method. 
This can be understood considering that whereas the background noise in the MUSE observations is not set to vary dramatically for objects further out (owing to how surface brightness is conserved with distance), AO-observations delivering a 0.4\arcsec wide point-spread function instead of the 0.8\arcsec width (typical for good-seeing conditions at Paranal) will lead to a four-fold increase in the central [\ion{O}{iii}] 5007 \AA{} peak flux and spectral amplitude, thus compensating for the flux attenuation observed for PNe twice as further away. 
By the same token, AO-observations of objects as far away as Virgo or Fornax will allow one to reach $-1.5$ mag further down from the PNLF bright cut-off, making it possible to further test the universality of the PNLF and better understand its origin. 


\begin{acknowledgements}
    We would like to thank the referee for their constructive responses, which helped to improved the content and clarity of this manuscript. Based on observations collected at the European Southern Observatory under ESO pro-gramme 296.B-5054(A). TS thanks R. Jackson, S. Kaviraj and A. Bittner for their help and ideas that aided in the construction and presentation of this paper. This work was supported by Science and Technology Facilities Council [grant number ST/R504786/1]. GvdV acknowledges funding from the European Research Council (ERC) under the European Union's Horizon 2020 research and innovation programme under grant agreement No 724857 (Consolidator Grant ArcheoDyn). RMcD acknowledges financial support as a recipient of an Australian Research Council Future Fellowship (project number FT150100333). J.~F-B and I.~M-N acknowledge support through the RAVET project by the grant PID2019-107427GB-C32 from the Spanish Ministry of Science, Innovation and Universities (MCIU), and through the IAC project TRACES which is partially supported through the state budget and the regional budget of the Consejer\'ia de Econom\'ia, Industria, Comercio y Conocimiento of the Canary Islands Autonomous Community. FP acknowledges support from grant PID2019-107427GB-C32 from The Spanish Ministry of Science and Innovation. EMC acknowledges support by Padua University grants DOR1885254/18, DOR1935272/19, and DOR2013080/20 and by MIUR grant PRIN 2017 20173ML3WW\_001.
    This project made use of the following software packages: LMfit \citep{newville_lmfit_2014,newville_lmfitlmfit-py_2019}, Astropy, a community-developed core Python package for Astronomy \citep{robitaille_astropy_2013}, scipy \citep{eric_jones_and_travis_oliphant_and_pearu_peterson_and_et_al_scipy_2001}, Numpy \citep{van_der_walt_numpy_2011}, matplotlib \citep{hunter_matplotlib_2007} and Pandas \citep{mckinney_data_2010}.
\end{acknowledgements}


\bibliographystyle{aa}
\bibliography{Bibliography}

\onecolumn
\begin{appendix}
\section{PNe Catalogue}
\renewcommand{\arraystretch}{1.1}
\begin{longtable}{llllllllll}
    \caption{Central PNe of FCC~083} \\
    \hline
    Source ID & RA  & Dec & $\rm m_{5007}$ & $\sigma \rm m_{5007}$ & A/rN & LOSV & $\sigma$ LOSV & label & index \\
    & (J2000) & (J2000) & & & & $\rm km \, s^{-1}$& & & \\
    \hline \hline
    \label{tab:FCC083_center}
    \endfirsthead
    \hline
    \caption{FCC\,083 continued }\\

    Source ID & RA  & Dec & $\rm m_{5007}$ & $\sigma \rm m_{5007}$ & A/rN & LOSV & $\sigma$ LOSV & label & index \\
    & (J2000) & (J2000) & & & & $\rm km \, s^{-1}$& & & \\
    \hline \hline
    \endhead
    \hline
    \endfoot
    F3D J033035.64-345145.52 & 03h30m35.64s & -34d51m45.52s & 28.08 & 0.06 & 4.9 & 1356.5 & 4.4 & PN & C-1 \\
    F3D J033035.95-345142.83 & 03h30m35.95s & -34d51m42.83s & 28.39 & 0.07 & 4.2 & 1566.9 & 5.6 & PN & C-2 \\
    F3D J033032.68-345140.24 & 03h30m32.68s & -34d51m40.24s & 28.09 & 0.09 & 4.0 & 1535.9 & 7.6 & PN & C-3 \\
    F3D J033036.38-345138.89 & 03h30m36.38s & -34d51m38.89s & 28.0 & 0.05 & 6.1 & 1495.4 & 3.8 & PN & C-4 \\
    F3D J033034.30-345138.50 & 03h30m34.30s & -34d51m38.50s & 28.53 & 0.07 & 3.9 & 1671.2 & 5.7 & PN & C-5 \\
    F3D J033035.98-345136.52 & 03h30m35.98s & -34d51m36.52s & 28.25 & 0.06 & 5.1 & 1278.9 & 4.9 & PN & C-6 \\
    F3D J033036.85-345131.73 & 03h30m36.85s & -34d51m31.73s & 27.48 & 0.05 & 9.8 & 1358.2 & 2.2 & PN & C-7 \\
    F3D J033033.65-345131.27 & 03h30m33.65s & -34d51m31.27s & 27.44 & 0.04 & 11.0 & 1475.4 & 2.1 & PN & C-8 \\
    F3D J033033.54-345131.08 & 03h30m33.54s & -34d51m31.08s & 27.58 & 0.05 & 10.1 & 1550.3 & 2.4 & PN & C-9 \\
    F3D J033033.16-345130.64 & 03h30m33.16s & -34d51m30.64s & 27.0 & 0.04 & 17.6 & 1638.3 & 1.4 & PN & C-10 \\
    F3D J033034.57-345129.77 & 03h30m34.57s & -34d51m29.77s & 28.34 & 0.07 & 4.2 & 1495.5 & 5.7 & PN & C-11 \\
    F3D J033036.82-345128.18 & 03h30m36.82s & -34d51m28.18s & 27.41 & 0.05 & 9.9 & 1429.7 & 2.5 & PN & C-12 \\
    F3D J033035.26-345125.84 & 03h30m35.26s & -34d51m25.84s & 27.65 & 0.06 & 5.8 & 1480.7 & 4.0 & PN & C-13 \\
    F3D J033036.61-345125.54 & 03h30m36.61s & -34d51m25.54s & 27.84 & 0.05 & 6.8 & 1527.5 & 3.6 & PN & C-14 \\
    F3D J033034.60-345124.13 & 03h30m34.60s & -34d51m24.13s & 27.16 & 0.05 & 9.9 & 1571.7 & 2.3 & PN & C-15 \\
    F3D J033036.31-345123.58 & 03h30m36.31s & -34d51m23.58s & 28.27 & 0.07 & 4.3 & 1446.9 & 5.7 & PN & C-16 \\
    F3D J033035.57-345123.23 & 03h30m35.57s & -34d51m23.23s & 27.78 & 0.06 & 4.7 & 1484.7 & 5.1 & PN & C-17 \\
    F3D J033033.05-345122.68 & 03h30m33.05s & -34d51m22.68s & 28.35 & 0.06 & 5.1 & 1667.5 & 4.9 & PN & C-18 \\
    F3D J033035.27-345121.94 & 03h30m35.27s & -34d51m21.94s & 27.37 & 0.05 & 6.2 & 1433.6 & 3.9 & PN & C-19 \\
    F3D J033032.26-345121.63 & 03h30m32.26s & -34d51m21.63s & 28.44 & 0.06 & 4.7 & 1665.4 & 4.7 & PN & C-20 \\
    F3D J033035.26-345120.23 & 03h30m35.26s & -34d51m20.23s & 27.61 & 0.07 & 4.1 & 1484.5 & 5.6 & PN & C-21 \\
    F3D J033035.56-345119.97 & 03h30m35.56s & -34d51m19.97s & 27.79 & 0.06 & 4.3 & 1584.0 & 5.4 & PN & C-22 \\
    F3D J033034.16-345119.00 & 03h30m34.16s & -34d51m19.00s & 27.2 & 0.05 & 10.2 & 1419.3 & 2.3 & PN & C-23 \\
    F3D J033035.50-345117.22 & 03h30m35.50s & -34d51m17.22s & 27.84 & 0.07 & 3.7 & 1439.9 & 6.4 & PN & C-24 \\
    F3D J033036.50-345116.67 & 03h30m36.50s & -34d51m16.67s & 28.06 & 0.06 & 6.0 & 1459.8 & 4.0 & PN & C-25 \\
    F3D J033034.23-345115.74 & 03h30m34.23s & -34d51m15.74s & 27.71 & 0.06 & 5.8 & 1825.9 & 4.2 & PN & C-26 \\
    F3D J033035.83-345115.12 & 03h30m35.83s & -34d51m15.12s & 27.42 & 0.05 & 7.9 & 1605.9 & 3.1 & PN & C-27 \\
    F3D J033036.09-345114.61 & 03h30m36.09s & -34d51m14.61s & 27.93 & 0.06 & 5.6 & 1411.0 & 4.1 & PN & C-28 \\
    F3D J033035.76-345112.37 & 03h30m35.76s & -34d51m12.37s & 28.08 & 0.07 & 4.3 & 1488.9 & 5.6 & PN & C-29 \\
    F3D J033036.67-345112.81 & 03h30m36.67s & -34d51m12.81s & 26.79 & 0.04 & 19.1 & 1540.4 & 1.3 & PN & C-30 \\
    F3D J033033.53-345110.09 & 03h30m33.53s & -34d51m10.09s & 27.84 & 0.05 & 6.3 & 1689.8 & 3.8 & PN & C-31 \\
    F3D J033034.46-345109.19 & 03h30m34.46s & -34d51m09.19s & 27.21 & 0.05 & 6.8 & 1728.8 & 3.5 & PN & C-32 \\
    F3D J033034.00-345110.54 & 03h30m34.00s & -34d51m10.54s & 26.97 & 0.04 & 11.6 & 1469.3 & 2.2 & imp & C-33 \\
    F3D J033034.06-345109.93 & 03h30m34.06s & -34d51m09.93s & 27.35 & 0.05 & 10.5 & 1472.6 & 2.5 & PN & C-34 \\
    F3D J033034.18-345109.27 & 03h30m34.18s & -34d51m09.27s & 27.91 & 0.06 & 4.4 & 1497.8 & 5.4 & PN & C-35 \\
    F3D J033033.99-345108.96 & 03h30m33.99s & -34d51m08.96s & 27.72 & 0.05 & 5.9 & 1520.4 & 3.7 & PN & C-36 \\
    F3D J033034.72-345108.46 & 03h30m34.72s & -34d51m08.46s & 27.07 & 0.05 & 6.8 & 1684.3 & 3.6 & PN & C-37 \\
    F3D J033037.25-345107.61 & 03h30m37.25s & -34d51m07.61s & 28.58 & 0.07 & 3.6 & 1426.4 & 6.4 & PN & C-38 \\
    F3D J033034.41-345107.57 & 03h30m34.41s & -34d51m07.57s & 26.94 & 0.05 & 9.6 & 1659.1 & 2.5 & PN & C-39 \\
    F3D J033033.90-345107.03 & 03h30m33.90s & -34d51m07.03s & 28.07 & 0.06 & 4.6 & 1546.0 & 5.3 & PN & C-40 \\
    F3D J033033.13-345106.98 & 03h30m33.13s & -34d51m06.98s & 28.47 & 0.07 & 4.1 & 1655.4 & 5.8 & PN & C-41 \\
    F3D J033034.79-345106.58 & 03h30m34.79s & -34d51m06.58s & 27.64 & 0.06 & 4.5 & 1707.8 & 4.9 & PN & C-42 \\
    F3D J033034.16-345106.30 & 03h30m34.16s & -34d51m06.30s & 27.7 & 0.06 & 5.5 & 1603.0 & 4.4 & PN & C-43 \\
    F3D J033035.26-345106.09 & 03h30m35.26s & -34d51m06.09s & 27.73 & 0.06 & 5.3 & 1591.8 & 4.6 & PN & C-44 \\
    F3D J033034.90-345106.01 & 03h30m34.90s & -34d51m06.01s & 27.65 & 0.06 & 4.6 & 1664.5 & 4.9 & PN & C-45 \\
    F3D J033034.62-345105.85 & 03h30m34.62s & -34d51m05.85s & 27.58 & 0.06 & 5.2 & 1690.3 & 4.6 & PN & C-46 \\
    F3D J033033.70-345105.61 & 03h30m33.70s & -34d51m05.61s & 27.66 & 0.05 & 7.1 & 1450.5 & 3.2 & PN & C-47 \\
    F3D J033034.78-345103.77 & 03h30m34.78s & -34d51m03.77s & 27.9 & 0.07 & 4.3 & 1660.1 & 5.4 & PN & C-48 \\
    F3D J033035.43-345103.45 & 03h30m35.43s & -34d51m03.45s & 27.96 & 0.06 & 5.1 & 1538.5 & 4.5 & PN & C-49 \\
    F3D J033035.27-345103.52 & 03h30m35.27s & -34d51m03.52s & 27.74 & 0.06 & 5.7 & 1516.3 & 4.1 & PN & C-50 \\
    F3D J033037.40-345102.78 & 03h30m37.40s & -34d51m02.78s & 27.28 & 0.05 & 9.5 & 1440.3 & 2.4 & PN & C-51 \\
    F3D J033035.99-345102.53 & 03h30m35.99s & -34d51m02.53s & 28.52 & 0.07 & 3.7 & 1529.0 & 6.2 & PN & C-52 \\
    F3D J033034.23-345102.49 & 03h30m34.23s & -34d51m02.49s & 28.37 & 0.08 & 3.1 & 1430.3 & 7.5 & PN & C-53 \\
    F3D J033035.57-345058.71 & 03h30m35.57s & -34d50m58.71s & 27.52 & 0.05 & 9.1 & 1455.8 & 2.6 & PN & C-54 \\
    F3D J033034.82-345058.16 & 03h30m34.82s & -34d50m58.16s & 27.99 & 0.06 & 5.2 & 1483.0 & 4.5 & PN & C-55 \\
    F3D J033033.74-345056.72 & 03h30m33.74s & -34d50m56.72s & 28.54 & 0.08 & 3.4 & 1778.5 & 7.1 & PN & C-56 \\
    F3D J033034.37-345056.12 & 03h30m34.37s & -34d50m56.12s & 27.33 & 0.05 & 9.7 & 1690.7 & 2.5 & PN & C-57 \\
    F3D J033033.01-345054.73 & 03h30m33.01s & -34d50m54.73s & 28.74 & 0.08 & 3.3 & 1624.3 & 7.0 & PN & C-58 \\
    F3D J033033.72-345054.61 & 03h30m33.72s & -34d50m54.61s & 27.67 & 0.05 & 7.8 & 1642.6 & 3.2 & PN & C-59 \\
    F3D J033035.75-345054.00 & 03h30m35.75s & -34d50m54.00s & 28.31 & 0.06 & 5.0 & 1526.8 & 4.5 & PN & C-60 \\
    F3D J033034.42-345053.77 & 03h30m34.42s & -34d50m53.77s & 28.49 & 0.07 & 3.8 & 1649.7 & 6.3 & PN & C-61 \\
    F3D J033035.73-345049.49 & 03h30m35.73s & -34d50m49.49s & 28.08 & 0.05 & 6.1 & 1571.7 & 3.5 & PN & C-62 \\
    F3D J033032.92-345044.44 & 03h30m32.92s & -34d50m44.44s & 28.82 & 0.08 & 3.2 & 1562.4 & 7.3 & PN & C-63 \\
    \hline \\
    \multicolumn{10}{l}{\tablefoot{PNe catalogue for the FCC\,083 central region: 'Source ID' (using the IAU standard, with the F3D prefix), 'RA (J2000)' and 'Dec (J2000)', '$m_{5007}$' as magnitude in [\ion{O}{iii}] 5007\AA{}, 'A/rN' for the signal-to-residual noise, 'LOSV (km s$^{-1}$)' for the observed LOS velocity ($\rm km \ s^{-1}$) and label each source under 'Identifier' with the appropriate label: PNe, SNR, \ion{H}{ii}, OvLu (over-luminous), or Interl (interloper).}} \\
\end{longtable}

\begin{longtable}{llllllllll}
    \caption{Halo PNe of FCC~083} \\
    \hline
    Source ID & RA  & Dec & $\rm m_{5007}$ & $\sigma \rm m_{5007}$ & A/rN & LOSV & $\sigma$ LOSV & label & index \\
    & (J2000) & (J2000) & & & & $\rm km \, s^{-1}$& & & \\
    \hline \hline
    \label{tab:FCC083_halo}
    \endfirsthead
    \hline
    \caption{FCC\,083 continued }\\
    
    Source ID & RA  & Dec & $\rm m_{5007}$ & $\sigma \rm m_{5007}$ & A/rN & LOSV & $\sigma$ LOSV & label & index \\
    & (J2000) & (J2000) & & & & $\rm km \, s^{-1}$& & & \\
    \hline \hline
    \endhead
    \hline
    \endfoot
    F3D J033033.89-345106.66 & 03h30m33.89s & -34d51m06.66s & 28.27 & 0.17 & 3.6 & 1549.5 & 5.5 & PN & H-1 / C-42 \\
    F3D J033033.69-345105.63 & 03h30m33.69s & -34d51m05.63s & 27.99 & 0.17 & 5.3 & 1455.7 & 3.4 & PN & H-2 / C-49 \\
    F3D J033031.81-345101.88 & 03h30m31.81s & -34d51m01.88s & 28.32 & 0.17 & 5.3 & 1682.5 & 3.4 & PN & H-3 \\
    F3D J033032.25-345058.51 & 03h30m32.25s & -34d50m58.51s & 28.71 & 0.17 & 3.9 & 1618.3 & 4.9 & PN & H-4 \\
    F3D J033032.99-345054.71 & 03h30m32.99s & -34d50m54.71s & 28.75 & 0.17 & 3.7 & 1643.4 & 4.9 & PN & H-5 / C-61 \\
    F3D J033033.72-345054.66 & 03h30m33.72s & -34d50m54.66s & 27.81 & 0.17 & 8.2 & 1651.2 & 2.3 & PN & H-6 / C-62 \\
    F3D J033033.75-345056.63 & 03h30m33.75s & -34d50m56.63s & 28.44 & 0.17 & 4.4 & 1783.5 & 4.4 & PN & H-7 / C-59 \\
    F3D J033031.58-345041.48 & 03h30m31.58s & -34d50m41.48s & 27.98 & 0.17 & 8.0 & 1549.0 & 2.4 & PN & H-8 \\
    F3D J033029.66-345037.11 & 03h30m29.66s & -34d50m37.11s & 28.1 & 0.17 & 6.9 & 1695.5 & 2.7 & PN & H-9 \\
    F3D J033030.57-345032.60 & 03h30m30.57s & -34d50m32.60s & 28.66 & 0.17 & 5.1 & 1597.1 & 3.7 & PN & H-10 \\
    F3D J033033.95-345031.17 & 03h30m33.95s & -34d50m31.17s & 28.1 & 0.17 & 7.1 & 1703.1 & 2.6 & PN & H-11 \\
    F3D J033031.87-345029.64 & 03h30m31.87s & -34d50m29.64s & 28.98 & 0.17 & 3.5 & 1699.2 & 5.5 & PN & H-12 \\
    F3D J033032.94-345027.21 & 03h30m32.94s & -34d50m27.21s & 27.45 & 0.17 & 12.9 & 1469.7 & 1.5 & PN & H-13 \\
    F3D J033033.94-345026.15 & 03h30m33.94s & -34d50m26.15s & 28.96 & 0.17 & 3.6 & 1644.8 & 5.1 & PN & H-14 \\
    F3D J033034.22-345018.64 & 03h30m34.22s & -34d50m18.64s & 27.54 & 0.17 & 13.8 & 1628.5 & 1.4 & PN & H-15 \\
    F3D J033032.06-345016.72 & 03h30m32.06s & -34d50m16.72s & 28.4 & 0.17 & 6.1 & 1717.4 & 3.0 & PN & H-16 \\
    F3D J033033.11-345014.43 & 03h30m33.11s & -34d50m14.43s & 28.19 & 0.17 & 7.4 & 1629.6 & 2.6 & PN & H-17 \\
    F3D J033034.67-345009.13 & 03h30m34.67s & -34d50m09.13s & 28.81 & 0.17 & 3.3 & 1563.8 & 5.6 & PN & H-18 \\
    F3D J033032.84-345008.18 & 03h30m32.84s & -34d50m08.18s & 28.51 & 0.17 & 4.8 & 1653.0 & 4.0 & PN & H-19 \\
    \hline \\
    
    \multicolumn{10}{l}{\tablefoot{Halo: Source catalogue for FCC\,083: Source ID (using the IAU standard, with the F3D prefix), RA and Dec, magnitude in [\ion{O}{iii}] 5007\AA{}, signal-to-residual noise, observed LOS velocity ($\rm km s^{-1}$) and object label (PNe, SNR, \ion{H}{ii}, OvLu (over-luminous), or Interl (interloper). This also an example of the rest of the Fornax PNe catalogues presented here.}} \\
\end{longtable}

\twocolumn

\section{Simulations for optimising both $\mu_\mathrm{{PNLF}}$ and $c_{2}$}
\label{app:sim}

\placefigsimsDMctwo

By exploiting the entire range of the observed m$_\mathrm{5007}$ values thanks to our precised understanding of our completeness function, our PNLF fitting methodology has the potential to also constrain possible variation in the PNLF shape. 
This is illustrated by Fig.~\ref{fig:simulations_dM_c2} which shows how the distance modulus $\mu_\mathrm{{PNLF}}$ and the PNLF shape parameter $c_{2}$ of Eq.~\ref{eq:PNLF} are left free to vary during our PNLF fit. In this case we note a trend for over- and under-estimating $\mu_\mathrm{{PNLF}}$ and $c_{2}$, respectively, for small PNe samples. The observed scatter in $c_{2}$ is quite remarkable, even for large number of PNe, but tend nonetheless to decrease with increasing numbers. As expected, also the scatter in the $\mu_\mathrm{{PNLF}}$ increases compared to the fits where $c_{2}$ was held fix, albeit only by 15\%.
As in the case of our single-parameter simulations, the level of completeness plays an important role in driving the accuracy in the $\mu_\mathrm{{PNLF}}$ and $c_{2}$ parameters estimation. Figs.~\ref{fig:simulations_dM_c2_FCC193} and ~\ref{fig:simulations_dM_c2_FCC147} parallel to top and lower panel of Fig.~\ref{fig:simulations_dM_FCC193_FCC147} and show how  deeper observations, reaching further down from the apparent magnitude of the PNLF cutoff, would lead to more accurate $\mu_\mathrm{{PNLF}}$ and $c_{2}$ measurements.

Even though in some instances such as FCC193, where we detect over 150 PNe, we could already arrive at essentially unbiased $c_2$ constraints within a 0.2 error, overall across the sample and after correcting for biases we cannot discern a systematic deviation from the canonical $c_2=0.307$ value.
Finally, we note that aside from a small bias at low PNe number $\mu_\mathrm{{PNLF}}$ would also be well recovered while letting $c_{2}$ free to vary, albeit with a somewhat larger uncertainty, which demonstrates the robustness of our results when $c_{2}$ is held fix.

\placefigsimsDMctwoA
\placefigsimsDMctwoB

\newpage
\section{PNe maps and luminosity functions}
\onecolumn

\placeFOVplotsgrid

\placefigPNLFgrid

\end{appendix}

\end{document}